\documentclass[letterpaper,11pt]{article}
\pdfoutput=1
\usepackage{jheppub}
\usepackage{slashed}
\usepackage{graphicx}
\usepackage{subfigure} 
\usepackage{amsmath}
\usepackage{multirow}
\usepackage{comment}
\usepackage{hyperref}
\usepackage{epstopdf} 
\usepackage{mathtools}

\newcommand{\beq}{\begin{equation}}
\newcommand{\eeq}{\end{equation}}
\newcommand{\bea}{\begin{eqnarray}}
\newcommand{\eea}{\end{eqnarray}}
\newcommand{\ba}{\begin{array}}
\newcommand{\ea}{\end{array}}

\def\m1{M_1}
\def\m2{M_2}
\def\m3{M_3}

\def\ch10{\tilde \chi^0_1}

\def\tev{\,{\rm TeV}}
\def\gev{\,{\rm GeV}}

\def\Re{\,{\rm Re}}
\def\to{\rightarrow}

\newcommand{\lsim}{\mathrel{\mathop{\kern 0pt \rlap
  {\raise.2ex\hbox{$<$}}}
  \lower.9ex\hbox{\kern-.190em $\sim$}}}
\newcommand{\gsim}{\mathrel{\mathop{\kern 0pt \rlap
  {\raise.2ex\hbox{$>$}}}
  \lower.9ex\hbox{\kern-.190em $\sim$}}}

\definecolor{pink}{RGB}{255,105,180}

\def\fbi{\,{\rm fb}^{-1}}
\def\abi{\,{\rm ab}^{-1}}

\newcommand{\Edit}[1]{\textcolor{black} {#1}}

\title{Challenges and opportunities for heavy scalar searches in the $t\bar t$ channel at the LHC}

\author[a,b,c]{Marcela Carena}
\author[a]{and Zhen Liu}

\affiliation[a]{Theoretical Physics Department, Fermi National Accelerator Laboratory,\\PO Box 500, Batavia, IL 60510, U.S.A.}
\affiliation[b]{Enrico Fermi Institute, University of Chicago, Chicago, IL 60637, U.S.A.}
\affiliation[c]{Kavli Institute for Cosmological Physics, University of Chicago, Chicago, IL 60637, U.S.A.}

\emailAdd{carena@fnal.gov}
\emailAdd{zliu2@fnal.gov}

\abstract{Heavy scalar and pseudoscalar resonance searches through the $gg\to S\to t\bar t$ process are  challenging due to the peculiar behavior of the large interference effects with the standard model $t\bar t$ background. Such effects generate non-trivial lineshapes from additional relative phases between the signal and background amplitudes.
We provide the analytic expressions for the differential cross sections to understand the interference effects in the heavy scalar signal lineshapes. We extend our study to the case of CP-violation and  further consider the effect of bottom quarks in the production and decay processes. We also evaluate the contributions from additional particles to the gluon fusion production process, such as stops and vector-like quarks, that could lead to significant changes in the behavior of the signal lineshapes.
Taking into account the large interference effects, we perform lineshape searches at the LHC and discuss the importance of the systematic uncertainties and smearing effects. 
We present projected  sensitivities for  two LHC performance scenarios  to probe the $gg\to S \to t\bar t$ channel  in various models.}

\keywords{Higgs, Top, LHC}

\preprint{
\begin{flushright}
FERMILAB-PUB-16-262-T
\end{flushright}
}

\begin{document}
\maketitle
\flushbottom 

\section{Introduction} 
The discovery of the Higgs boson is a great triumph of the Standard Model (SM) and has opened a new era in particle physics. 
Being the first fundamental scalar particle ever observed, the existence of the Higgs boson substantiates  the questioning of basic  concepts in particle physics, such as  the hierarchy problem, the naturalness problem and  the true nature of neutrino masses. It also opens a window for possible connections to dark matter and the origin of the matter-anti matter asymmetry. Many of these conundrums can be (partially) addressed by some of the best motivated models currently under exploration, such as  supersymmetry (SUSY)~\cite{Flores:1982pr,Gunion:1984yn,Djouadi:2005gj}, composite Higgs models~\cite{Gripaios:2009pe,ArkaniHamed:1998nn,Randall:1999ee,Randall:1999vf}, extended gauge symmetries -- e.g.  grand unification theories~\cite{Georgi:1974sy}, and extended Higgs models such as two-Higgs doublet models (2HDM)~\cite{Branco:2011iw}. Most of these extensions of the SM require additional scalar bosons. This poses two basic questions: would there be additional scalar bosons at the electroweak scale?; how can they be sought at the LHC?

It will be challenging to discover a heavy  scalar at the LHC, in particular if  its couplings to electroweak gauge bosons are  small compared to its couplings to third generation fermions,  as  occurs in many extensions of the SM. Hence, we will focus on the decays of a heavy scalar into  the $t \bar t$ final state.
Hierarchical couplings of the heavy scalars to light quarks  lead to low production rates through tree-level processes and the $t \bar t$ final state has
large backgrounds from SM hadronic processes. 
In addition, as has been noticed in an earlier work~\cite{Dicus:1994bm} and recently discussed in a  related context~\cite{Barcelo:2010bm,Barger:2011pu,Bai:2014fkl,Jung:2015gta,Craig:2015jba}, the production of a  heavy Higgs boson through top-loop induced gluon-gluon-fusion with its subsequent decay into $t\bar t$ has a very large interference effect with the SM background. This large interference effect is further augmented by a non-trivial relative phase between the signal and the SM background amplitude, leading to a complex structure of the signal lineshape as a function of the $t\bar t$ invariant mass. Possible lineshapes vary from a pure bump to bump-dip, dip-bump and pure dip structures depending on the  different heavy scalar masses and the possible additional effects of other new particles in the loop.  Authors in Ref.~\cite{Barcelo:2010bm} studied the $gg\to S \to t\bar t$ channel in supersymmetric and Little Higgs models at  the LHC and considered a parton level analysis without taking into account the effects of smearing  on the reconstructed $t\bar t$ invariant mass and the  systematic uncertainties. 
More recent  works~\cite{Jung:2015gta,Craig:2015jba} have considered such effects on the signal  total rates. In many cases, however, it is necessary to go beyond a parametrization in terms of the total rate since this may overlook  cancelations between the peak and the dip structures after smearing. The previous studies triggered the interest of the community in further investigating  the discovery potential for heavy scalars in $t\bar t$ final states.

In this work  we concentrate on the  unique features of the  interference effects in the $gg\to S \to t\bar t$,  to investigate the feasibility of heavy scalar searches at the LHC. 
In  Sec.~\ref{sec:minimal} we provide a detailed study  in the baseline model with only top-quark loops contributing to the production vertex. In Sec.~\ref{sec:models} we  expand our study to consider 
 additional effects in extensions  of the baseline model. In particular, we investigate  the effects of two nearly degenerate Higgs bosons, as in 2HDMs, both for CP eigenstates and in the case of CP-violation in the Higgs sector.
Moreover, in Sec.~\ref{sec:models} we also study the effects of additional particles, beyond the top quark, contributing to the production vertex. These include effects from bottom quarks that become relevant in a Type II 2HDMs with sizable ratio of the two Higgs vacuum expectation values ($\tan \beta$), heavy colored particles such as stops  in SUSY models and Vector-Like Quarks (VLQs) that naturally appear in composite Higgs scenarios. Also in Sec.~\ref{sec:models}, we present a study to highlight the relevance of  interference effects in the $t\bar t$ final state for a prospective 750 GeV scalar that could account for the excess in the di-photon channel observed at the LHC experiments~\cite{ATLAS-CONF-2015-081,CMS-PAS-EXO-15-004}.
In Sec.~\ref{sec:LHC} we perform detailed collider analyses to investigate the  reach  at the 13 TeV LHC in the search for $t\bar t$ resonances in the presence of large interference effects,   emphasizing the importance of smearing effects and systematic uncertainties. We propose a  a lineshape search at the LHC, taking into account both the excess and deficit as part of the signal for two LHC  performance scenarios. We demonstrate the physics potential of this new search in examples of the baseline model and a 2HDM, including the possibility of  nearly degenerate bosons with and without  CP-violation. We reserve  Sec.~\ref{sec:conclusion} to
summarize, and briefly discuss possible future directions for  scalar resonance searches in the $t\bar t$ final state.

\section{The baseline model:  a single resonance from top quark loops}
\label{sec:minimal}

The importance of the  $gg\to S \to t\bar t$ channel  well justifies a  comprehensive study of all the subtleties inherent to this signal, in particular the interference effects.   In this section, we analyze the baseline model that only takes into account  the top quark  contribution to  the gluon fusion production process and considers   the effects of one additional single heavy scalar at a time. 

\subsection{The interference effects anatomized}

In the following we focus on heavy neutral scalars that are not charged under the standard model gauge groups after electroweak symmetry breaking (color and electrically neutral).
In many beyond the standard model extensions, the additional scalar couplings to fermions are hierarchical, according to the fermion masses. We adopt such simple set-up for the heavy scalar couplings to the SM fermion sector, which renders the production rate from $q\bar q$ fusion process small and, at the same time, makes the gluon fusion process the dominant production mode.

In addition, for example in CP-conserving 2HDMs, one can study the effects of the  CP-even or CP-odd heavy Higgs bosons  produced via gluon fusion and decaying into top pairs, that destructively interfere with the SM $ t\bar t$ background.  The baseline model considers only  top quark contributions to the gluon fusion production process, and  this is appropriate, e.g. for a Type II 2HDM at low $\tan\beta$, but could be otherwise for moderate to large  values $\tan\beta$, for which the bottom loop becomes relevant.  Moreover, generic 2HDMs  usually assume no additional relevant colored particles other than the standard model fermions and gauge bosons. 

The above consideration motivates us to write down the following interaction terms of a general Lagrangian for a heavy scalar after electroweak symmetry breaking:
\beq
\mathcal{L}^{\rm Yukawa}\supset \frac {y_{i}^s}{\sqrt 2} \bar t t S + i \frac {\tilde y_{i}^s} {\sqrt 2} \bar t \gamma_5 t S~~.
\eeq
The top-loop in the triangle diagram induces an effective gluon-gluon-scalar vertex. This can  also be expressed by  effective  interactions,
\beq
\mathcal{L}^{\rm Yukawa}\xRightarrow[{\rm 
}]{\rm loop-induced} -\frac {1} {4} g_{_Sgg}(\hat s) G_{\mu\nu} G^{\mu \nu} S -  \frac i 2 \tilde g_{_Sgg}(\hat s) \tilde G_{\mu \nu} G^{\mu \nu} S,
\eeq
where $\tilde G_{\mu\nu}\equiv \frac 1 2 \epsilon_{\mu\nu\alpha\beta}G^{\alpha\beta}$. This expression is given in terms of  form factors of the loop-induced vertices that  explicitly depend on $\hat s$.

We concentrate on the flavor diagonal Yukawa-like couplings between the heavy scalar $S$ and the chiral fermion fields, since only these diagonal terms  contribute to the loop-induced $Sgg$ couplings. The $Sgg$ couplings depend on the Yukawa interactions and corresponding fermion masses,
\beq
g_{_Sgg}(\hat s)=\frac {\alpha_s} {2\sqrt 2\pi} \frac {y_t^s} {m_t} I_{\frac 1 2}(\tau_t),~~~~\tilde g_{_Sgg}(\hat s)=\frac {\alpha_s} {2\sqrt 2\pi} \frac {\tilde y_t^s} {m_t} \tilde I_{\frac 1 2}(\tau_t),
\label{eq:gSgg}
\eeq
where $I_{\frac 1 2}(\tau_t)$ and $\tilde I_{\frac 1 2}(\tau_t)$ are the corresponding loop-functions and\footnote{Alternatively, these more conventional loop-functions can be written in terms of kinematic variable $\beta$ as shown and discussed in the Appendix. The kinematic factor $\beta$ of the final state top quarks is defined as $\sqrt {1-\frac {4 m_t^2} {\hat s}}$. This kinematic factor $\beta$  is unrelated to $\tan\beta\equiv v_2/v_1$, the ratio of the vacuum expectation values of the two Higgs doublets, to be used later on in this paper. }
\bea
\tau_t=\frac {\hat s} {4 m_t^2},&&\ f(\tau) =\left\{
\begin{array}{lcl}
\arcsin^2(\sqrt\tau)&~~~~~~~~& {\rm for\ } \tau\leq 1,\\ \nonumber
-\frac 1 4 \left(\log\frac {1+\sqrt {1-1/\tau}} {1-\sqrt {1-1/\tau}}-i \pi\right)^2\ &\ &{\rm for\ } \tau> 1
\end{array}\right.\\
\label{eq:ggh}
I_{1/2}(\tau)&&= \frac 1 {\tau^2} (\tau + (\tau-1)f(\tau)),~~~\tilde I_{1/2}(\tau)= \frac {f(\tau)} {\tau}. 
\eea
In the above, $y_t^s$ is  the Yukawa coupling of the heavy scalar to the top quark, whose mass is denoted by  $m_t$.

\begin{figure}[t]
\centering
\centering
\includegraphics[width=0.55\textwidth]{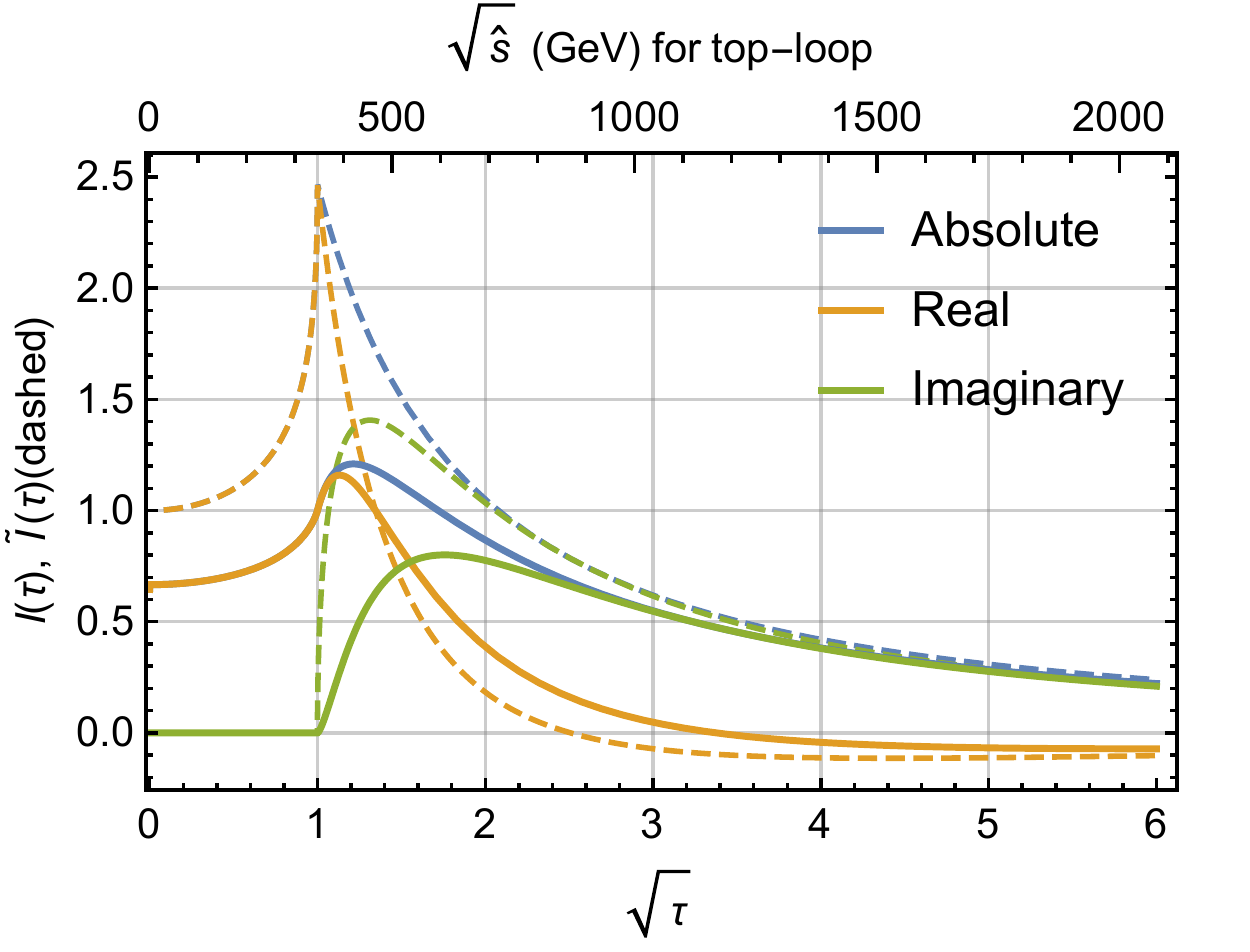}
\caption[]{Loop functions of the fermion induced gluon-gluon-scalar vertex as a function of the 
parameter $\sqrt{\tau}\equiv\sqrt{\hat s}/(2m_f)$, for a CP-even scalar (solid line) and a CP-odd scalar (dashed lines), respectively.
The blue, yellow and green lines correspond to the absolute value, real component and imaginary component of the loop functions, respectively.  For convenience, we  show the corresponding center of mass energy $\sqrt{\hat s}$ in units of GeV for the case of a top quark loop on the upper edge of the figure.  
}\label{fig:loopggh}
\end{figure}

 In Fig.~\ref{fig:loopggh} we show the numerical values of the loop functions. For convenience, we also label the upper edge of the $x$-axis in the figure with the corresponding center of mass energy $\sqrt {\hat s}$  for the case of a top quark loop. 
Although we are writing these effective form factors considering only the top quark in the loop, they can be generalized for other fermions by replacing $y_t^s$ and $m_t$ by  $y_f^s$ and $m_f$ in Eqs.~\ref{eq:gSgg} and ~\ref{eq:ggh}. 
In Fig.~\ref{fig:loopggh} one observes a clear  jump in the behavior of the values of the loop functions when $\sqrt{\hat s}  \simeq  2m_f$, associated with the threshold effect from the on-shell top pairs. For the region far below the threshold, $\tau_f\equiv\sqrt{\hat s}/(2m_f)\ll 1$, the function is real and very slowly varying (almost  constant).  

A direct application of the loop function behavior for $\tau \ll 1$ is the derivation of the 
  heavy (chiral) fermion decoupling theorem for the SM Higgs.  For any heavy chiral fermion that  acquires mass through its coupling to the SM Higgs, the ratio of the Yukawa coupling to the mass depends on $v\simeq246~\gev$ -- the Vacuum Expectation Value (VEV) of the SM Higgs -- $\frac {y_{f}} {\sqrt 2 m_{f}}=\frac 1 v$.
Considering the case of the SM Higgs, we observe that each generation of heavy chiral fermions, will contribute to the Higgs-gluon coupling, Eq.~(\ref{eq:gSgg}), like \footnote{For completeness, the expansion for a pseudoscalar at low $\tau_f$ follows,
\beq
\tilde g_{_agg}(\hat s)=\frac {\alpha_s} {2\pi v}   \tilde I_{\frac 1 2}(\tau_f)\approx \frac {\alpha_s} {2\pi v} (1+ \frac 1 3 \tau_f + \frac 8 {45} \tau_f^2 + O(\tau_f^3)).
\eeq}

\beq
g_{_hgg}(\hat s)=\frac {\alpha_s} {2\pi v}   I_{\frac 1 2}({\tau_f})\approx \frac {\alpha_s} {3\pi v} (1+\frac 7 {30} \tau_f + \frac 2 {21} \tau_f^2 + O(\tau_f^3)).
\eeq
Neglecting corrections of higher order in 
$\tau_f$, each chiral fermion generation contributes the same amount $\frac {\alpha_s} {3\pi v}$ to the SM Higgs-gluon  coupling. 

Just after crossing the fermion pair  threshold,  $\tau \geq 1$, the imaginary part of the loop functions (as shown in Fig.~\ref{fig:loopggh}) rises quickly, and
then decreases  slowly for increasing values of $\tau$.
 The real part, instead,  decreases monotonically slightly above the fermion pair threshold and flips its sign for sufficiently large $\tau$. This implies that the phase of the loop function rapidly grows after crossing the threshold and remains large (of order $\pi/2$) for any value of $\sqrt{\tau} \gsim 2$. This special behavior drives the unconventional BSM phenomenology discuss in this paper and we will come back to this in more detail later on.

\begin{figure}[t]
\centering
\centering
\includegraphics[width=0.55\textwidth]{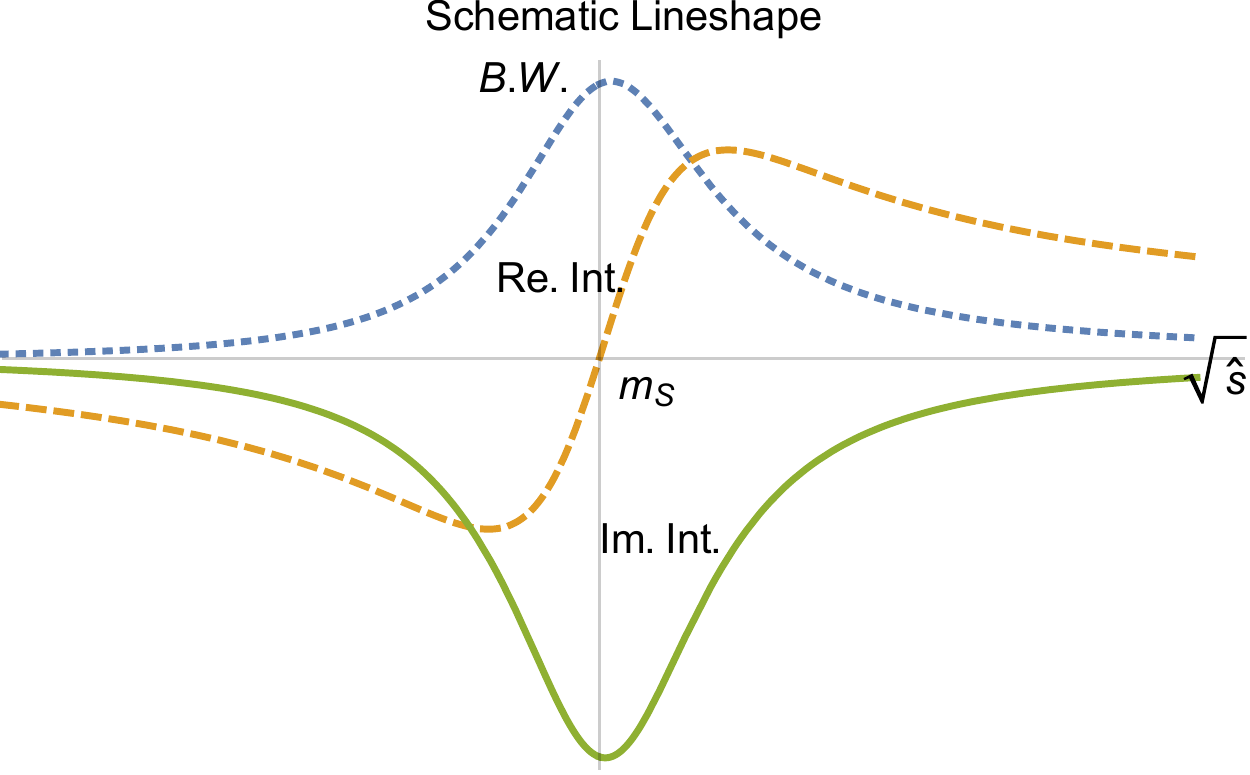}
\caption[]{The schematic lineshapes of \Edit{the components of the signal, namely,} the Breit-Wigner resonance (blue, dotted line), \Edit{the} interference with background proportional to the real component of the propagator (orange, dashed line), and \Edit{the} interference with background proportional to the imaginary component of the propagator (green, solid line) \Edit{as a function of the center of mass energy $\sqrt{\hat s}$}.  
}\label{fig:propagator}
\end{figure}

In Fig.~\ref{fig:propagator} we illustrate three components of the lineshapes for the scalar signal, namely the Breit-Wigner piece (blue, dotted line), the interference piece proportional to the real component of the scalar propagator (orange, dashed line) and the interference piece proportional to the imaginary component of the scalar propagator (green, solid line). 
To understand the interference effects in a more explicit way, we can parameterize the scalar propagator,  conveniently normalized by a factor $\hat s$, as:
\bea
\frac {\hat s} {(\hat s - m_S^2)+i\Gamma_S m_S}&&\approx \frac {m_S} {\Gamma_S} \frac {2\Delta-i} { 4\Delta^2+1}
\label{eq:propagator}
\\
{\rm with~~~}\Delta&&\equiv \frac {\hat s -m_S^2} {2 m_S \Gamma_S}\approx \frac {\sqrt{\hat s}-m_S} {\Gamma_S}~{\rm for~} \frac {\hat s} {m_S^2}-1\ll 1. \nonumber
\eea
In the above,  $\Delta$ basically parameterizes the deviation of the center mass energy $\sqrt{\hat s}$ from the scalar mass $m_S$ in units of the scalar width $\Gamma_S$. 
The denominator of the propagator in the above equation is positive definite and increases as the deviation $|\Delta|$ increases. This provides an arc-type profile  around values of $\sqrt{\hat s}$ close to the scalar mass, since   the denominator is minimized for $\Delta = 0$. After  squaring and with small modifications from the numerator, this generates the Breit-Wigner lineshape as shown by the blue, dotted line in Fig.~\ref{fig:propagator}.
The real part of the numerator, $2 \Delta$, flips its sign when crossing the scalar mass pole, while  the imaginary part of the numerator remains negative.  Multiplying the numerator by the arc-type profile  of the denominator, this leads in Fig.~\ref{fig:propagator} to the lineshapes schematically shown as a dip-bump (orange, dashed line) and a dip (green, solid line) for the real and imaginary parts, respectively. The contributions to the signal lineshapes from the real and imaginary parts of the propagator can be further modified by the detailed dynamics of the underlying physics. In particular if the overall sign is flipped, these lineshapes will change into a bump-dip structure or a pure bump, instead.

In  standard analyses of  tree-level BSM particle resonant production and decays, the BSM amplitudes  are real up to an imaginary contribution from the propagator. Given that the SM backgrounds are real as well, the only part of the propagator that survives is the real one. Moreover,  the real part of the propagator is odd around the resonance mass -- as illustrated by the orange, dashed line in Fig.~\ref{fig:propagator} -- 
implying  that the interference effect  does not contribute to the total  signal rate.
If the BSM amplitude  acquires an imaginary piece  in addition to the imaginary part of the propagator, e.g., from loop functions, a new interference piece will emerge. This new interference contribution is even around the resonance mass --as illustrated by the green, solid line  in Fig.~\ref{fig:propagator} -- and {\it does change} the  total signal rate. \Edit{The relevance of this interference contribution does not depend on the precise magnitude of the width of the resonance.} 

The signal amplitudes for the specific case of  $gg\to S\to t \bar t$,  both for a  CP-even and  CP-odd heavy scalar $S$, are proportional to:\footnote{For  simplicity of notation, from here on we drop the superscript $S$ from  the top Yukawa couplings to heavy scalars.}
\beq
\mathcal{A}^{\rm even}\propto y_t g_{_Sgg}=y_t^{2} I_{\frac 1 2}(\tau_t), \;\;\;\;\;\;\; \mathcal{A}^{\rm odd}\propto \tilde y_t \tilde g_{_Sgg}= \tilde y_t^{2} \tilde I_{\frac 1 2}(\tau_t),
\label{eq:minimalamp}
\eeq
where we have omitted the scalar propagator, color factor and strong coupling constant dependence for simplicity.
We can then define 
the phase of the resonant signal amplitude in terms of the reduced amplitude $\mathcal{\bar A}$  and the normalized propagator as,\footnote{The background amplitude is defined to be positive, as one can always rotate the phase of the signal and background amplitudes simultaneously without changing the physical results. This uniquely fixes the definition of the  phase $\theta_{\mathcal {\bar A}}$.} 
\beq
\mathcal A =  \frac {\hat s} {\hat s-m_S^2+i \Gamma_S m_S} | \mathcal{\bar A} | e^{i \theta_{\mathcal{\bar A}}}, \;\;\;\;\;\;\;\;{\rm with } \;\;\;\theta_{\mathcal {\bar A}}\equiv \arg (\mathcal {\bar A}).
\eeq
When $\theta_{\mathcal{\bar A}}$ is 0 (or $\pi$), only  the real part of the propagator contributes to the interference term yielding a dip-bump (or bump-dip) structure.
This is the standard case mostly studied in the literature, that does not affect  the total signal rate.
When $\theta_{\mathcal{\bar A}}$ is $\pi/2$ (or $3\pi/2$), instead, only  the imaginary part of the propagator contributes to the interference term, yielding a pure dip (or a pure bump) structure that can significantly change the total signal rate.

\begin{figure}[t]
\centering
\centering
\includegraphics[width=0.55\textwidth]{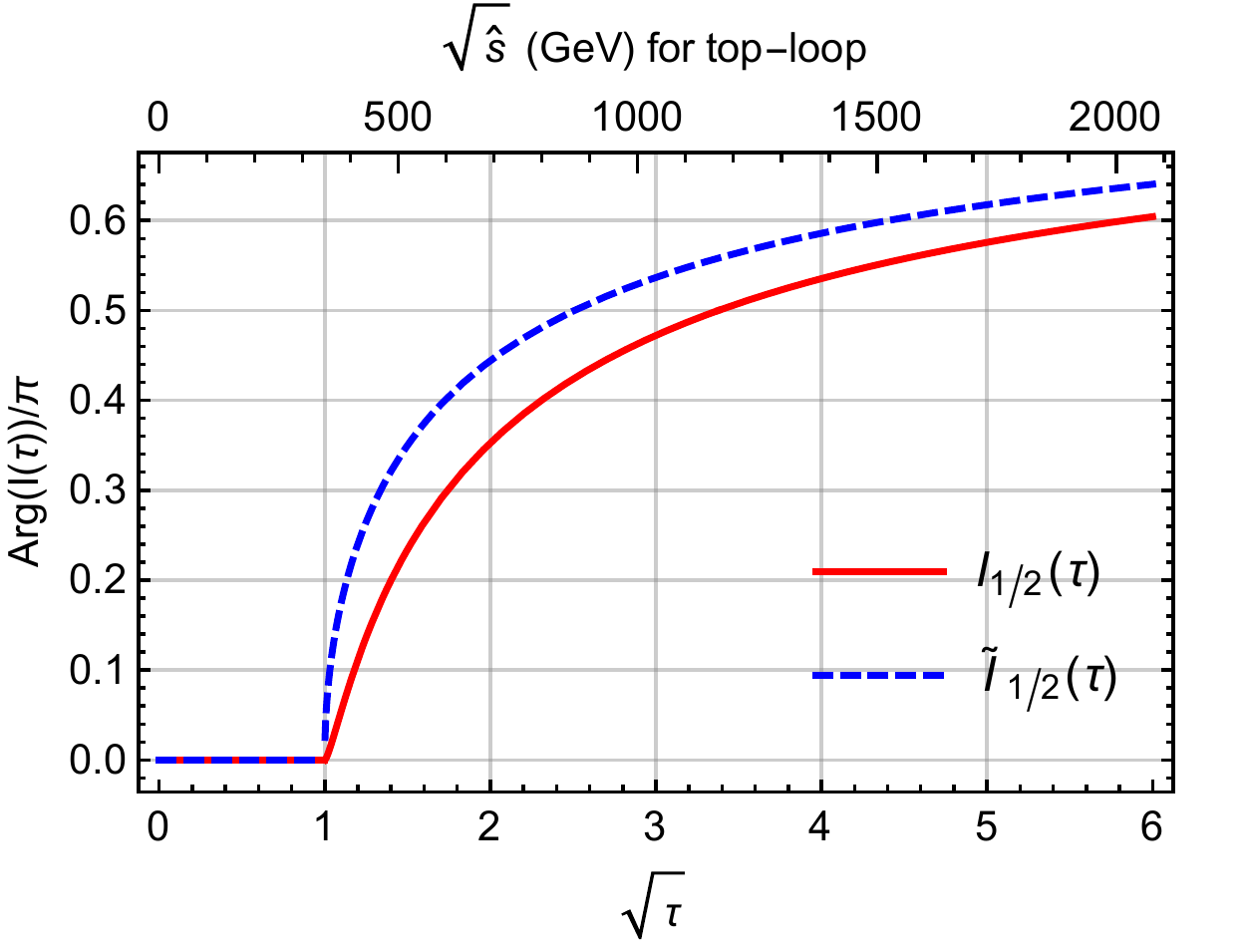}
\caption[]{The phase (argument)  of the loop functions \Edit{in units of $\pi$} as a function of $\tau$ for a scalar (red line) and a pseudoscalar (blue, dashed line), respectively.  We label the upper edge  of the x-axis with the  corresponding center of mass energy $\sqrt{\hat s}$ in GeV for the case of a top quark loop. 
}\label{fig:loopgghphase}
\end{figure}

For the process $gg\to S\to t\bar t$ in consideration, the loop functions ($I(\tau_t)$ and $\tilde I(\tau_t)$) are the only sources of the additional phase $\theta_{\mathcal{\bar A}}$ ( $\theta_{\mathcal{\bar A}}=\arg I(\tau)$ or  $\theta_{\mathcal{\bar A}}=\arg{\tilde I(\tau)}$). We show in Fig.~\ref{fig:loopgghphase} the phase of the fermion loop functions  both for the scalar (red line) and pseudoscalar (blue, dashed line) cases. These phases follow the numerical values of the loop functions discussed in Fig.~\ref{fig:loopggh}, and they  will be useful in analyzing the signal lineshapes later on.  Similarly to Fig.~\ref{fig:loopggh}, we  label the upper edge  of the x-axis with the  corresponding center of mass energy $\sqrt{\hat s}$ in GeV for the case of a top quark loop. Throughout the whole $\tau$ range, the phase for the pseudoscalar is larger than that of the scalar. A phase of  $\pi/4$ occurs for a scalar around 550~GeV and for a pseudoscalar around 450~GeV, respectively. In this case the real and imaginary parts of the loop function are the same and as a result both interference terms are comparable. The phase reaches $\pi/2$ for a scalar around 1.2~TeV and  a pseudoscalar around 850~GeV, respectively. In this case only the interference term proportional to the imaginary part of the propagator survives, highlighting  the relevance of the pure dip interference structure.

It is worth mentioning that the more  complex interference behavior presented above is well established in  hadronic physics~\cite{Pumplin:1970kp,Bauer:1970hk,Basdevant:1977ya,Basdevant:1978tx}, and it may be useful to further  investigate the treatment of these lineshape structures 
in the hadronic physics studies.
 
\subsection{The heavy scalar lineshapes}
\label{sec:baseline_lineshapes}
 
After analyzing the generic features of different lineshape contributions in the previous section, we now concentrate on the baseline model.
The background amplitude from QCD $t\bar t$ production is much larger in magnitude than the baseline signal amplitude. As a result, the interference terms often are larger in size and more important than the BSM Breit-Wigner term.  
Furthermore, as discussed in the previous section, the phase generated by the loop function grows rapidly after crossing the threshold. This phase enhances  the interference contribution proportional to the imaginary part of the scalar propagator, rendering it   much larger than that proportional to the real part. Although the sign of the interference is not fixed in the general case, the baseline model ensures this interference contribution to be destructive. 
Three factors are important here. Firstly, the loop function rapidly becomes (positive) imaginary after crossing the $t\bar t$ threshold. Secondly, the propagators near the resonance have a constant (negative) imaginary part. Thirdly, there is an overall minus sign from the fermion-loop in the signal amplitude relative to the background. These three factors lead to the overall negative sign of the signal amplitude near the resonance relative  to the background amplitude, generating the destructive interference. This feature makes the search for heavy Higgs bosons  in this channel rather unconventional and challenging. 

Specifically, the partonic cross sections for the signals for the CP-even scalars read,
\bea
\hat \sigma^{\rm even}_{\rm BSM}(\hat s; y_t) (gg\to S\to t\bar t)&&=\hat \sigma^{\rm even}_{\rm B.W.}(\hat s; y_t)+\hat \sigma^{\rm even}_{\rm Int.}(\hat s; y_t) \nonumber \\
\frac {d\hat \sigma^{\rm even}_{\rm B.W.}(\hat s; y_t)} {d z} &&=\frac {3\alpha_s^2 \hat s^2} {4096 \pi^3 v^2} \beta^3 \left| \frac {y_t^2 I_{\frac 1 2}(\tau_t) } {\hat s - m_S^2 + i m_S \Gamma_S (\hat s)} \right|^2\nonumber \\ 
\frac {d \hat \sigma^{\rm even}_{\rm Int.}(\hat s; y_t)} {dz} &&= - \frac {\alpha_s^2} {64\pi} \frac {\beta^3} {1-\beta^2 z^2}  \Re\left[ \frac {y_t^2 I_{\frac 1 2}(\tau_t)} {\hat s - m_S^2 + i m_S \Gamma_S (\hat s)} \right],\label{eq:csparton_even}
\eea
while for the CP-odd scalars are,
\bea
\hat \sigma^{\rm odd}_{\rm BSM}(\hat s;\tilde y_t) (gg\to S\to t\bar t)&&=\hat \sigma^{\rm odd}_{\rm B.W.}(\hat s; \tilde y_t)+\hat \sigma^{\rm odd}_{\rm Int.}(\hat s; \tilde y_t) \nonumber \\
\frac {d \hat \sigma^{\rm odd}_{\rm B.W.}(\hat s; \tilde y_t)} {dz} &&=\frac {3\alpha_s^2 \hat s^2} {4096 \pi^3 v^2} \beta \left| \frac {\tilde y_t^2 \tilde I_{\frac 1 2}(\tau_t)} {\hat s - m_S^2 + i m_S \Gamma_S (\hat s)} \right|^2\nonumber \\ 
\frac {d \hat \sigma^{\rm odd}_{\rm Int.}(\hat s; \tilde y_t)} {dz} &&= - \frac {\alpha_s^2} {64\pi} \frac {\beta} {1-\beta^2 z^2} \Re\left[ \frac {\tilde y_t^2 \tilde I_{\frac 1 2}(\tau_t)} {\hat s - m_S^2 + i m_S \Gamma_S (\hat s)} \right],
\label{eq:csparton_odd}
\eea
where $\Gamma_S(\hat s)$ is the energy dependent width for the scalar, detailed in the Appendix in Eq.~\ref{eq:width}, and the variable $z$ is the cosine of the scattering angle between an incoming parton and the top quark. \Edit{The leading-order expression for the background partonic cross sections from $gg\to t\bar t$ and $q\bar q \to t\bar t$ are outlined in the Appendix in Eq.~\ref{eq:bkgparton}.
} For collider analyses with detector acceptance, not the full phase space of $z$ can be used equally, we thus provide the differential distribution. However, as the top quark is not very boosted and even forward ones with $z=\pm 1$ can be detected after they decay, we integrate $z$ over the range of $[-1,1]$ for our simplified analysis throughout this paper. In all expressions the factors $y_t^2 I(\tau_t)$ and $\tilde y_t^2 \tilde I(\tau_t)$ are basically the dynamical part of the reduced amplitudes $ A^{\rm even, odd}$ in Eq.~\ref{eq:minimalamp}, written here explicitly for direct connection with the phase $\theta_{\mathcal{\bar A}}$ from the loop functions. For generalized cases with additional contributions, the reduced amplitudes are more useful. The superscripts even and odd refer to the CP properties of the heavy scalar. 

For a single heavy scalar being non-CP eigenstate, e.g., coupling to top quarks as $y_t + i\tilde y_t$, the resulting parton level cross sections are given by,
\bea
\hat \sigma^{CPV}_{\rm BSM}(\hat s;y_t, \tilde y_t) (gg\to S\to t\bar t)=&&\hat \sigma^{\rm CPV}_{\rm B.W.}(\hat s; y_t,\tilde y_t)+\hat \sigma^{\rm CPV}_{\rm Int.}(\hat s; y_t,\tilde y_t) \label{eq:CPVparton} \\ 
\frac {d \hat \sigma^{\rm CPV}_{\rm B.W.}(\hat s; y_t,\tilde y_t)} {dz}=&&\frac {3\alpha_s^2 \hat s^2} {4096 \pi^3 v^2} \beta (y_t^2 |I_{\frac 1 2}(\tau_t)|^2+\tilde y_t^2 |\tilde I_{\frac 1 2}(\tau_t)|^2)(\beta^2 y_t^2+\tilde y_t^2)\nonumber\\
&&\left| \frac {1} {\hat s - m_S^2 + i m_S \Gamma_S (\hat s)}\right|^2\nonumber\\
\hat \sigma^{\rm CPV}_{\rm Int.}(\hat s; y_t,\tilde y_t)=&&\sigma^{\rm even}_{\rm Int.}(\hat s;y_t) (gg\to S\to t\bar t)+\sigma^{\rm odd}_{\rm Int.}(\hat s;\tilde y_t) (gg\to S\to t\bar t),\nonumber
\eea
where the even and odd  interference pieces follow Eqs.~\ref{eq:csparton_even} and ~\ref{eq:csparton_odd}, respectively. 
\Edit{The Breit-Wigner component receives a contribution proportional to $y^2_t \tilde y^2_t$ as a result of CPV.}
With CP-violation~\cite{Branco:2011iw,Carena:2000yi,Carena:2015uoe} in the heavy scalar-top sector, the coupling between the scalar $S$ and the top quarks can be expressed as,
\beq
y_t+i\tilde y_t= |Y_t|(\cos\theta_{\rm CP}+i\sin\theta_{\rm CP}).
\eeq
The maximal CP-violation (CPV$_{\rm max}$) in this sector is for $\theta_{\rm CP}=\pi/4$. 

\begin{figure}[t]
\subfigure{
\centering
\includegraphics[width=0.5\textwidth]{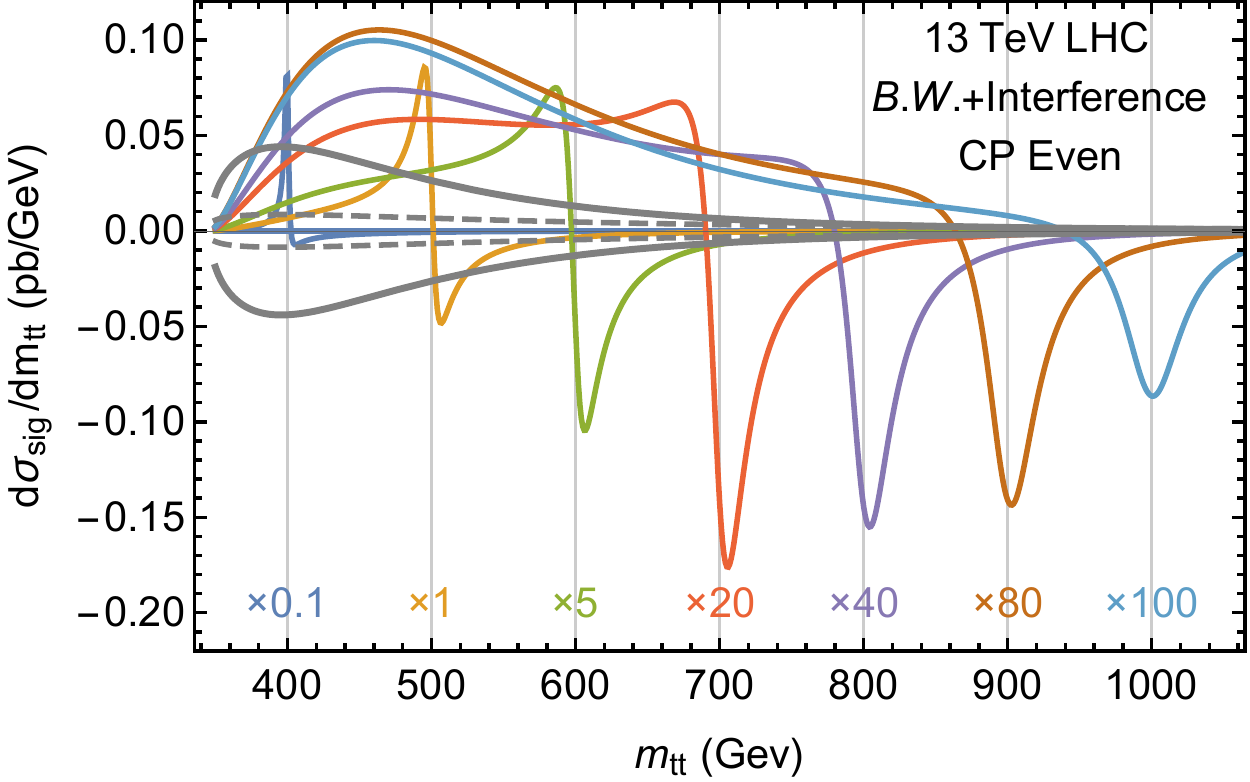} }
\subfigure{
\centering
\includegraphics[width=0.49\textwidth]{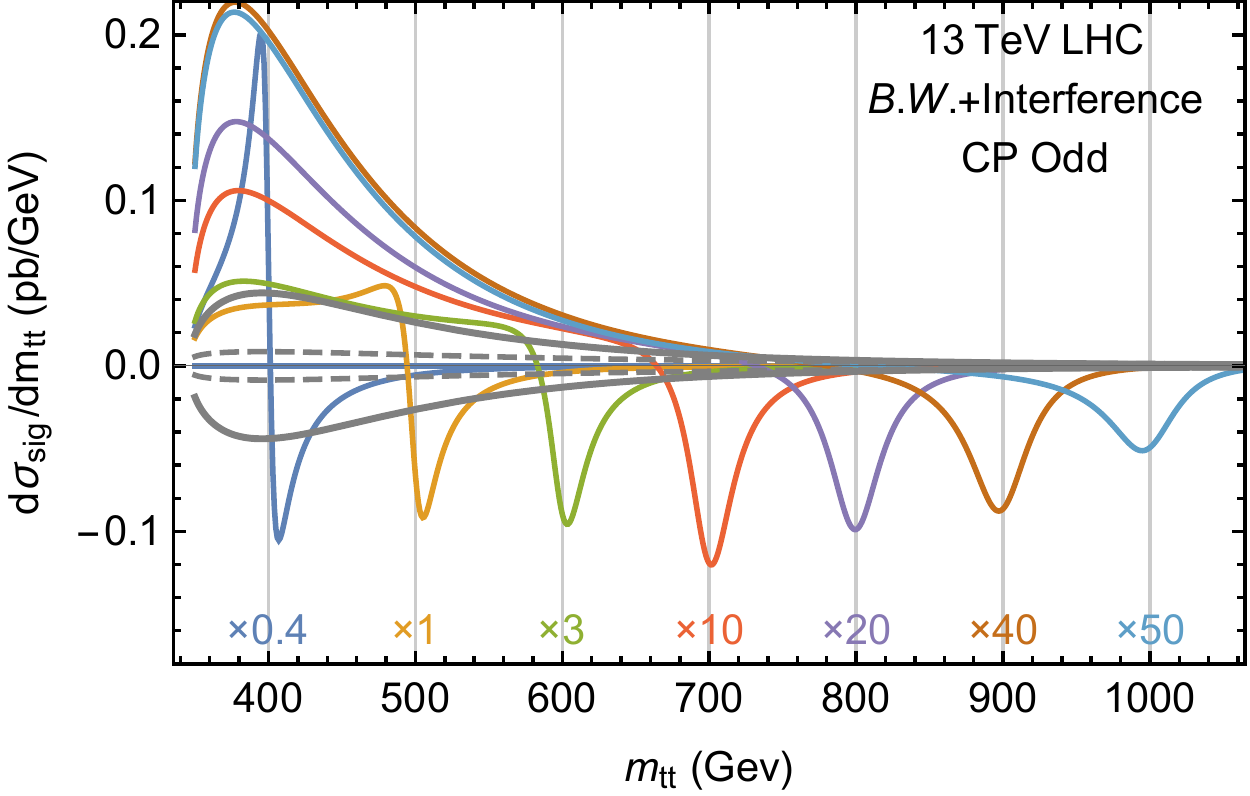} }
\caption[]{Differential cross sections of heavy CP-even (left panel) and CP-odd (right panel) scalar signals as a function of the $t\bar t$ invariant mass at the 13 TeV LHC. Each signal is for a specific value of the scalar mass and includes  both the contribution from the Breit-Wigner lineshape as well as the contributions from the signal-background interference. The vertical grid lines indicate the location of the each of the heavy scalar  masses, which range from  400 to 1000 GeV in  steps of 100 GeV. The signal lineshapes are multiplied by factors indicated in the lower part of the figure to  render the signal lineshapes visible on a same  scale. The solid and dashed gray lines represent a systematic uncertainty of the background at the  $\pm 2\%$  level and a statistical uncertainty evaluated at $300~\fbi$ assuming a  10\% selection efficiency, respectively. 
}\label{fig:minimalshape}
\end{figure}

In Fig.~\ref{fig:minimalshape} we show the typical signal differential cross section for $gg\to S \to t\bar t$ as a function of the $t\bar t$ invariant mass, $m_{tt}=\sqrt {\hat s}$,  for  $y_t=1$. \Edit{The width of the heavy scalar in this model varies from 3 GeV to 48 GeV (12 GeV to 55 GeV) for a 400 GeV and a 1 TeV CP-even (CP-odd) scalar, respectively.} Throughout this paper, we use NNPDF3.0LO~\cite{Ball:2014uwa} for the parton distribution functions and set the factorization scale to be  the same as the $t\bar t$ invariant mass. We show the CP-even and CP-odd scalar lineshapes at LHC 13 TeV in the left panel and right panel, respectively. To make the lineshapes for different masses visible, we multiply the signal lineshapes by various factors, indicated in the lower part of both panels. We further show the statistical uncertainty at 300$~\fbi$ with 10\% selection efficiency and systematic uncertainties of $\pm 2\%$ of the SM background in  dashed and solid gray lines, respectively. Both uncertainties include the QCD background from $gg\to t\bar t$ and $q\bar q\to t\bar t$. 

From  Fig.~\ref{fig:minimalshape} it follows that for the $t\bar t$ invariant mass above $\sim$500~\gev ($\sim$400~\gev), the interference effects are dominant
for the CP-even scalar (CP-odd scalar), as indicated by the size of deviation from the Breit-Wigner lineshape. The loop function behaviors shown in Fig.~\ref{fig:loopggh} and Fig.~\ref{fig:loopgghphase} determine the lineshape structures. For increasing values of  the $t\bar t$ invariant mass, the imaginary component of the loop functions grows with respect  to its real component, inducing a larger phase $\theta_{\mathcal{\bar A}}$. This behavior of the imaginary part explains the increasingly  pronounced dip structure in the lineshapes for larger values of the  $m_{t\bar t}$. Furthermore, the $\theta_{\mathcal{\bar A}}$ phase grows faster for the pseudoscalar than the scalar case, yielding the lineshape pure dip structure for  smaller values of the scalar mass in the former case. Another important feature is the off-shell interference effect, and especially for an off-shell heavy scalar at $t\bar t$ invariant mass around 400~GeV this effect is quite visible. This off-shell interference is more prominent for the pseudoscalar because of the $s$-wave nature of the cross section, compared to the $p$-wave ($\beta^2$ suppressed) nature of the scalar case, and is further augmented by the slightly larger width of the pseudoscalar.

\begin{figure}[t]
\subfigure{
\centering
\includegraphics[width=0.50\textwidth]{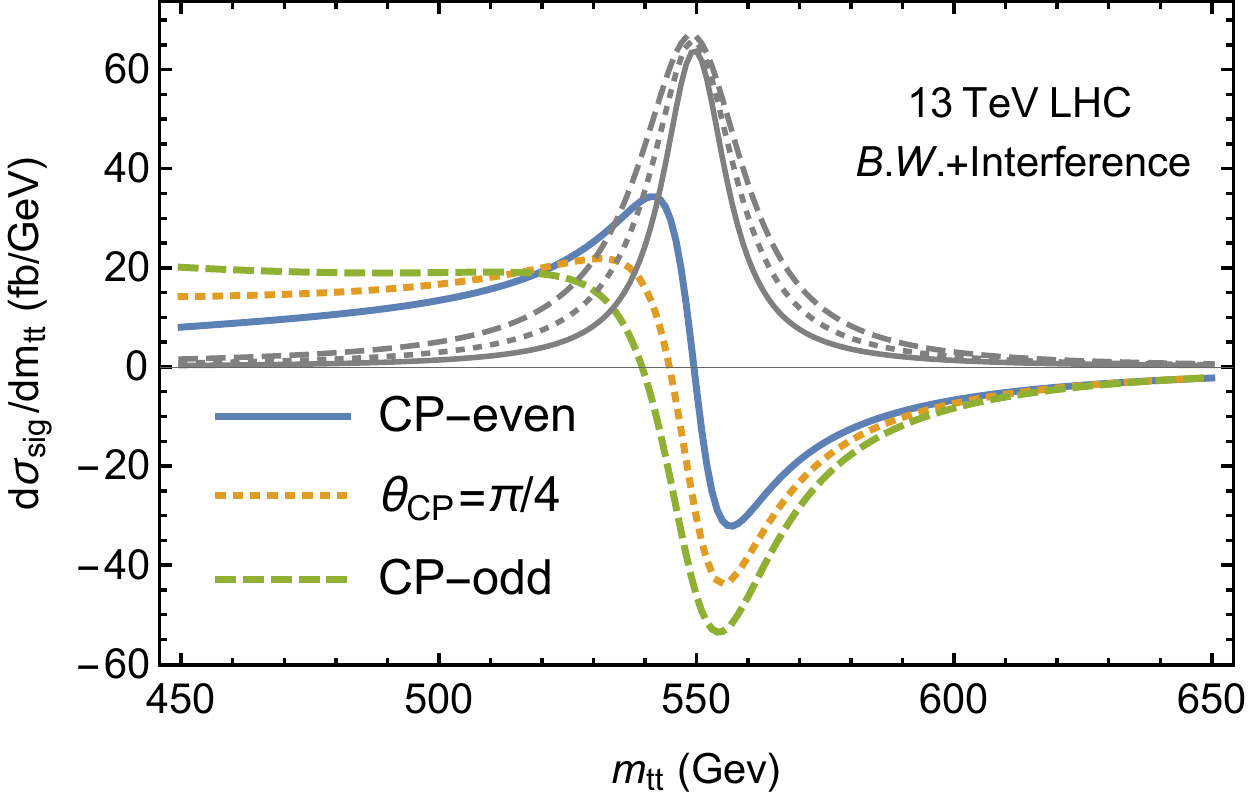} }
\subfigure{
\centering
\includegraphics[width=0.49\textwidth]{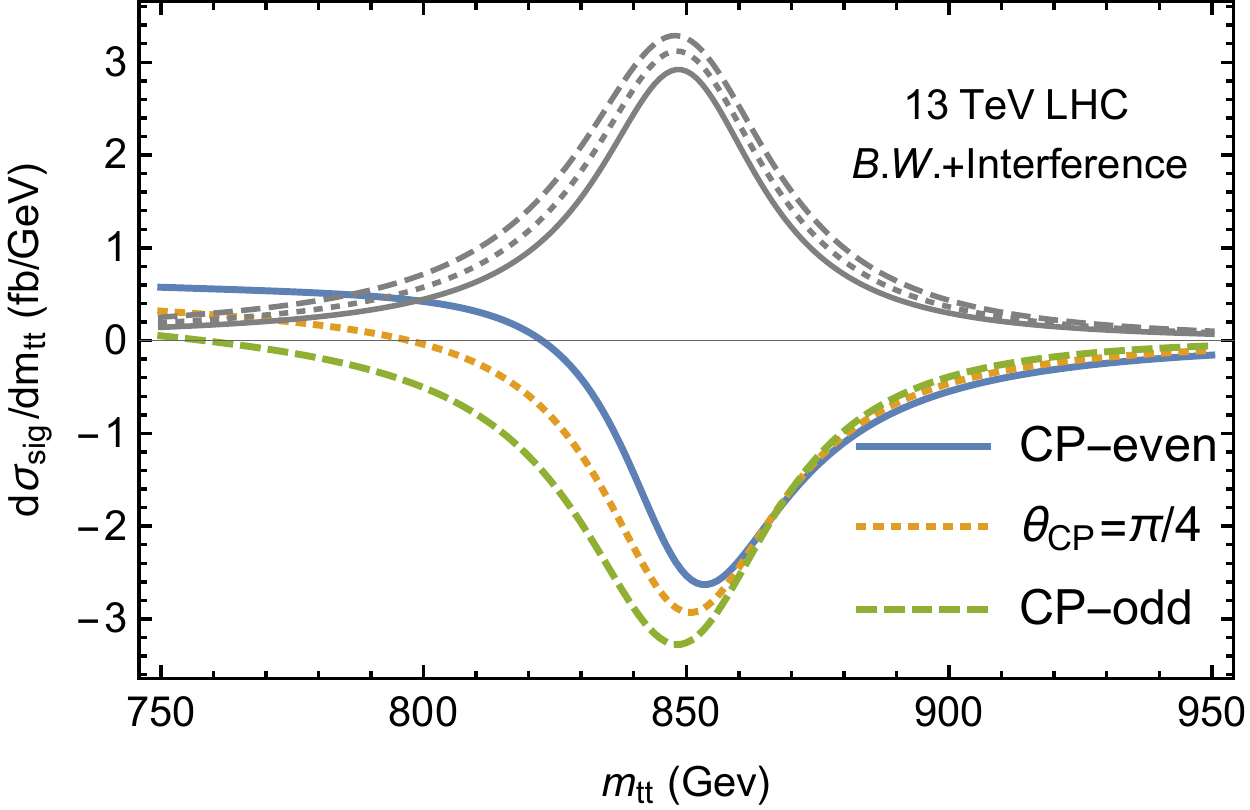} }
\caption[]{The signal lineshapes as the sum of the Breit-Wigner contribution and the interference contributions for the baseline model as a function of the $t\bar t$ invariant mass at the 13 TeV LHC. The blue, orange dotted and green dashed lines indicate the total BSM lineshapes for CP phases of 0 (CP-even), 1/4$\pi$(CPV$_{\rm max}$) and 1/2$\pi$ (CP-odd), respectively. The gray curves are the  Breit-Wigner contributions to the total lineshapes alone, with the corresponding CP phases. The heavy scalar masses are set at 550 GeV  and 850 GeV for the left and right panels, respectively.
}\label{fig:CPbenchmark}
\end{figure}

We show in Fig.~\ref{fig:CPbenchmark} detailed lineshapes for two representative scalar masses of 550 GeV and 850 GeV. 
 For a 550 GeV CP-even scalar, the phase $\theta_{\mathcal{\bar A}}$ is approximately $\pi/4$ while for a 550 GeV CP-odd scalar the phase is approximately $3\pi/8$,  as can be read from Fig.~\ref{fig:loopgghphase}.
For an 850 GeV CP-odd scalar, instead, the phase $\theta_{\mathcal{\bar A}}$ is approximately $\pi/2$ while for an 850 GeV CP-even scalar the phase is approximately $2\pi/5$.
%
These two benchmarks highlight the cases of the baseline model for which  {\it i)} the interferences proportional to the real and imaginary part of the propagator are comparable in size (left panel)
and {\it ii)} the interferences are dominantly from the piece proportional to the imaginary part of the propagator, resulting in a pure dip structure (right panel).

In Fig.~\ref{fig:CPbenchmark} the blue, solid lines; green, dashed lines and orange, dotted lines are the total lineshapes for a CP-even scalar; a CP-odd scalar and a scalar in the CPV$_{\rm max}$ ($\theta_{\rm CP}=\pi/4$) case, respectively.  These colored lines are the total BSM effects, including both the Breit-Wigner contribution and the interference with the SM background for  a scalar-top quark coupling  $y_t=1$. The  corresponding  the Breit-Wigner contributions alone are shown by the gray lines.
For the 550 GeV scalars,  the Breit-Wigner contribution is narrower for the CP-even scalar than for the CP-odd one,  due to the $\beta^2$ suppression in the former case. For the 850 GeV scalars, the $\beta^2$ suppression is negligible, resulting in almost identical widths for the CP-even and CP-odd scalars. In addition, as shown in Fig.~\ref{fig:loopggh}, the absolute value of the loop function for the CP-even scalar is smaller than the CP-odd one. Consequently, the CP-odd scalar Breit-Wigner lineshapes are higher than the CP-even ones. 
For both benchmark masses the total lineshapes given by the colored curves show a more pronounce dip structure for  the CP-odd case than for the  CP-even one.
The growth and the 
larger phase $\theta_{\mathcal {\bar A}}$ of the CP-odd loop function discussed in the previous section generates this feature.
For the CPV case, the lineshapes can be viewed as a properly weighted combination of the CP-even and CP-odd lineshapes,
following Eq.~\ref{eq:CPVparton}.

\section{Beyond the baseline model}
\label{sec:models} 

The channel  $gg\to S\to t\bar t$ at hadron colliders is crucial for heavy Higgs searches, especially in the alignment limit~\cite{Carena:2013ooa} (with or without decoupling) favored by current Higgs boson measurements at the LHC. Gluon-gluon-fusion is the dominant production mode of the heavy scalar and $t\bar t$ is likely to be the dominant decay mode. 

The baseline model introduced in the previous section helps us to understand the challenges of the $gg\to S\to t\bar t$ search. However, general BSM models usually contain more ingredients, adding new features to the baseline case.\footnote{Some alternative channel have been proposed and studied~\cite{Dev:2014yca,Hajer:2015gka,Craig:2015jba,Chen:2015fca,Gori:2016zto,Craig:2016ygr,Goncalves:2016qhh}, for gauge extensions, see e.g. Ref~\cite{Dobrescu:2013gza,Dobrescu:2015yba}.}
Firstly, there could be more than one heavy scalar particle, as in 2HDMs.
 If their masses are almost degenerate, as for example in the MSSM, these scalars will provide new contributions to the signal. Secondly, in addition to the top quark,  one can consider  the effects of other  colored fermions or scalars contributing to  the  gluon-gluon-scalar vertex. This could importantly modify 
 the phase $\theta_{\mathcal{\bar A}}$ in several different ways.
Specifically, there could be effects from loops involving bottom quarks and/or additional BSM colored particles, such as squarks and VLQs. There could also be CPV effects due to the direct couplings between the heavy scalar and  SM fermions as well as other particles in the loop. These modifications allow for partial cancellations or enhancements among the  different components of the gluon-gluon-scalar vertex.
We shall discuss all these possibilities in the following sections.



\subsection{Multiple scalar bosons}
\label{sec:2scalars}

In this section we study the case of two neutral heavy Higgs bosons with similar masses, a situation that occurs in various models. In a 2HDM, large splittings between these scalar bosons are  disfavored by low energy measurements such as the oblique parameters~\cite{RamseyMusolf:2006vr}. In  the Minimal-Supersymmetric-Standard-Model (MSSM), in particular, the heavy Higgs bosons ($H,~A,~H^\pm$) are nearly degenerate because of the specific supersymmetric structure of  the quartic couplings. 
Even after radiative corrections, the mass difference between the heavy CP-even and CP-odd scalars in the MSSM is at most of a few tens of GeV for  heavy scalar masses in the  500-1000 GeV range.

In the CP-conserving case,  the CP-even and CP-odd Higgs bosons do not interfere and the resulting partonic cross section is simply given as the sum of both,
\beq
\sigma_{\rm BSM}(\hat s)(gg\to H/A\to t\bar t)=\sigma^{\rm even}_{\rm BSM}(\hat s)(gg\to H\to t\bar t)+\sigma^{\rm odd}_{\rm BSM}(\hat s)(gg\to A\to t\bar t),
\eeq
where the terms in the above expression are given in Eqs.~\ref{eq:csparton_even} and~\ref{eq:csparton_odd}, with proper replacement of the coupling strengths. On the other hand, the results becomes slightly more complex and interesting if the actual scalar mass eigenstates contain an admixture of CP-even and CP -odd components. 
In terms of the mass eigenstates $S_1 $ and $S_2$, the  cross section reads,
\bea
\sigma_{\rm BSM}^{\rm CPV}(\hat s)(gg\to S_1, S_2\to t\bar t)=&&\sigma^{\rm CPV}_{\rm BSM}(\hat s)(gg\to S_1\to t\bar t)+\sigma^{\rm CPV}_{\rm BSM}(\hat s)(gg\to S_2\to t\bar t)\nonumber\\
&&+\sigma^{\rm S_1-S_2}_{\rm Int.}(\hat s) (gg\to S_1, S_2\to t\bar t),
\eea
where the cross sections for $S_1$ and $S_2$ follow the expressions for CPV scalars given in Eq.~\ref{eq:CPVparton}, whereas the additional  interference term  between the scalars $S_1$ and $S_2$ is given by,\\
\bea
\frac {d \sigma^{\rm S_1-S_2}_{\rm Int.}(\hat s) (gg\to S_1, S_2\to t\bar t)} {dz}=\frac {3\alpha_s^2 \hat s^2} {2048\pi^3 v^2} ~~~~~~~~~~~~~~~~~~~~~~~~~~~~~~~~~~~~~~~~~~~~~~~~~~~~~~~~~~~~&&\\
\Re\left[\frac {(y_{t}^{S_1}  y_{t}^{S_2}  |I_{\frac 1 2}(\tau_t)|^2+\tilde y_{t}^{S_1} \tilde y_{t}^{S_2} |\tilde I_{\frac 1 2}(\tau_t)|^2) (\beta^2 y_{t}^{S_1}  y_{t}^{S_2}+\tilde y_{t}^{S_1} \tilde y_{t}^{S_2})} {(\hat s - m_{S_1}^2 + i m_{S_1} \Gamma_{S_1}(\hat s)) (\hat s - m_{S_2}^2 - i m_{S_2} \Gamma_{S_2}(\hat s))}\right].&&\nonumber
\eea
The coefficient in the above equation can be further simplified in the alignment limit of a Type II 2HDM,
\bea
(y_{t}^{S_1}  y_{t}^{S_2}  |I_{\frac 1 2}(\tau_t)|^2+\tilde y_{t}^{S_1} \tilde y_{t}^{S_2} |\tilde I_{\frac 1 2}(\tau_t)|^2) (\beta^2 y_{t}^{S_1}  y_{t}^{S_2}+\tilde y_{t}^{S_1} \tilde y_{t}^{S_2})=\nonumber\\
\frac {\sin^2 2\theta_{\rm CP}}  {4} \left(\frac {y_t^{\rm SM}}  {\tan\beta} \right)^4 (|\tilde I_{\frac 1 2}(\tau_t)|^2-|I_{\frac 1 2}(\tau_t)|^2)(1-\beta^2).
\label{eq:2hdm_2scalar_simplified}
\eea
The corresponding CP-violating couplings in the alignment limit satisfy,
\bea
y_{t}^{S_1} + i \tilde y_{t}^{S_1} &&= -\frac {y_t^{\rm SM}} {\tan\beta} (\cos\theta_{\rm CP} + i\sin\theta_{\rm CP}),\nonumber \\
y_{t}^{S_2} + i \tilde y_ {t}^{S_2} &&= -\frac {y_t^{\rm SM}} {\tan\beta} (-\sin\theta_{\rm CP} + i\cos\theta_{\rm CP}).
\eea
From Eq.~\ref{eq:2hdm_2scalar_simplified}, it is clear  through its dependence on  $\sin^2 2\theta_{\rm CP}$ that the  interference piece between the two scalars is only relevant in the presence of CPV. Moreover, due to the propagator suppression, this contribution is sizable for almost degenerate masses and mostly in the region between the two scalar masses. 

\begin{figure}[t]
\subfigure{
\centering
\includegraphics[width=0.5\textwidth]{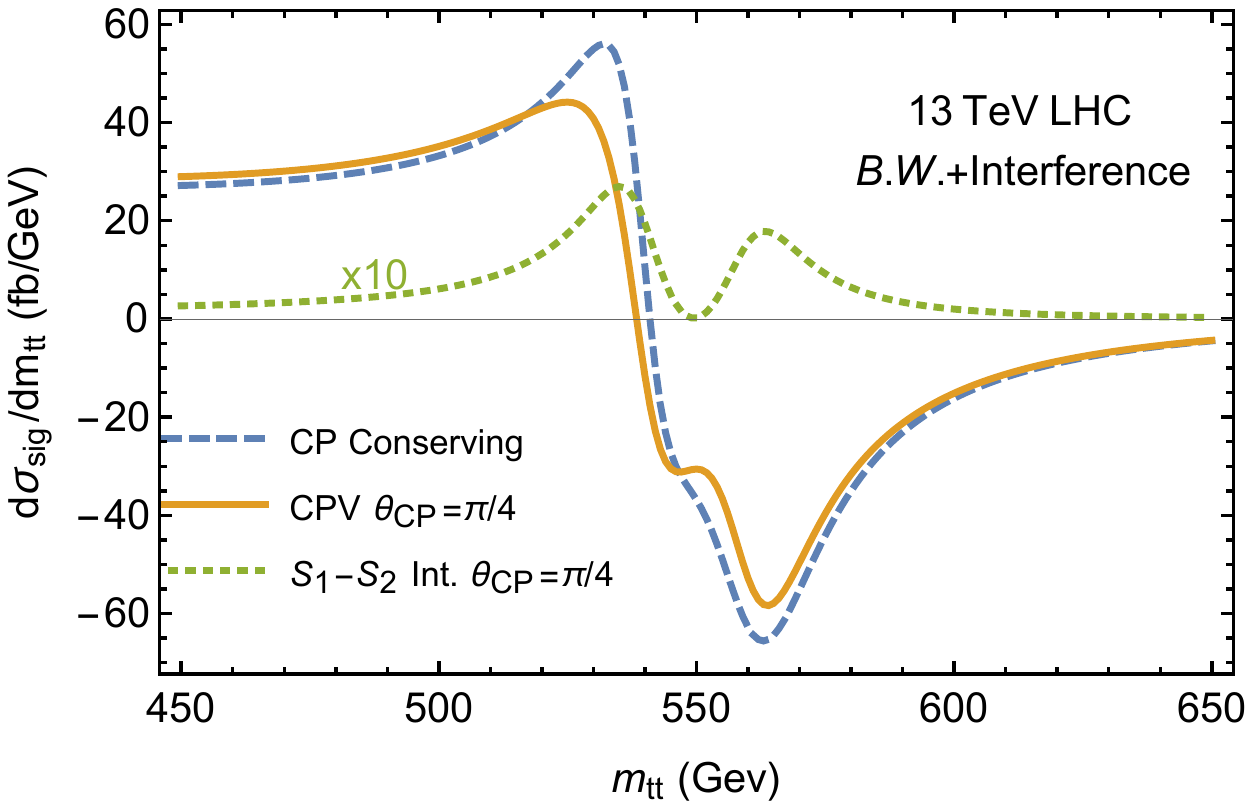} }
\subfigure{
\centering
\includegraphics[width=0.49\textwidth]{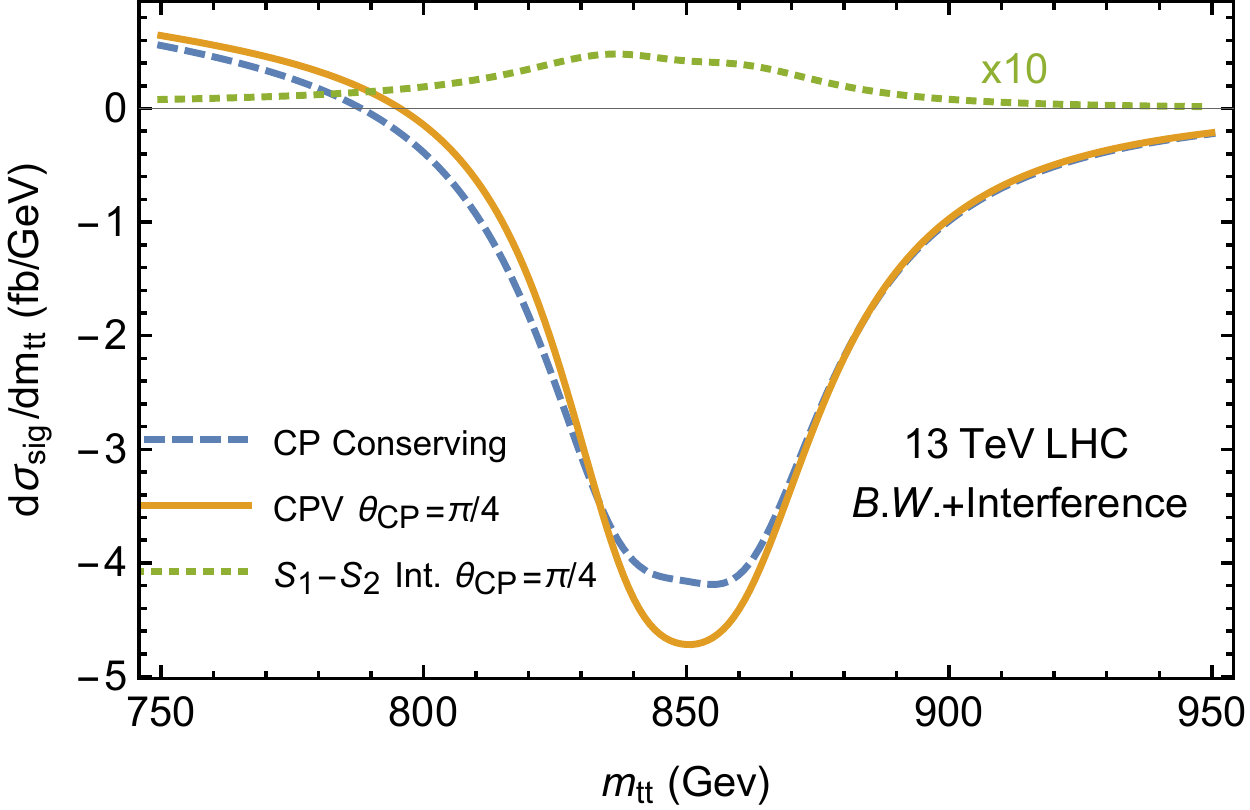} }
\caption[]{The signal lineshapes as the sum of the Breit-Wigner contribution and the interference contributions for nearly degenerate heavy scalars as a function of the $t\bar t$ invariant mass at the 13 TeV LHC. The orange, solid and blue, dashed lines correspond to lineshapes of the CP-violation case with $\theta_{\rm CP}$ $\pi/4$ and the CP conserving case, respectively.
The green, dotted lines are the interference between the two heavy scalars for the  CP-violating case. 
The heavy scalar masses are taken to be 540 GeV and 560 GeV for the left panel, and  840 GeV and 860 GeV for the right panel.
}\label{fig:CP2S2hdm}
\end{figure}

The $t \bar t $ signal from the decay of two  nearly degenerate scalars allows for a rich phenomenology. The resulting lineshape now depends on the masses, the separation between the mass values, the widths, and the CPV phase of the scalars. In Fig.~\ref{fig:CP2S2hdm}, we show the total signal  lineshapes  for the two nearly degenerate scalars,  both for the CP-conserving (blue, dashed lines)  and the maximally  CP-violating (orange, solid lines) cases. We consider scalars masses  of $540~\gev$ and $560~\gev$ for the left panel, and $840~\gev$ and $860~\gev$ for the right panel, where we take the CP-odd scalar $A$ to be 20 GeV heavier than the CP-even scalar $H$. 
The green, dotted lines single out the effect of the additional interference term between the scalars. To make this new interference term easily visible in the figure, we multiplied it by a factor of ten.

The main features of the two nearly degenerate  heavy scalars yielding a $t \bar t $ signal are: {\it i)} the signals of the two heavy scalars add to each other, almost ``doubling'' the height of the bumps and dips; {\it  ii)} a new contribution from  the $S_1$ and $S_2$ signal amplitude interference appears  in the CPV case.  
In  the left panel of Fig.~\ref{fig:CP2S2hdm},  the mass separation between the two scalar masses is somewhat larger than their respective widths and a 
 ``double dip'' structure  for the nearly degenerate scalars at around 550 GeV appears. In the right panel, we consider  scalar masses around 850 GeV and again a mass separation of 20 GeV.  In this case the widths of the two scalars are larger than the mass separation and a single, centrally flat, dip region appears, instead of  the  previous ``double dip''. 
The CPV lineshapes differ from the CP-conserving ones, and in particular they receive the contribution from  the new interference term between the two scalars. From Fig.~\ref{fig:CP2S2hdm} we observe that the new interference term is mainly in the region between the two scalar masses, and this is easily understood due to the kinematic suppression from the two scalar propagators.
Moreover, this new interference term is proportional to the real component of the product of the two scalar propagators, approximately, $1+ 4\Delta_1 \Delta_2$, where $\Delta_{1,2}$ are\footnote{Here $\Delta_{1,2}$ is defined analogously to $\Delta$ in Eq.~\ref{eq:propagator}} the mass differences between the $t\bar t$ system and the pole masses of each of the two scalars, $S_{1,2}$, respectively. The product $\Delta_1 \Delta_2$ is negative whenever $\sqrt {\hat s}$ is between the two scalar masses and positive otherwise. Moreover, when the mass splitting of the two scalars is smaller than the average of their widths, $\Delta_1 \Delta_2$ is a small negative quantity, which is not sufficient to flip the sign of the interference term.  As a result, the new interference term is positive for both examples. 
Furthermore, in the benchmark model shown in the right panel of this  figure, the CPV case has a deeper overall dip structure, which may open the possibility of differentiating CPV from CP-conserving scenarios in future high precision measurements.

\subsection{Scenarios with additional contributions to the gluon-fusion process}

Models with heavy scalar bosons often occur in association with additional colored particles yielding new contributions to the loop-induced gluon-gluon-scalar vertex. In addition bottom quark effects, not taken into account in the baseline model, may also contribute in specific regions of parameter space.

Before proceeding with a detailed discussion of lineshapes, let us  comment on  some essential differences between new particle contributions to the SM Higgs boson gluon fusion production with respect to the same production mode for heavy scalars. For the SM Higgs boson, one is entitled to make use of the low energy theorem to include the effects of heavy BSM particle contributions to loop-induced couplings. In such case one can add the new physics loops directly to the SM top quark loop, since around the SM Higgs boson mass   all these loop-functions are below the thresholds of the heavy particles, and therefore real. For heavy scalars, instead, the top quark loop-function is no longer real, and the heavy BSM particle contributions could have various phases depending on the kinematics. Consequently, a relative phase  will be generated between the SM fermion contributions and the BSM particle contributions. This effect could lead to drastic changes in the lineshapes for the heavy scalar and demands a careful treatment of the  inclusion of BSM effects in the heavy scalar production.

In the following we discuss several well-motivated scenarios with additional  colored particle effects. We focus on heavy scalar lineshapes considering the  new contributions from fermions and scalars that arise in general 2HDMs as well as in models with VLQs or SUSY models with squarks.

\subsubsection{Standard Model light quark contributions}

In the framework of 2HDMs, it is interesting to revisit the relevance of 
top quark-loops in the heavy Higgs-gluon fusion production process. The complete 2HDM is only defined after considering the interaction of the Higgs fields to fermions. In a Type I 2HDM, all SM fermions couple to a single Higgs field and hence the bottom quark-loop scales in the same way as the top quark-loop. Therefore the dominant contributions will always come from  the top-loop and the subsequent $t\bar t$ decay, regardless of the $\tan\beta$ value. Consequently, the bottom quark contribution is merely a small correction to the  phase of the gluon-gluon-scalar vertex and will minimally perturbe our previous discussions. 
In a type II 2HDM, instead, the contribution from bottom quark-loops can be sizable for moderate to large values of $\tan\beta$, and it is also directly correlated with the additional partial decay width into $b\bar b$.  
More specifically, the heavy Higgs-bottom Yukawa coupling, and hence the bottom  quark-loop contribution, scales as $\tan\beta$, while  the top quark one scales as $1/\tan\beta$. The interplay between these two competing contributions leads to a  rich phenomenology. In fact, in the large $\tan\beta$ regime, where bottom-loop induced gluon-gluon-fusion production and $b\bar b$ decay are dominant, the search strategy changes, and alternative channels such as  those with  $\tau^+\tau^-$ final states become more sensitive. Still in the low to intermediate $\tan\beta$ regime it is of interest to explore the $gg\to S\to t \bar t$ channel and consider the effects of the bottom quarks. 
Due to kinematics, the bottom-loop induced $ggS$ coupling will be in the large $\tau_b$ regime of Eq.~\ref{eq:ggh}, leading to very slowly varying  loop functions $I_{1/2}(\tau_b)$ and $\tilde{I}_{1/2}(\tau_b)$. The bottom quark- and  top quark-loop  contributions could then interfere constructively or destructively, depending on the relative sign between the two  corresponding Yukawa couplings to the heavy scalars.

\begin{figure}[t]
\subfigure{
\centering
\includegraphics[width=0.46\textwidth]{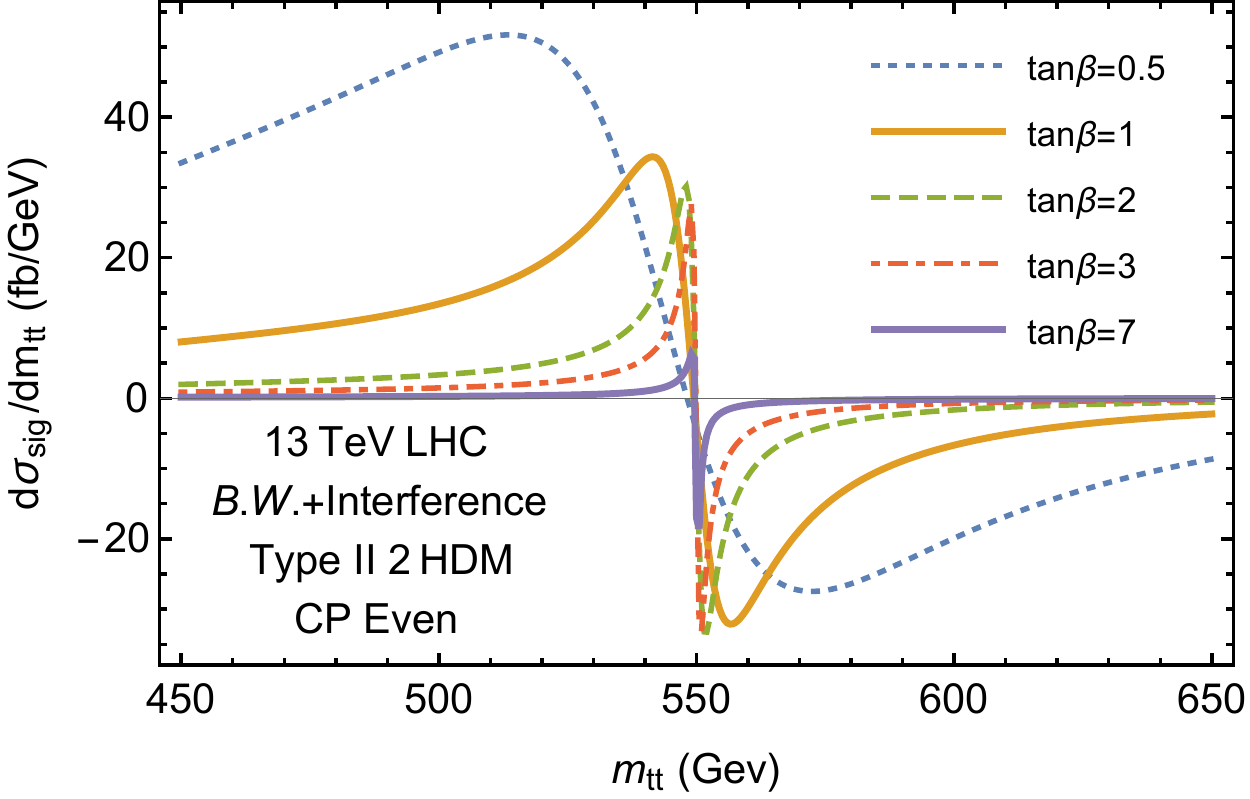}}
\subfigure{
\centering
\includegraphics[width=0.45\textwidth]{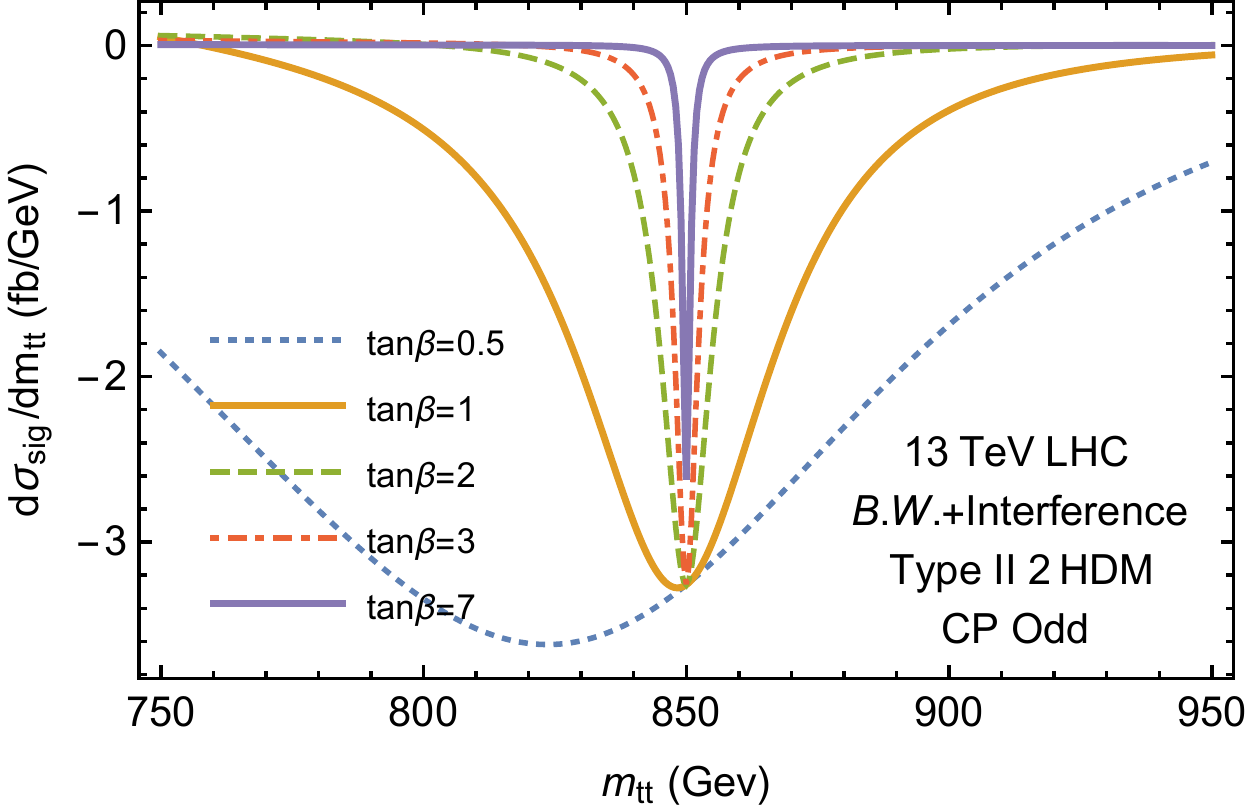}}
\caption[]{Signal lineshapes as the sum of the Breit-Wigner contribution and the interference contributions in Type II 2HDMs as a function of the $t\bar t$ invariant mass at the 13 TeV LHC,  for various values of $\tan\beta$, and  including the bottom quark contributions. The left and right panels correspond to a heavy CP-even scalar with mass 550 GeV and a CP-odd scalar with mass 850 GeV, respectively.
}\label{fig:2hdm}
\end{figure}

In the following, for simplicity, we only consider the CP conserving Type II 2HDM in the alignment limit. The $\tan\beta$ enhanced bottom quark contribution to the gluon-fusion production of the 125 GeV Higgs boson can be tuned away in the alignment or decoupling limit, therefore avoiding the corresponding  precision  measurement constraints. The CPV case can be considered in  a similar way  as the CPV discussion in Sec.~\ref{sec:baseline_lineshapes}. 
Including the contributions from both top and bottom quarks, the  gluon-gluon-scalar interaction for the CP-even Higgs boson from Eq.~\ref{eq:gSgg} now reads,
\beq
g_{Sgg}(\hat s) = \frac {\alpha_s} {2 \pi v} \left(-\frac 1 {\tan\beta} I_{\frac 1 2}(\tau_t)+\tan\beta I_{\frac 1 2}(\tau_b)\right),
\eeq
and analogously for the CP-odd Higgs. 

In Fig.~\ref{fig:2hdm} we show two  benchmark scenarios for a CP-conserving type II 2HDM, one for a CP-even scalar of mass 550 GeV (left panel) and the other for a CP-odd scalar of mass 850 GeV (right panel), while considering  various values of $\tan\beta$.   From Fig.~\ref{fig:CPbenchmark}, it follows that changing the CP-properties of the scalar for a similar mass window results in similar lineshapes as those shown in  each of the corresponding panels of Fig.~\ref{fig:2hdm}.
We choose to vary $\tan\beta$ between 0.5 to 7, where 0.5 yields an enhanced scalar top-quark coupling and 7 represents the case where the top- and bottom- quark loop induced gluon-gluon-scalar couplings are  minimized. Beyond $\tan\beta=7$, the $t\bar t$ decay will be substantially suppressed due to the large couplings of the scalar to bottom quarks. 
The lineshapes in this figure include both the Breit-Wigner and interference terms for both the bottom-  and top-quark contributions to the loop function.
For the $\tan\beta$ range considered, a lower value of $\tan\beta$ indicates a larger width and a  larger signal cross section. 
From Fig.~\ref{fig:2hdm} we observe  that the resulting signal phase changes more visibly with respect to the SM background for a lighter Higgs boson. 
This can be understood because for heavier scalars the kinematics is such that the phases of the top  and bottom-quark contributions  are closer to the asymptotic behavior for large values  of $\tau_{t,b}$,
as shown in Fig.~\ref{fig:loopgghphase}. 
Such feature is unique to light quark contributions to the loop function. Heavy particles, instead, will only contribute  to the real component of the loop-function. Finally, it is also interesting to notice that the height of the peaks does not change much for the  $\tan\beta$ regime under consideration.
In this regime the height of the peak has two contributing factors that cancel each other: the on-resonance amplitude  is proportional to $1/\Gamma$ from the propagator and the production rate  is proportional to $\Gamma_t$, which in turn dominates the total width  $\Gamma$. For higher values of $\tan\beta$ than those considered in this paper, the height will be further suppressed by the 
increasing contribution of  $\Gamma_b$ to the total width.

\subsubsection{Vector-like quark contributions}

Vector-like quarks are well motivated in many BSM theories, e. g. composite Higgs models~\cite{vonGersdorff:2015fta,Fichet:2016xvs,Fichet:2016xpw}, flavor models, grand unified theories. 
The heavy scalar effective couplings to gluons can receive sizable contributions from these vector-like quarks,  resulting in important changes to the phenomenology. We shall discuss some of the most relevant features in  this section by considering the minimal case of one vector-like $SU(2)_L$ quark doublet, $Q_L = (\psi_L  \;\;\; N_L)^T$  and $Q_R = (\psi_R \;\;\; N_R)^T$, and one vector-like $SU(2)_L$  quark singlet, $ \chi_R$ and $\chi_L$, respectively.
In the context of 2HDMs, the heavy scalar couplings to vector-like quarks are linked to their chiral masses.

The  vector-like fermion mass matrix, after electroweak symmetry breaking, can be expressed as,  

\beq
\left(\bar \psi_L,~\bar \chi_L\right) M_{\Psi}  \begin{pmatrix}
\psi_R  \\
\chi_R \\
 \end{pmatrix}=\left(\bar \psi_L,~\bar \chi_L\right) 
 \begin{pmatrix}
M_{\psi}  & y_\Psi \frac {v} {\sqrt 2}  \\
y_\Psi \frac {v} {\sqrt 2} &  M_{\chi} \\
 \end{pmatrix}
 \begin{pmatrix}
\psi_R  \\
\chi_R \\
 \end{pmatrix},
\eeq
where for simplicity we assume the off-diagonal entries to be identical. The subscript $L$ and $R$ always label chirality.
The mixing angle, defined for the mass eigenstates of Dirac spinors $\Psi_1$ and $\Psi_2$, follows
\beq
\Psi_1=\begin{pmatrix}
\Psi_{1,L}  \\
\Psi_{1,R} \\
 \end{pmatrix}=
 \cos\theta_\Psi \begin{pmatrix}
\psi_L  \\
\psi_R \\
 \end{pmatrix} +
\sin\theta_\Psi \begin{pmatrix}
\chi_L  \\
\chi_R \\
 \end{pmatrix}.
\eeq
with  $\Psi_2$ given by the orthogonal combination.
Due to the simplified identical chiral mass term, the mixing angles $\theta_\Psi$s are identical for the chiral-left  and right components, $\Psi_{i,L}$ and $\Psi_{i,R}$, and satisfy:
\beq
\sin2\theta_\Psi=-\frac {\sqrt{2}(M_{\psi}+M_{\chi}) y_\Psi v} {m_{\Psi_1}^2-m_{\Psi_2}^2}.
\eeq
In the alignment limit of a type II 2HDM, the heavy scalar coupling to the vector-like quarks $g_{\Psi_i}$ can be expressed as:
\begin{equation}
g_{\Psi_i} =
\mp \frac {1} {\tan\beta} \frac {y_\Psi} {\sqrt 2} \sin2\theta_\Psi~.
\end{equation}
Consequently, the sum of the vector-like quark contributions to the gluon-gluon-heavy scalar coupling reads
\bea
g_{ggH}^{\Psi}=\frac {\alpha_s} {2 \pi}\left(\frac {g_{\Psi_1}} {m_{\Psi_1}} I_{\frac 1 2}(\frac {\hat s} {4 m_{\Psi_1}^2})+\frac {g_{\Psi_2}} {m_{\Psi_2}}I_{\frac 1 2}(\frac {\hat s} {4 m_{\Psi_2}^2})\right),
\eea
while the corresponding result for the heavy CP-odd scalar is very similar. 
In the heavy mass limit of $m_{\Psi_1},~m_{\Psi_2}\gg  m_H$, the above contribution can be approximated as,
\beq
g_{ggH}^{\Psi}\approx \frac {\alpha_s} {2 \pi \tan\beta} \frac {(M_{\Psi_L}+M_{\Psi_R})y_\Psi^2 v} {m_{\Psi_1} m_{\Psi_2} (m_{\Psi_1}+m_{\Psi_2})} I(\frac {\hat s} {4 m_{\Psi_1}m_{\Psi_2}})
\approx \frac {\alpha_s} {3 \pi \tan\beta} \frac {y_\Psi^2 v} {m_{\Psi_1} m_{\Psi_2}}.
\label{eq:VLQggS}
\eeq
We can see from Eq.~\ref{eq:VLQggS} that the  loop-induced contribution to gluon-gluon-scalar couplings takes a form very similar to that one obtained from the low energy theorem of the SM Higgs~\cite{Carena:2012xa}. Although the heavy Higgs doublet does not have a VEV, its couplings to the heavy vector-like fermions are proportional to that of the SM doublet.

\begin{figure}[t]
\subfigure{
\centering
\includegraphics[width=0.46\textwidth]{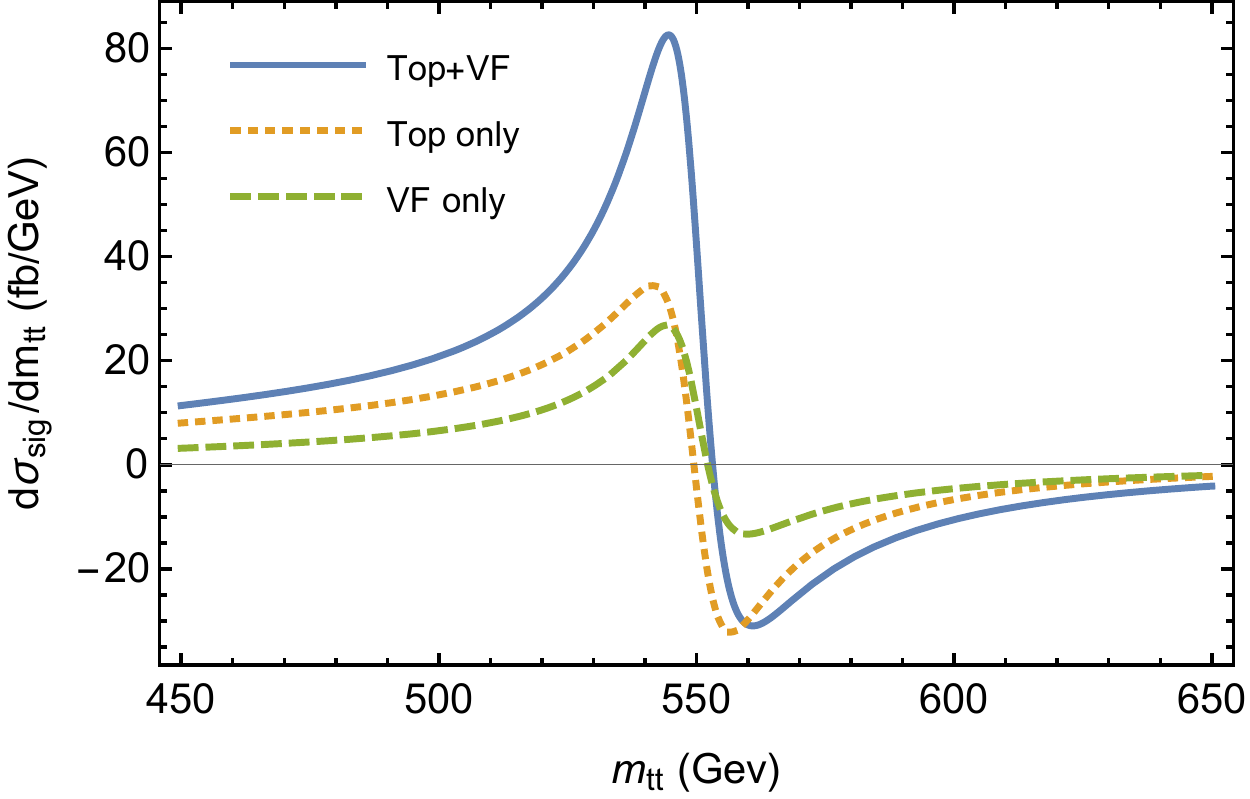}}
\subfigure{
\centering
\includegraphics[width=0.45\textwidth]{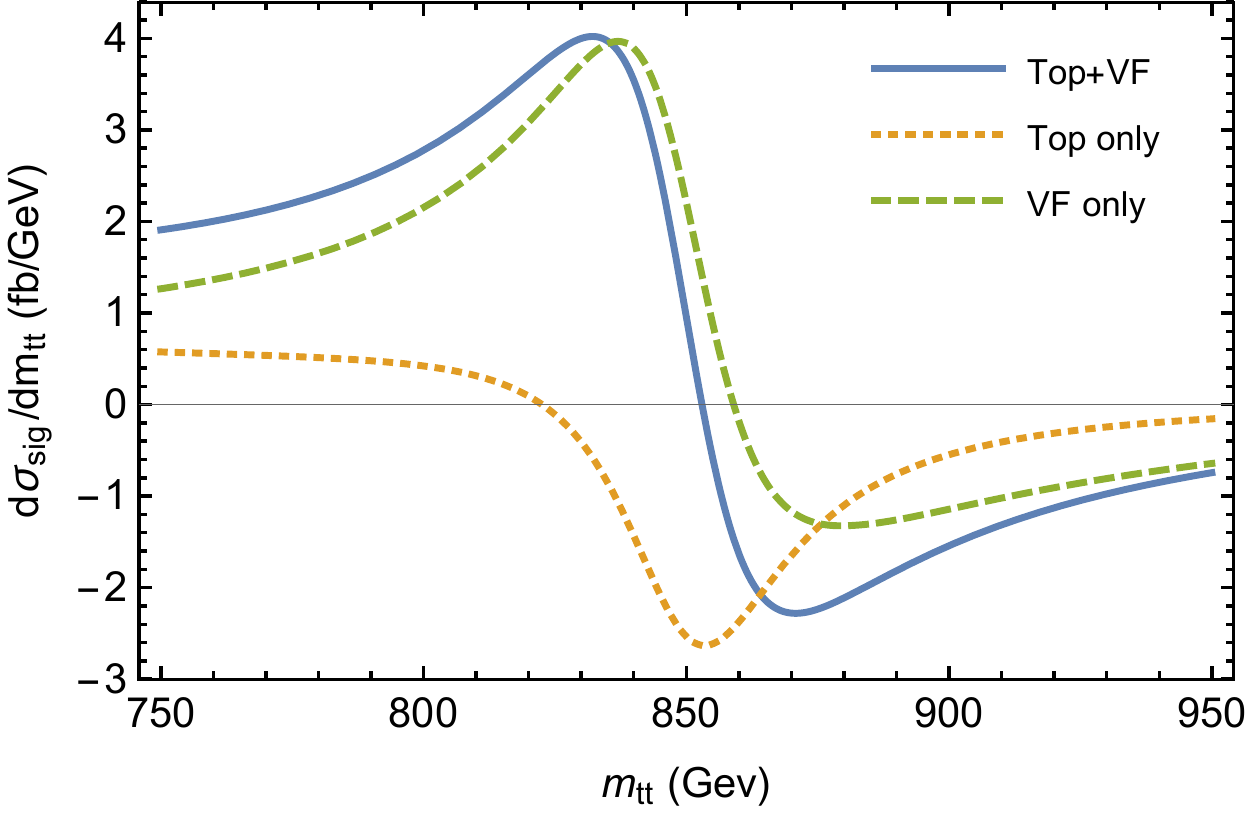}}
\caption[]{Signal lineshapes as the sum of the Breit-Wigner contribution and the interference contributions 
as a function of the $t\bar t$ invariant mass at the 13 TeV LHC. The blue (solid), orange (dotted) and green (dashed) lines correspond to the sum of top quark and vector-like quark loop contributions, the top quark contribution alone and the vector-like quark contribution alone, respectively. The vector-like quark contribution is computed for benchmark parameters  $M_\psi=$600~GeV, $M_\chi=$1200~GeV and Yukawa $y_{\Psi}=2$ as defined in the text. The left and right panels correspond to heavy CP-even scalar masses  of 550 GeV and 850 GeV, respectively.
}\label{fig:VF}
\end{figure}

In Fig.~\ref{fig:VF}, left and right panels,  we present the heavy CP-even scalar lineshapes with contributions from the vector-like fermions for benchmark scalar masses of 550 GeV and 850 GeV, respectively. We show the lineshapes from considering only the  top quark contribution  (orange, dotted lines), only the VLQ contribution (green, dashed lines) and the coherent sum of both contributions (blue, solid lines). The resulting changes to the lineshapes are sizable. The vector-like fermions may enhance the production of the heavy scalars with respect to the SM top-quark loop contribution. At the same time, due to the fact that the VLQ induced loop function is real, there will be no destructive interference  with the SM background. We choose a benchmark point with mass parameters $M_\psi$ and $M_\chi$ of 600 GeV and 1200 GeV, respectively. The Yukawa coupling is chosen as $y_\Psi=2$. In such case the masses of the eigenstates are 440~GeV and 1360~GeV, respectively. Consequently, the 850 GeV scalar is closer to the  threshold of the lighter vector-like quark and receives relatively larger corrections to the lineshapes in comparison to the 550 GeV one. 
We note that in 2HDMs, the VLQ will also contribute to the SM Higgs couplings to gluons, and therefore, the current measurement of the SM-like Higgs properties will constrain the size of the allowed contributions from these new fermions. However, due to the $m_h/m_{\Psi}$ suppression and the current level of accuracy in the  Higgs boson measurements, such constraints do not play a relevant role at present.

If the  intermediate colored particles are heavy, effective operators will be sufficient to describe the physics. In such case our loop-induced gluon-gluon-scalar form factor in Eq.~\ref{eq:VLQggS} becomes a constant, and can be identified as the Wilson coefficient of the  effective field theory (EFT) operators $\frac 1 \Lambda SGG$ or $\frac 1 \Lambda SG\tilde G$. We give an example in Sec.~\ref{sec:750}.

\subsubsection{SUSY scalar quark contributions}

The SUSY partners of the SM colored fermions may also contribute to the  gluon-gluon-scalar effective coupling. These scalar quarks  also modify the predictions for the  observed $\sim$125 GeV Higgs boson measurements, however, for sufficiently heavy stops  as those considered here current data does not impose any relevant constraints.
The squark contributions  to  the heavy scalar Higgs production are of the form:
\beq
g^{\tilde q}_{_Sgg}(\hat s)=-\frac {\alpha_s} {8 \pi} \sum_{q;i=1,2} \frac {g^{\tilde q}_{i} v} {m_{\tilde q_i}^2} \frac {1} {\tau_i^{\tilde q}} \left(1-\frac {1} {\tau^{\tilde q}_i} f(\tau^{\tilde q}_i)\right),
\label{eq:Higgs-squark}
\eeq
where the subscript $i$ labels the two scalar mass eigenstates with masses  $m_{\tilde q_i}$, that are  the superpartners of the corresponding SM fermion $q$. Only the diagonal Higgs-squark-squark couplings in the mass basis contribute to Eq.~\ref{eq:Higgs-squark}, and thus the Higgs-squark-squark couplings $g^{\tilde q}_{ij}$ are labeled $g^{\tilde q}_i$ . For the case of $\tau^{\tilde q}_i \ll 1$   the above equation becomes a slowly varying function of the scale ratio parameter $\tau^{\tilde q}_i$, and the EFT approach  is sufficient to describe the physics results in this channel. However, the scalars we consider are relatively heavy, and could be close to the   squarks  threshold.  In this case the phenomenology is rich and interesting and we shall keep the full scale dependence  to properly account for such possibility. 

\begin{figure}[t]
\subfigure{
\centering
\includegraphics[width=0.50\textwidth]{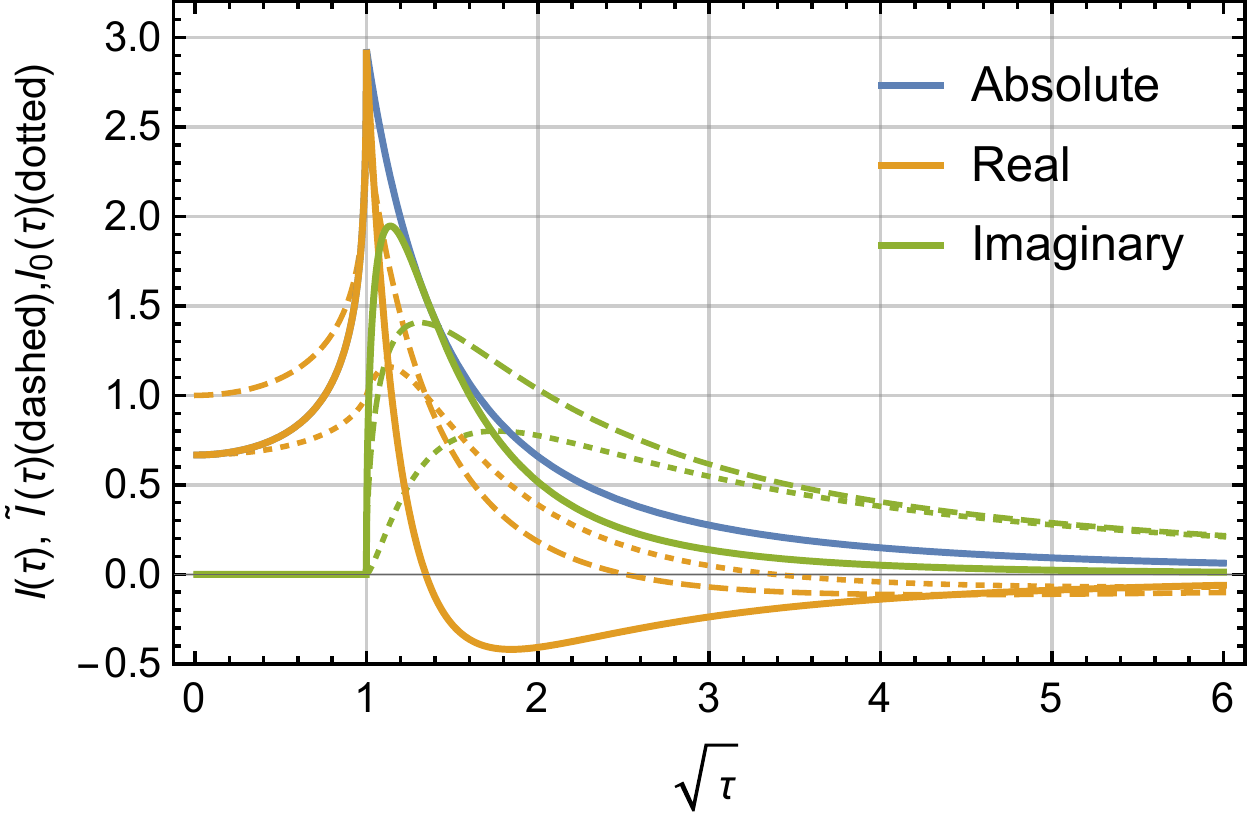} }
\subfigure{
\centering
\includegraphics[width=0.49\textwidth]{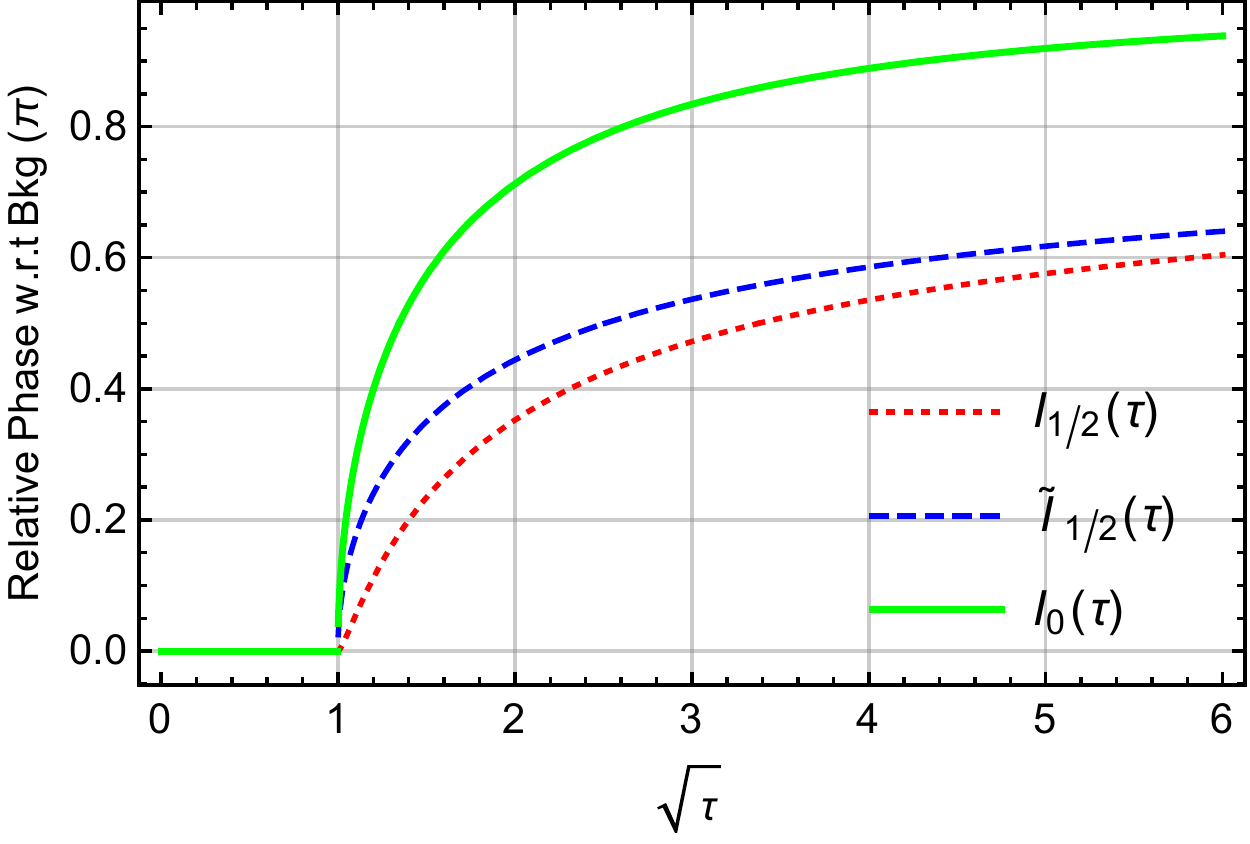} }
\caption[]{{\bf Left panel:} loop functions of the scalar-gluon pair vertex as a function of  $\sqrt {\tau}\equiv\sqrt{\hat s}/(2m)$, with $m$  the mass of the new particle in the loop. The orange, green and blue lines correspond to the real, imaginary and absolute values of these functions. The solid lines represent the values of the squark loop function,  while for the  fermion loop contribution 
 the real and imaginary parts are shown in dotted and dashed lines for the scalar and pseudoscalar case, respectively. The squark-loop function is multiplied by a factor of four to be visible in a common scale with the fermion loop functions. {\bf Right panel:}  induced relative phase with respect to the SM background \Edit{in units of $\pi$} for the sfermion loop (green line), fermion loop for a scalar (dotted red line) and a pseudoscalar (dashed blue line).
}\label{fig:loopgghscalar}
\end{figure}

For scalar masses such that  $2 m_t < m_S < 2 m_{\tilde t}$, the loop function for gluon-gluon Higgs coupling from top-quark loop is dominantly imaginary, while that from scalar quarks is real. As a result these two contributions do not interfere with each other, in sharp contrast to the SM Higgs boson case, where $m_h < 2 m_t < 2m_{\tilde q}$.  The squark contributions allow for an additional adjustment of the relative phases between the  $ggS$ production vertex  and the $t\bar t S$ decay vertex, enriching the phenomenology. 

In the  left panel of Fig.~\ref{fig:loopgghscalar} we show in blue, orange and green, solid  lines the absolute, real and imaginary values of the corresponding loop-functions for scalar quarks, respectively. Comparing to spin-1/2 loop-functions shown by the dashed and dotted lines for the scalar and pseudoscalar cases, respectively, the squark loop-function rises and falls much more abruptly near the threshold. Its real component becomes negative right above threshold. We multiply the squark function by a factor of four to make it more visible.
In the right panel of Fig.~\ref{fig:loopgghscalar} we show the phase generated by the different loop functions as a function of the scale parameter $\sqrt{\tau}$. 
As discussed in Section \ref{sec:minimal}, the closer the phase is to $\pi/2$, the more important is the interference proportional to the imaginary part of the propagator with the SM background, rendering the dip structure more prominent. We show the evolution of such phase for the fermion loop for a scalar (dotted red line) and a pseudoscalar (dashed blue line), as well as for the squark loop (green line). The phase of the squark loop raises much faster comparing to the fermion-loop cases, and at large $\sqrt \tau$ the phase is close to $\pi$. The phases from the fermions approaches $\pi/2$ instead, which is the cause for a pure dip structure at high scalar masses for the baseline model.

In the following we will concentrate in the more intriguing case in which the scalar quark mass is only slightly above half the scalar mass. In this situation the threshold effect can create additional structures in the line shapes.

Consider the squark mass matrix:
\beq
M_{\tilde q} =
 \begin{pmatrix}
M_{\tilde Q}^2 +m_q^2 + D_L^q  & m_q X_q\\
 m_q X_q &  M_{\tilde q_R}^2 +m_q^2 + D_R^q \\
 \end{pmatrix}     ~~,
\eeq
and the mixing angles (defined as $\tilde q_1=\cos\theta_{\tilde q} \tilde q_L + \sin \theta_{\tilde q} \tilde q_R$)  that satisfy:
\beq
\sin2\theta_{\tilde q}=\frac {2 m_q X_q} {m_{\tilde q_1}^2-m_{\tilde q_2}^2},
\eeq
with  $X_q$, $Y_q$, $D_{L}^q$ and $D_{R}^q$ for $q=u,d$ defined in Appendix A, Eq.~\ref{eq:SUSYdef}. In the alignment limit and  considering only the dominant  stop contributions (setting $q=t$ in the above equations), $g_i^{\tilde t}$ can be expressed as:
\begin{equation}
g^{\tilde t}_{1,2}(S) \frac  {v} {\sqrt{2}}=\left\{\ba{lcl}
m_t^2 + \cos2\beta ( D_{L/R}^t \sin^2\theta_{\tilde t} + D_{R/L}^t \cos^2\theta_{\tilde t} )\pm \frac 1 2 m_t X_t \sin2\theta_{\tilde t} &,&~{\rm for}~S=h\\
-\frac {m_t^2} {\tan\beta} - \sin2\beta (D_{L/R}^t \sin^2\theta_{\tilde t} + D_{R/L}^t \cos^2\theta_{\tilde t}) \mp \frac 1 2 m_t Y_t \sin2\theta_{\tilde t}&,&~{\rm for}~S=H\\
\mp 
\frac 1 2 m_t Y_t \sin2\theta_{\tilde t}&,&~{\rm for}~S=A\ea\right.
\label{eq:SUSYcouplings}
\end{equation}
In the above expressions the terms proportional to $X_t$ and $Y_t$  correspond to the off-diagonal couplings of the light CP-even Higgs and heavy CP-even Higgs to L-R stops, respectively. 
While the phenomenological studies on the light Higgs boson focus on $X_t$, which is directly connected to stop masses and mixing, and correspondingly to the  Higgs mass radiative corrections, the heavy Higgs boson coupling  mainly depends on an orthogonal quantity $Y_t$. The stop  L-R mixing contribution to the heavy Higgs boson coupling to gluons are proportional to $Y_t \sin2\theta_t $,  which in turn is proportional to the product of $X_t Y_t$,
\beq
X_t Y_t = \frac {A_t^2} {\tan\beta} - \frac {\mu^2} {\tan\beta} -  A_t\mu (1-\frac 1 {\tan^2\beta}).
\label{eq:SUSYXY}
\eeq

\begin{figure}[t]
\subfigure{
\centering
\includegraphics[width=0.50\textwidth]{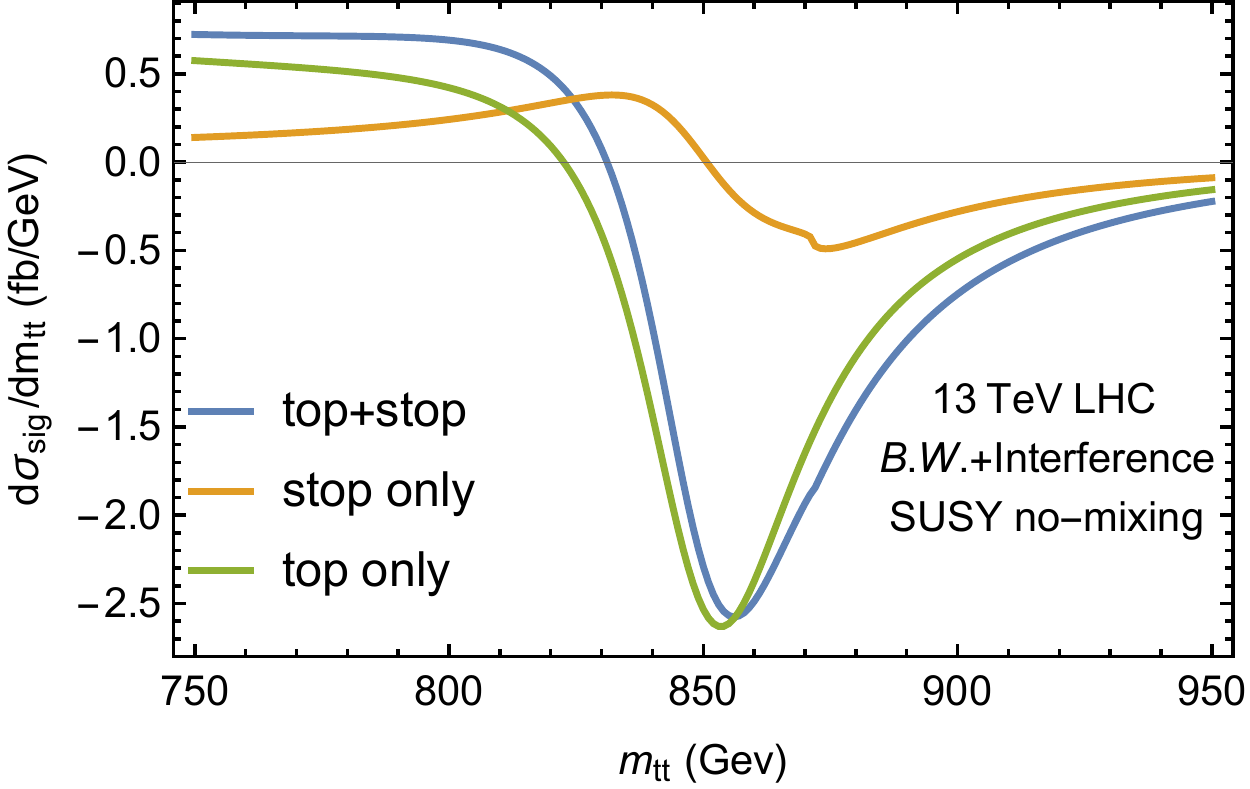}}
\subfigure{
\centering
\includegraphics[width=0.485\textwidth]{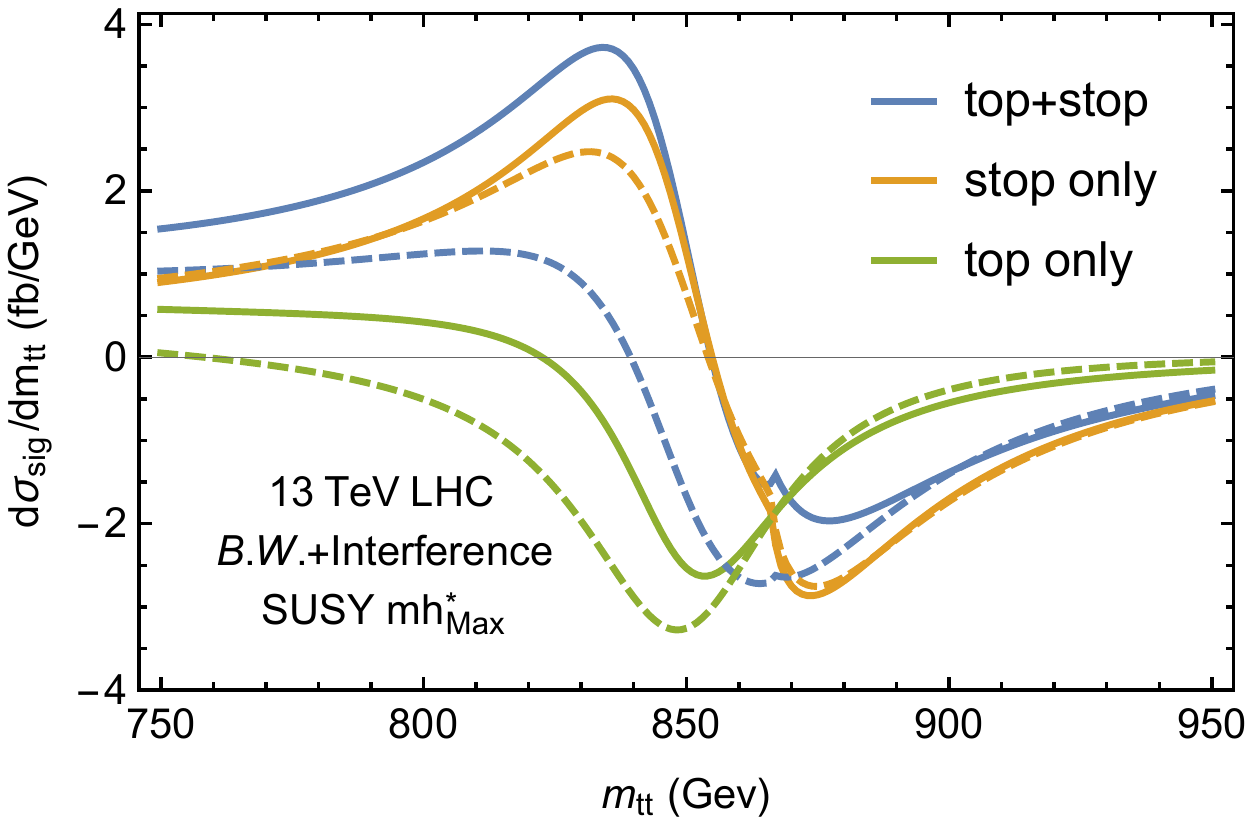}}
\caption[]{Signal lineshapes as the sum of the Breit-Wigner contribution and the interference contributions, including the effects of SUSY stops in the loop for  850 GeV CP-even  scalars,   as a function of the $t\bar t$ invariant mass at the 13 TeV LHC. The left panel corresponds to the SUSY stop scenario with zero L-R mixing, while the right panel corresponds to the SUSY stop scenario  $mh^*_{\rm max}$. The green and yellow lines represent the cases with only top quark loops or stop loops, respectively. The blue lines are the total lineshapes including all contributions. In the right panel we show both the case for a CP-even and a CP-odd scalar for the  solid and dashed lines, respectively.}\label{fig:SUSY}
\end{figure}

In Fig.~\ref{fig:SUSY} we show the comparison of the lineshapes for a heavy scalar of mass 850 GeV considering stop contributions to the loop function, and for two scenarios for the stop mixing parameters $X_t$ and $Y_t$. 
One is the zero L-R mixing case with vanishing  $X_t$. The other is a variation of the mh$_{\rm max}$ scenario~\cite{Carena:2002qg,Carena:2005ek,Carena:2013ytb} in which we take  $X_t=\sqrt {6} M_{\rm SUSY}\approx \sqrt{6 m_{\tilde Q_3} m_{\tilde t_R}}$ and $Y_t=2 X_t$. We named this modified maximal mixing scenario mh$_{\rm max}^*$  such that  for $\tan\beta=1$, it  corresponds to $A_t=3\mu$. 
The channel $gg\to H,A\to t\bar t$ in supersymmetry could be a dominant channel in discovering the heavy Higgs bosons in the low $\tan\beta$ regime. Despite that the observed 125 GeV Higgs mass disfavors the low $\tan\beta$ ($<3$) regime in the MSSM, extensions of the minimal model, such as the next-to-minimal-supersymmetric standard-model can work well in this regime. Therefore, for the purpose of demonstrating the $t\bar t$ channel's physics potential and for easier comparison with previous non-SUSY discussions, we choose a benchmark value of $\tan\beta=1$ in these figures.
The green and orange lines correspond to the production of heavy scalars with only the SM top quark loop contribution and only the SUSY stop loop contribution, respectively. The blue lines represent the lineshapes with all contributions taken into account. 
In both scenarios we choose the lighter stop mass to be close to half of the the heavy Higgs boson mass  and the heavier stop to be  around 1 TeV. The detailed numerics of our benchmark stop parameters are listed in the Appendix in Eq.~\ref{eq:SUSYpara}. 

The stops could change the heavy scalar  lineshapes in a distinct way depending on the L-R stop mixing. For the case with zero L-R mixing shown in the left panel of Fig.~\ref{fig:SUSY}, the stop contribution (orange line) is relatively small compared to the top contribution (green line), due to the smaller value of the squark loop function. In spite of the fact that the stop loop function is real and only produces interference through the real part of the propagator, the small value of the Breit-Wigner contributions implies that the interference piece is dominant, leading to a bump-dip structure crossing zero at the scalar pole mass. Once both the top and stop loop contributions are summed up the effect of the stop is hardly noticeable.
Moreover, in the zero L-R mixing case  the CP-odd scalar does not couple to the stops, and hence  we do not show those lineshapes for the CP-odd Higgs. 
For the mh$_{\rm max}^*$ scenario shown in the right panel of Fig.~\ref{fig:SUSY}, the stop contribution could be sizable. We show both the lineshapes for the CP-even Higgs boson and the CP-odd Higgs boson in solid and dashed lines, respectively. The Breit-Wigner contribution from the stop loop shifts the value of $t \bar t$  invariant mass where the signal rate is zero slightly above the heavy scalar pole mass, as illustrated  by the orange lines. The contribution from the L-R mixing term dominates and changes the pure dip structures from the top only contribution (green lines) into  a bump-dip structure (blue lines). 
We purposefully choose the parameters such that the heavy scalar is only slightly below the light stop pair production threshold, with a light stop mass of about 435~GeV. 
We observe that the stop threshold effect is only minimally visible in the orange and blue lineshapes in both panels, through the small discontinuity at a $t\bar t$ invariant mass of around 870~GeV.
The above discussion shows that a relatively light stop, depending on the L-R mixing parameters, could have a relevant impact on the search strategy and the sensitivity reach of heavy scalars in the $t\bar t$ decay channel.

\subsection{Special discussion: A (pseudo)scalar from a putative di-photon excess}
\label{sec:750}

At  the end of 2015 both the ATLAS and CMS collaborations  reported a diphoton excess at about 750 GeV that could have been a truly striking signal of new physics beyond the standard model~\cite{ATLAS-CONF-2015-081,CMS-PAS-EXO-15-004}. 
This excess drew significant attention from the theory community.\footnote{For a relatively comprehensive study and quasi-review, see e.g., Ref~\cite{Franceschini:2015kwy,Franceschini:2016gxv,Strumia:2016wys}, and references therein.}
Many theoretical descriptions to explain a putative diphoton excess also implied the existence of the $t\bar t$ signal~\cite{Franceschini:2015kwy,Angelescu:2015uiz,Buttazzo:2015txu,Ellis:2015oso,Bellazzini:2015nxw,Low:2015qep,Gupta:2015zzs,Kobakhidze:2015ldh,Bian:2015kjt,Aloni:2015mxa,Altmannshofer:2015xfo,Craig:2015lra,Dev:2015vjd}.  
Moreover, many of the explanations, involved sizable contributions from heavy particles, vector-like fermions and scalars, in the loop functions for both the gluon-gluon-scalar production vertex and the diphoton-scalar decay vertex.
In the following, we focus on some detailed features of the $t\bar t$ signal lineshapes  from a heavy scalar in the framework of an EFT,  where  heavy particle loop contributions to the gluon-gluon-scalar coupling compete with the  top quark loop one. We further introduce a convenient rescaling factor to quantify the signal rate after smearing effects to correctly translate  current bounds on a $t\bar t $ resonance search, also taking into account the important interference effects. We consider as an example a  750 GeV scalar with no special relevance of the precise mass value as far as it is in the several hundred GeV range.

As it is well-known, the $t\bar t$-scalar coupling induces at one-loop level the gluon-gluon-scalar and gamma-gamma-scalar effective vertices. If this $t\bar t$-scalar coupling is the dominant source of the diphoton process, although the production rate will be  sizable, the diphoton branching fraction will be  too small to accommodate a sizable diphoton signal at the reach of the LHC. Indeed, the tree-level two-body decay of a several hundred GeV heavy scalar  to top quark pairs is orders of magnitude too large compared to the electromagnetic, loop suppressed scalar to diphoton decay. A possibility is to increase the production rate to compensate such small decay branching fraction to diphotons, however, other searches on the hadronic channels will strongly disfavor such scenario. Instead, an intriguing possibility for a heavy scalar  diphoton  signal could be from heavy charged particle dominance in the gluon production as well as in the diphoton decay modes, with suppressed but still very sizable decay to $t\bar t$. A very straightforward example is a neutral heavy scalar that mainly receives its coupling to gluon pairs and photon pairs through multiple heavy top partner loops, while the coupling of the new heavy scalar to top quarks is controlled by the mixings of the top partners with the top quark.
 We consider the following  minimal interaction Lagrangian for a pseudoscalar $S$,\footnote{We changed  notation to the standard one for composite scalar models.
Analogous treatment holds for the scalar case.}
\beq
\mathcal{L}_{int}\supset \frac {S} {f} (i\tilde y_t \bar Q_L \tilde H t_R + h.c.) + c_{_G} \frac {\alpha_s} {8\pi f_{_G}} S G \tilde G,
\eeq
where the coefficient $c_{_G}$ captures contributions to the gluon-gluon-scalar coupling by integrating out the heavy colored particles. The total gluon-gluon-fusion rate for the scalar production also receives contribution, from the top-quark loop and reads
\bea
\sigma(gg\rightarrow S)&&=\sigma(gg\to H^{\rm SM}_{750\gev})\frac {v^2} {f^2} \frac {\left|\tilde I_{\frac 1 2}(\tau_t)+\frac {f} {\tilde y_t f_{_G}} c_{_G}\right|^2} {\left|I(\tau_t)\right|^2}, \nonumber \\
&&=31 \left(\frac {\tev} {f} \right)^2 \left|\tilde I_{\frac 1 2}(\tau_t)+c_g\right|^2~{\rm fb},{\rm~with~}c_g\equiv \frac {c_{_G} f} {\tilde y_t f_{_G}}.
\label{eq:750rate} 
\eea
The rate for the SM Higgs is approximately 740 fb at the 13 TeV LHC~\cite{Dittmaier:2011ti} and the loop functions $I_{\frac 1 2}(\tau_t)$ and $\tilde I_{\frac 1 2}(\tau_t)$ are as defined in Eq.~(\ref{eq:ggh}). 
However, it is very important to emphasize that using $\sigma(gg\to S)$   from Eq.~(\ref{eq:750rate}) multiplied by the ${\rm Br}(S\to t\bar t)$ is no longer a valid approach, since the large interference effects should be appropriately taken into account, as discussed in the previous sections.

\begin{figure}[t]
\centering
\subfigure{
\includegraphics[width=0.485\textwidth]{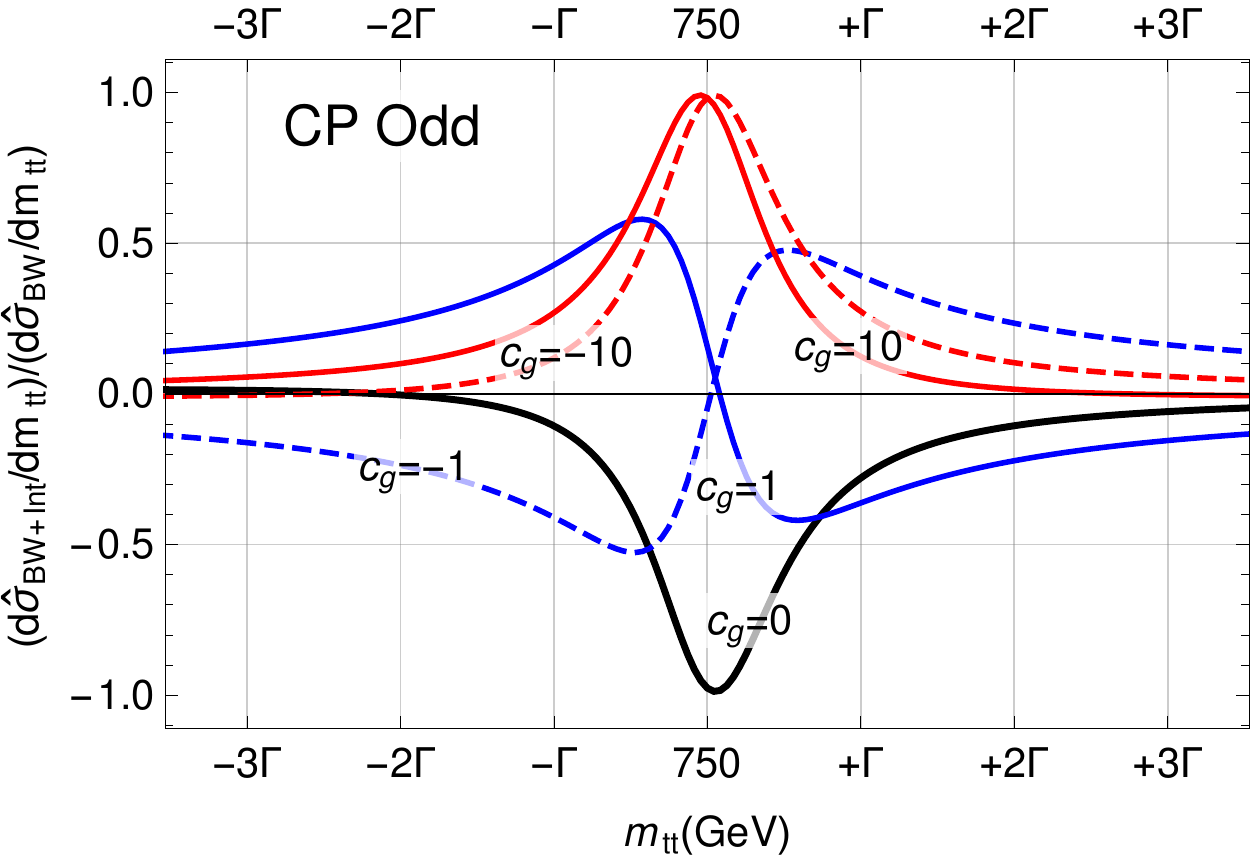}}
\subfigure{
\includegraphics[width=0.485\textwidth]{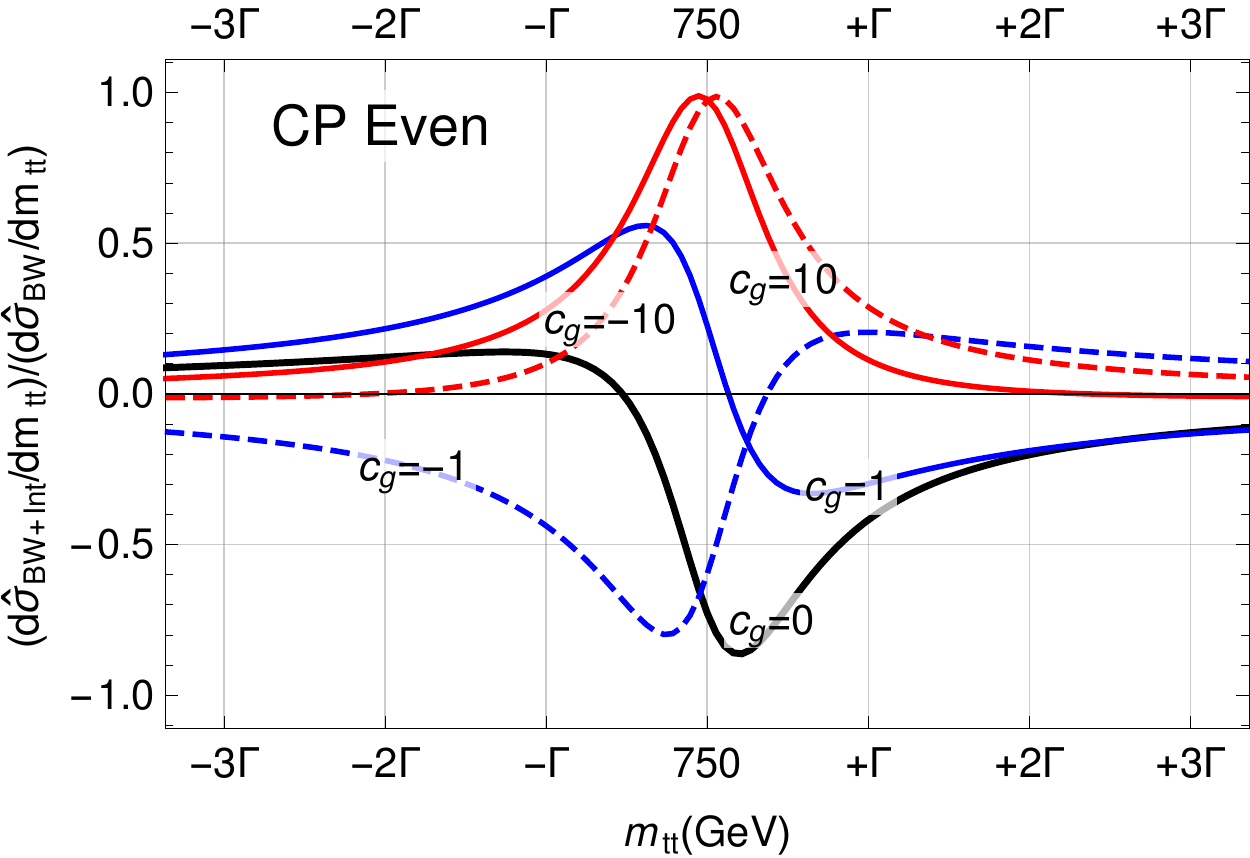}}
\subfigure{
\includegraphics[width=0.485\textwidth]{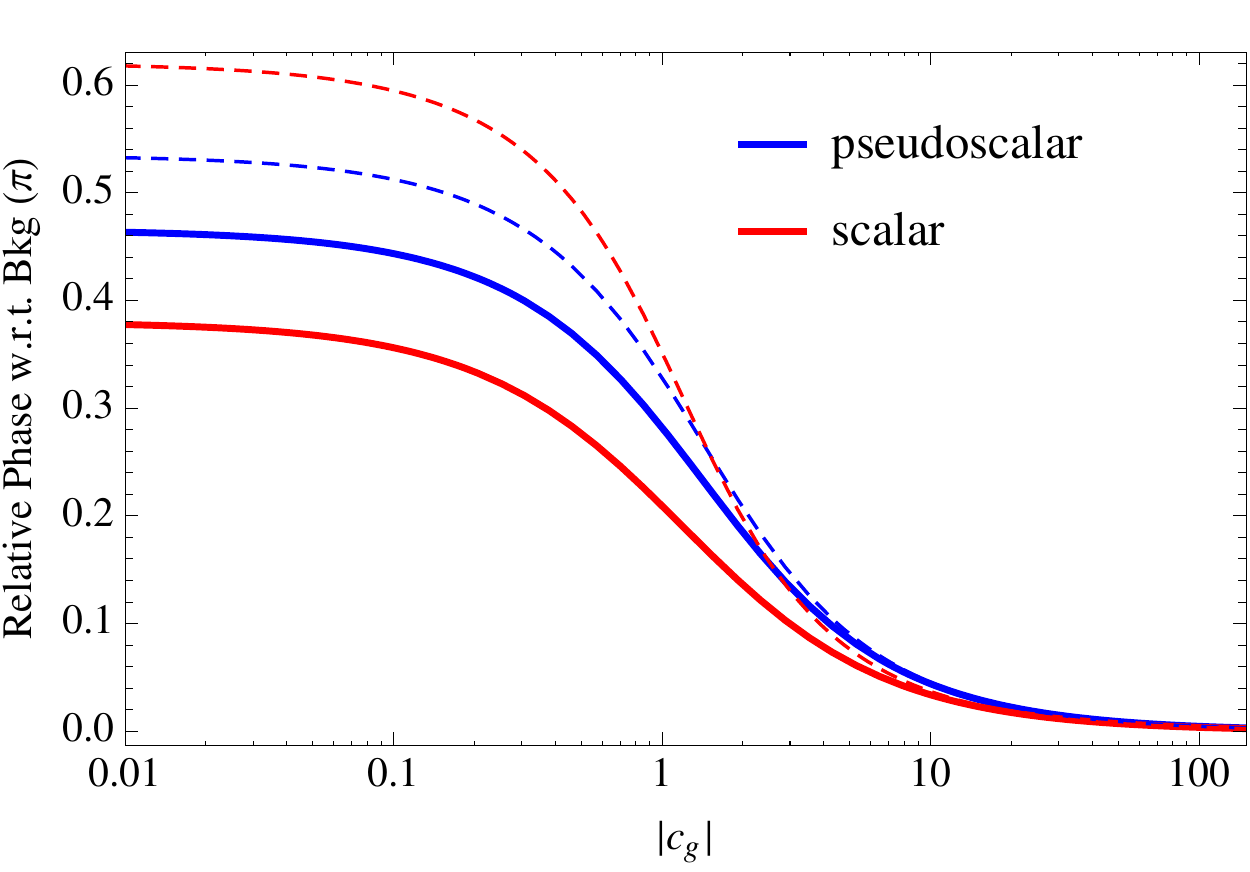}}
\subfigure{
\includegraphics[width=0.49\textwidth]{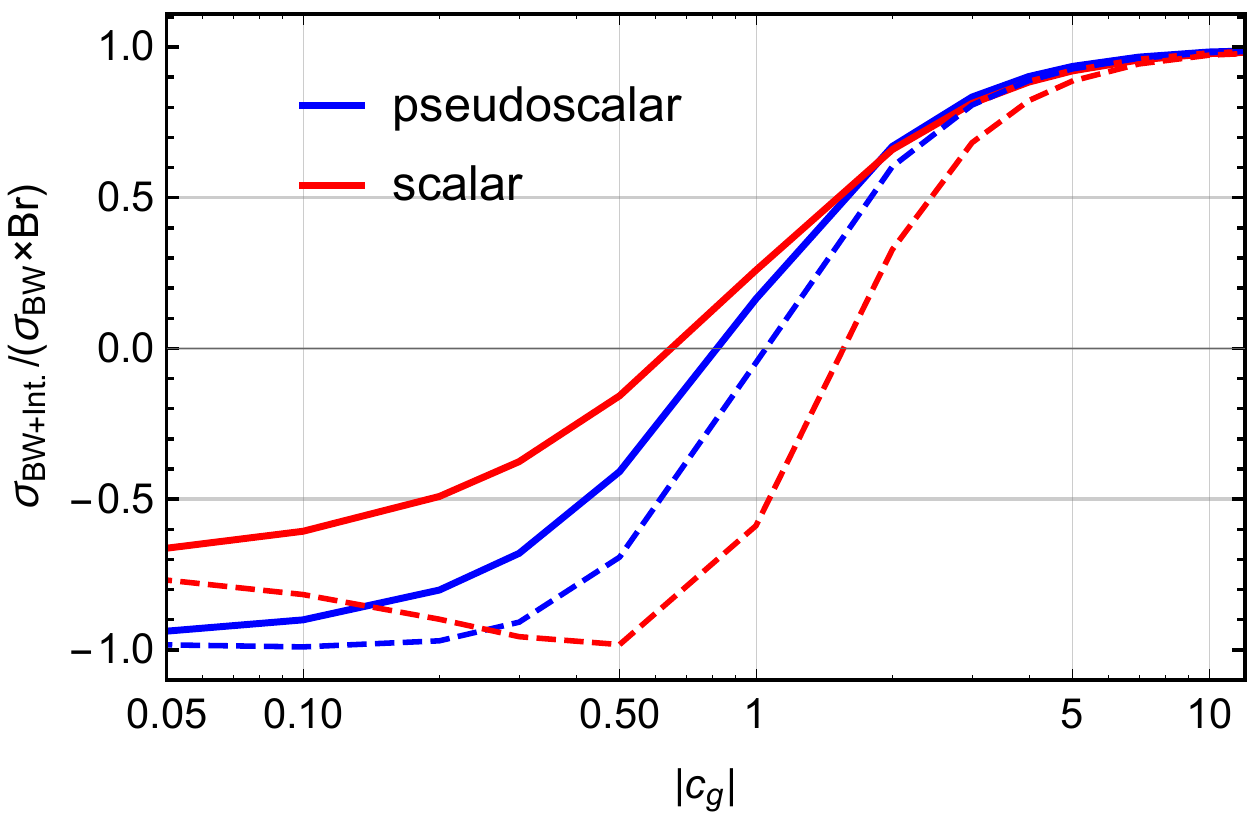}}
\caption[]{
{\bf Upper panel:} Representative lineshapes of differential distributions at parton level for the process $gg\to S\to t\bar t$ normalized to  the value of the Breit-Wigner contribution at the scalar mass pole,  as a function of the $t\bar t$ invariant mass, for various values of $c_g$. The corresponding values of $c_g$ are labelled on the lines and characterize the relative contribution from heavy colored particles to the gluon-heavy scalar vertex with respect to the top quark one.  The heavy particle S is a CP-odd  or a CP-even  scalar for the left and right panels, respectively. {\bf Lower left panel:} Relative phase $\theta_{\mathcal{\bar A}}$ \Edit{in units of $\pi$} as a function of  $c_g$. The dashed lines represent $(\pi-{\rm relative~phase)}$ for negative $c_g$.
{\bf Lower right panel:} ratio of the total signal rate within a $\pm 3\Gamma_S$ window for a  750 GeV scalar, including both resonance and interference contributions, over the naive resonance rate for the $gg\to S\to t\bar t$ process, see details in the text. For both lower panels, the CP-even and CP-odd cases are represented by the red and blue lines, respectively, while  the positive and negative values of $c_g$ are represented by the solid and dashed lines, respectively. For all the figures, the new physics scale $f$ is chosen to be 1 TeV.
 }\label{fig:750}
\end{figure}

In the lower left panel of Fig.~\ref{fig:750} we show how the relative phase $\theta_{\mathcal{\bar A}}$ with respect to the SM $gg\to t\bar t$ background varies as a function of $c_g$, as define in Eq.~\ref{eq:750rate}. 
The phases for the scalar and pseudoscalar are represented by the red and blue lines, respectively. The solid lines represent the relative phase for positive $c_g$, while the dashed lines represent  $\pi $ minus the relative phase for negative $c_g$. 
In the case of dominant $t\bar t$ contribution (low $c_g$), the relative phase is near $\pi/2$ ($2\pi/5$) for pseudoscalar (scalar). For comparable contribution from top-loop and heavy colored particle loop the phase is still as large as $\pi/4$, while when $c_g$ is greater than 10 the relative phases becomes negligible. 

In the upper panel of Fig.~\ref{fig:750} we show several lineshapes for the differential distribution for the $gg\to S\to t\bar t$ cross section as a function of  the $t\bar t$ invariant mass, for various benchmark values of $c_g$. The example cases of a 750 GeV pseudoscalar and scalar are displayed in the upper left and upper right panels, respectively. For clarity of presentation, we normalize the lineshapes to the Breit-Wigner parton level cross section at the scalar mass pole. We assume the total width is dominated by the partial decay to $t\bar t$, $\Gamma_{\rm total}\approx \Gamma_{t\bar t}$. The resulting lineshape behavior is independent of the precise normalization of the interaction strength $v/f$, and therefore we plot the lineshapes in units of the total width $\Gamma\propto v^2/f^2$. 
This can be understood since the signal amplitude does not depend on $v/f$ near the scalar mass pole: the numerator of the signal amplitude scales as scalar-top pair coupling squared, proportional to $v^2/f^2$, due to the production and decay vertex while the denominator is proportional to the total width, which is also proportional to $v^2/f^2$. Moreover, the overall lineshape is determined by the relative importance between the Breit-Wigner contribution and the interference contribution, which is characterized by the relative strength of the signal amplitude to the background amplitude, independent of $v^2/f^2$.
 
To better understand Fig.~\ref{fig:750}, let us discuss the different lineshape behaviors for different values of $c_g$.
For large values of $c_g$, for which the heavy colored particle loop dominates in the gluon-gluon-fusion production,\footnote{note that $t\bar t$ could still be the dominant decay channel in comparison with the loop-suppressed (e.g., $\alpha_s^2/(8\pi)^2$) decays to gluon pairs.} the resulting lineshape for the $t\bar t$ signal is governed by the Breit-Wigner contribution with a smaller contribution from the interference effect proportional to the real part of the propagator. This is shown by the red and red, dashed lines for $c_g=10$ and $c_g=-10$, respectively. 
For negligible  values of $c_g$, for which the top-loop dominates the production, the resulting lineshapes for the $t\bar t$ signal are pure dips as shown by the black curves for $c_g=0$. 
In the limit of large statistics, the bounds from bump search and dip search could be treated more or less equivalently. However, in these two limits, the constraints from the $t\bar t$ resonance search  should be interpreted with caution.\footnote{We note that the bump search itself is dominated by the systematic uncertainties and thus projections on this channel should be done in a careful way, otherwise, overly aggressive results can be obtained by blindly assuming statistical uncertainty dominance.  A detailed discussion follows in the next section.} 
A very different behavior occurs when the top-loop and heavy particle-loop contribution are comparable,
resulting in a bump-dip or a dip-bump structure, as shown in the blue lines and blue dashed lines for $c_g=1$ and $c_g=-1$, respectively. In such case, the smearing effects from the $t\bar t$ invariant mass reconstruction will flatten the dips and bumps in the lineshapes and render the experimental search much more challenging, as we shall see in the next section. 

In the lower right panel of Fig.~\ref{fig:750} we plot the ratio of the total $gg\to S\to t\bar t$ BSM rate to the naive rate obtained from
$\sigma(gg\to S)\times {\rm Br}(S\to t\bar t)$. The total rate includes the interference effect and is  defined by integrating the signal lineshape over the $\pm 3 \Gamma_S$ region.
In this figure we show the ratios for a heavy scalar and a pseudoscalar, with both signs of $c_g$, with the same line coding as the lower left panel. For low $|c_g|$, all cases are more of a dip structure and this ratio could be as small as $-1$ ($-0.7$) for the pseudoscalar (scalar). For sufficiently large $|c_g|$ ( $>$ 5) , the signal is  Breit-Wigner like and the ratio tends to be one as expected. For $c_g$ around unity, large cancellations occur. Furthermore, the sign of $c_g$ also plays a role in the  exact value of $c_g$ for which this ratio approaches  zero. The negative $c_g$ usually requires larger values to be dominant, as the new physics contributions must  first  cancel the real component from the top quark-loop. The ratio of the total $gg\to S\to t\bar t$ BSM rate to the naive 
$\sigma(gg\to S)\times {\rm Br}(S\to t\bar t)$ rate provides a crude estimate of the current collider constraints for a given 750 GeV scalar model in the $t\bar t$ channel. One can divide the current constraints on the $t\bar t$ production rate, which neglect the interference effects, by the absolute value of this ratio to obtain an  estimate of the constraints on the total production rate. 

For the process of $gg\to S\to VV$ and $gg\to S \to aa$, where $V$ represents SM electroweak gauge bosons ($\gamma$, $W$, $Z$) and $a$ is the light particle that later fakes the photon, using $\sigma(gg\to S)\times {\rm Br}(S\to VV, aa)$ is appropriate for the total BSM effect because of the smallness of interfering SM background. Still, the detailed lineshapes could be useful to determine the properties of the scalar~\cite{Jung:2015etr,Craig:2016iea,Martin:2016bgw}, although the effect is not very sizable and quite large statistics is needed.


\section{LHC Sensitivity}
\label{sec:LHC}
\subsection{Signal and background considerations}

The search for a new heavy scalar signal in the $gg\to S\to t\bar t$ channel at the hadron collider is challenging in various ways. The first challenge comes from the non-conventional dip, bump-dip, or dip-bump structures for which the normal bump search is not optimized. The second  is related to  the top-quark invariant mass reconstruction that smears the signal by a large amount. The bump and dip become less pronounced due to events in the bump that will populate the dip via mis-reconstruction of the invariant mass and vice versa. Indeed, the fact that events in one region are interpreted as events in the other one produces the smearing that results in a reduced excess or deficit  of events and  diminishes considerably the significance of the lineshape analysis.
 The third  significant challenge is due to  the systematic uncertainty associated with the large production cross section of the  SM top quark pairs, which is the irreducible background for $t\bar t$ resonance searches. The background cross section starts to increase quickly once the process is kinematically allowed, reaching its peak at an invariant mass near 400 GeV at 13 TeV LHC.
  In  Fig.~\ref{fig:minimalshape}, we show that the background statistical uncertainty (dashed gray contour) is very small compared to the systematic uncertainty (solid gray contour) that hides the signal lineshapes.  Consequently, reducing the systematical uncertainty is a key task in order to achieve sensitivity in this channel. 
Due to the difficulties  just mentioned, the search for a new heavy scalar in the $gg\to S\to t\bar t$ channel is  basically not constrained in the entire mass range slightly above the $t\bar t$ threshold. 
 In the following  we shall  re-evaluate the above challenges considering various techniques, and discuss their impact on the LHC reach.


\begin{figure}[t]
\centering
\subfigure{
\includegraphics[width=0.52\textwidth]{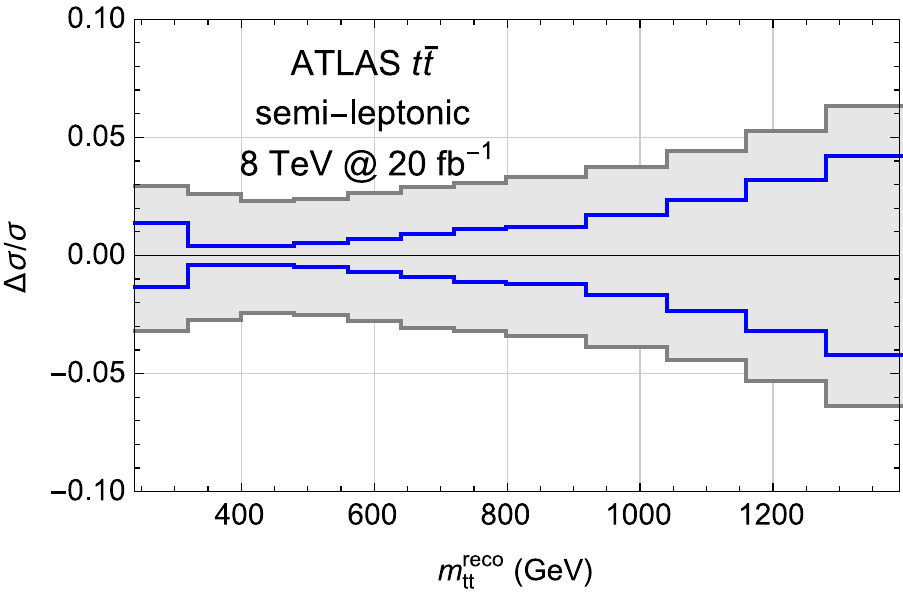}}
\caption[]{The total and statistical only bin-by-bin relative error as a function of the the $t\bar t$ invariant mass from the ATLAS 8 TeV analysis~\cite{Aad:2015fna} shown in gray and blue histograms, respectively. For further details, see the discussion in the text. 
}\label{fig:mtt_ATLAS}
\end{figure}
 
 The current result for a $t\bar t$  resonance search  performed by ATLAS~\cite{Aad:2015fna} results in approximately \Edit{$8\%$ (6\%)} smearing of  the reconstructed $t\bar t$ invariant mass distribution \Edit{at around 400 GeV (1 TeV)}. 
For our regions of interest, the signal mainly lies in the resolved-topology selection of the ATLAS search, for which the decay products of the hadronically decaying top quark are expected to be reconstructed in three small radius jets, in contrast to the boosted case. The resolved-topology is of relevance for our study since we  focus on the phenomenologically interesting  region below one TeV. \Edit{The CMS $t\bar t$ resonance search at 8 TeV has similar invariant mass resolution of around $10\%$~\cite{Khachatryan:2015oqa}. }

In Fig.~\ref{fig:mtt_ATLAS}\footnote{The numbers are obtained from the auxiliary material of  the ATLAS analysis~\cite{Aad:2015fna} available at \url{http://hepdata.cedar.ac.uk/view/ins1373299}. The current 13 TeV search has already shown better systematic control~\cite{ATLAS-CONF-2016-014} but the smallness of the systematics prohibits us from extracting the numbers accurately from the plot. CMS 8 TeV analysis has similar but slightly worse systematic uncertainties~\cite{Khachatryan:2015oqa}.} we show the current total uncertainty (gray band) and statistical uncertainty (blue band) achieved by the ATLAS 8 TeV analysis~\cite{Aad:2015fna}. The systematic uncertainty can be controlled at the level of about 2\% to 4\% in the mass range between  240~\gev~and 1~TeV. This search exploits the large data sample available from the LHC by marginalizing the nuisance parameters that characterize the systematic uncertainties. The uncertainties derived from this method  use the data more extensively than other more traditional treatments. 
The systematics for a lineshape search that correlates adjacent bins, such as the one we are considering  in our study {\it should} be comparable or  better than that of a  single bin. Therefore we expect that the systematic uncertainty values from the ATLAS study can be applied to our analysis. 
With higher integrated luminosity, we expect that the systematic uncertainties will improve.
On one hand, the large amount of $t\bar t$ events can be used to better understand the detector performance and reduce the systematic uncertainties. On the other hand, the large data set  also means that one  can afford a lower signal selection efficiency allowing for $t\bar t$ events with higher quality in terms of invariant mass reconstruction accuracies \Edit{and systematic uncertainties}.  Moreover, alternatively to the Monte-Carlo based method for background modeling used by the ATLAS study, one could consider the widely used data driven  background subtraction method that  tends to improve with  larger data sets. Many applications of this method
 show great advantage in  complex experimental environments. In addition, development in the analysis techniques may help further reduce the systematics~\cite{Kaplan:2008ie,Rehermann:2010vq,Plehn:2010st}.
 \Edit{The above arguments enable us to define scenarios for our study.}
 
\begin{table}[t]
\caption{Benchmarks for two LHC performance scenarios for the $t\bar t$ lineshape search at 13 TeV, motivated by current results from 8 TeV searches and assuming 30 fb$^{-1}$ and 3 ab$^{-1}$ of data, respectively.  Scenario A is based on a conservative assumption for the projected   $t\bar t$ invariant mass resolution and systematic uncertainties,  while Scenario B is based on a more aggressive assumption for both experimental parameters.}
\label{tab:LHCbenchmark}
\begin{center}
\begin{tabular}{|c|c|c|c|c|}
\hline
 & $\Delta m_{t\bar t}$ & Efficiency & Systematic Uncertainty \\ \hline
 Scenario A & 15\% & 8\% & ~4\% at 30$~\fbi$, halved at $3\abi$\\ \hline
 Scenario B & 8\% & 5\% & 4\% at 30$~\fbi$, scaled with $\sqrt{L}$\\ \hline
\end{tabular}
\end{center}
\end{table}%

\Edit{In Table~\ref{tab:LHCbenchmark}, we consider two scenarios for the $t\bar t$ lineshape search using the semi-leptonic $t\bar t$ sample.}
Scenario A is more conservative, \Edit{both for the invariant mass resolution and} the high luminosity projection, \Edit{while scenario B is more aggressive}.\footnote{
\Edit{In scenario B, we take an invariant mass resolution of $8\%$ throughout the mass range,  as quoted by ATLAS. In scenario A we take a very conservative value of 15\%, slightly above the value quoted by CMS.}
}  
Another relevant parameter is the signal selection efficiency. We chose 8\% signal selection efficiency (branching fraction included) for Scenario A. 
\Edit{
For scenario B, instead, we consider a lower signal efficiency of 5\%, allowing for a possible more strict requirement on data quality to allow for more optimistic assumptions on the smearing effects and the systematic uncertainties.}
As discussed earlier,  the current values of the systematic uncertainties can be as low as 2\% with the LHC 8 TeV data.  We assume a flat 4\% systematic uncertainty for the whole range 400$-$1000~GeV at 30$~\fbi$. In Scenario A we assume the systematics being halved with the full HL-LHC luminosity and in Scenario B we assume the systematics being scaled with the squared root of the total integrated luminosity.  We also choose a binning size of 5\% of the scalar mass in the $t\bar t$ invariant mass distribution. In most cases the experimental search  uses  the full information on each event,  hence binning is not necessary. However, in our simplified statistical treatment  binning is important, and given the size of the smearing effect, we consider a bin size of 5\% of the scalar mass  appropriate. For illustration purposes we show in  Fig.~\ref{fig:binned_smeared_compare} of the Appendix the signal lineshape before and after smearing and binning,  for the case of a pseudoscalar  of mass 550 GeV  with a Yukawa coupling $y_t=1$.

As discussed in earlier sections, many models contain a heavy scalar with different features and may also include two scalars of similar masses but different CP properties. The resulting lineshapes are very diverse, depending on the relative phase, new contributions to the effective gluon-gluon-scalar coupling and the precise separation between heavy scalar masses. As a first step, we propose a search for a single scalar  on the lineshape of the $t\bar t$ system,  performing a template fit in the differential distribution of the $t\bar t$ invariant mass. 

\begin{figure}[t]
\centering
\subfigure{
\includegraphics[width=0.46\textwidth]{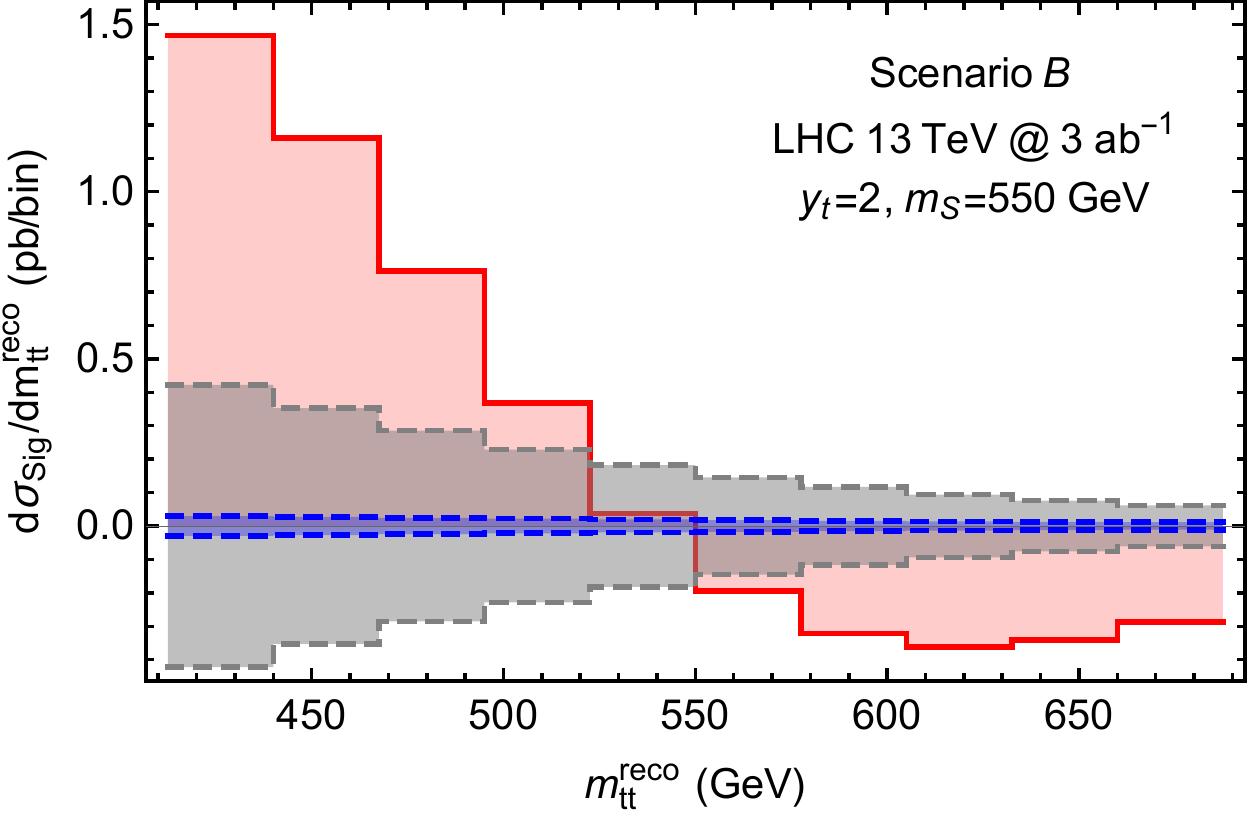}}
\subfigure{
\includegraphics[width=0.481\textwidth]{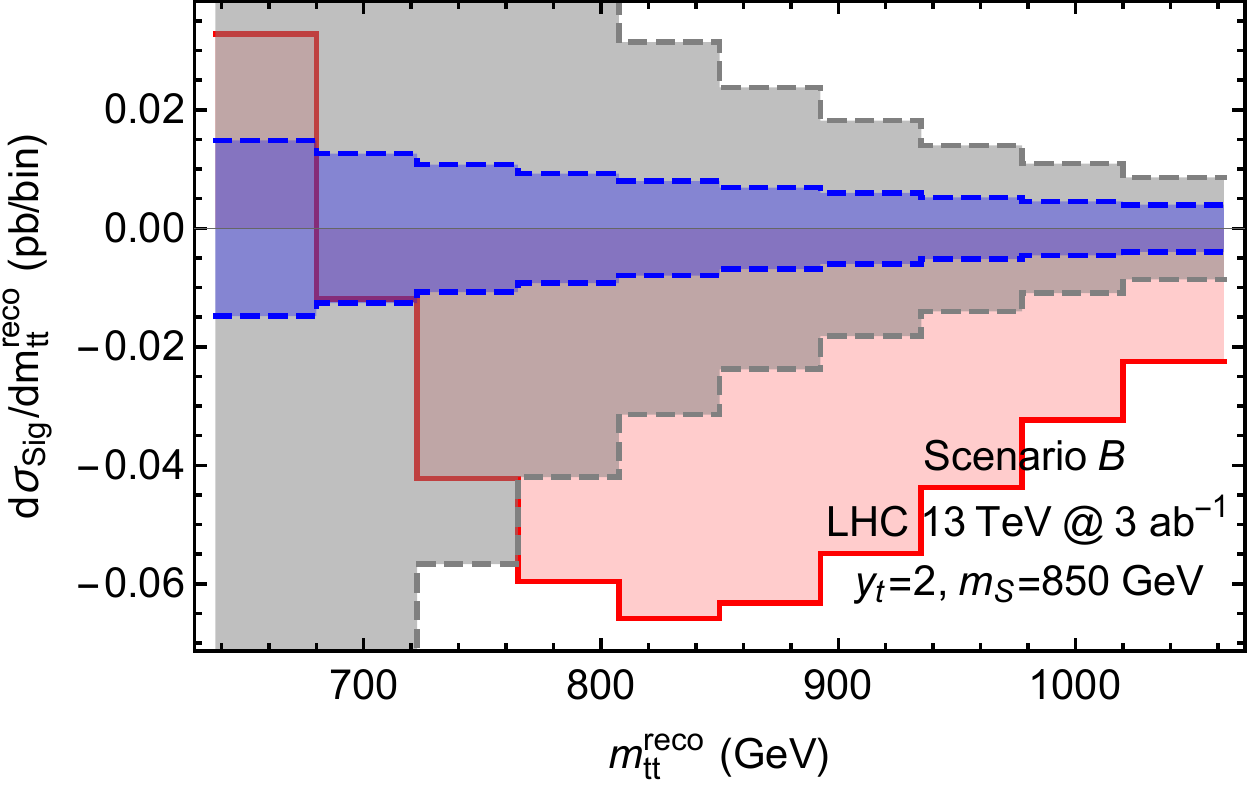}}
\caption[]{The binned differential distribution of the signal and background uncertainties in units of pb/bin. 
The red histograms are the binned signal histograms after background subtraction with the heavy scalar-top quark Yukawa coupling $y_t=2$. The blue and gray bands are the background statistical uncertainties and  total uncertainties after smearing and binning, respectively, for 3 $\abi$ of total  integrated luminosity in the semi-leptonic channel for the  performance scenario B. The left and right panels show results for a  heavy CP-odd scalar mass of 550 GeV and 850 GeV, respectively.
}\label{fig:mtt_smeared}
\end{figure}

In Fig.~\ref{fig:mtt_smeared} we show,  after smearing and binning, the resulting signal lineshapes  for a CP-odd scalar with masses of 550~GeV and 850~GeV,  in the left and right panels, respectively, for the baseline model with Yukawa coupling $y_t=2$.  The signal distributions feature, as shown by the red histograms,  a bump-dip structure for  the 550 GeV case and almost a pure dip structure for the 850 GeV case.  The statistical uncertainty and total uncertainty at 3~$\abi$ are shown in blue and gray histograms for scenario B, respectively. 
 As discussed in earlier sections, the systematic uncertainty is the dominant effect and reducing it by upgrading the detector, using data to calibrate the machine to the best achievable level, and improving the $t\bar t$ system mass reconstruction are crucial for further improvements and possible discovery in this channel. 

Based on the distributions and uncertainties shown in Table.~\ref{tab:LHCbenchmark} and Fig.~\ref{fig:mtt_smeared} and assuming a null BSM result  in the future data, we can project which region of BSM parameter space can be probed. We calculate the \Edit{significance squared} of the lineshape in the $(1\pm 0.25) m_S$ range, that is equivalent to considering a sum over 10 bins with a bin size of 5$\%$ of the scalar mass
 \beq
-\log(p)=\sum_{\rm 10 bins} \frac {n^2_{\rm sig}} {{n_{\rm bkg}+\delta_{\rm sys}^2 n^2_{\rm bkg}}}.
\label{eq:significance}
\eeq
In the above,  $n_{\rm sig}$ is the number of signal events (could be both positive and negative), $n_{\rm bkg}$ is the number of background events and $\delta_{\rm sys}$ is the systematic uncertainty. The p-value for the signal is then the sum of the significance in quadrature of the bins in the mass window of $(1\pm 0.25) m_S$.  This is the large background limit of the median expected significance for the likelihood ratio, where we have dropped two small corrections of order $|n_{\rm sig}|/n_{\rm bkg}$ and $\delta_{\rm sys}^2 n_{\rm bkg}$ according to the Asimov approximation~\cite{Cowan:2010js,statistics}. This treatment basically corresponds to a template fit in the invariant mass distribution of the $t\bar t$ system. We then translate this p-value into significance for a given signal model lineshape. We derive the projected limits as a function of the parameter space for specific models by generating  a grid of p-values and finding (multi-dimensional) contours of $2\-\sigma$ exclusion. Generating a grid of signal lineshapes with respect to model parameters is necessary for this search, even for the simplest baseline model, since the lineshape is a combination of the interference part proportional to $y^2_{t}$ and the Breit-Wigner contribution that, when off peak, is proportional to $y^4_{t}$.

It is worth to highlight that in the region where the SM background shape departs from simple polynomials, for example near the SM threshold peak around 400 GeV, additional uncertainties on the shape will enter. Simulation driven background estimations may become a better handle and different systematic uncertainties arise.
In addition to considering data driven estimation for the background,  high precision SM calculations are evidently of great importance. Indeed, in the case of sizable values of the  heavy scalar width,  there is important  interference between the signal and background at far off the peak, and this  might change the overall slope of the background estimation using side bands. Such effects could have an impact on  the sensitivity derived using the simplified procedure described in this study.

\Edit{
To summarize, in this section we propose to perform a lineshape search using the semi-leptonic $t\bar t$ channel in the resolved sample. We include the two leading effects, namely, smearing and the background normalization systematic uncertainties, and adopt an approximated statistical treatment given in Eq.~(\ref{eq:significance}). Further inclusion of the merged channel and other decay modes in the $t\bar t$ searches could improve the sensitivity, whereas the background shape uncertainties may affect our sensitivity estimation and need to be taken into account in future analyses.}

\subsection{Projections in model space}
\label{sec:results}

\begin{figure}[t]
\centering
\subfigure{
\includegraphics[width=0.485\textwidth]{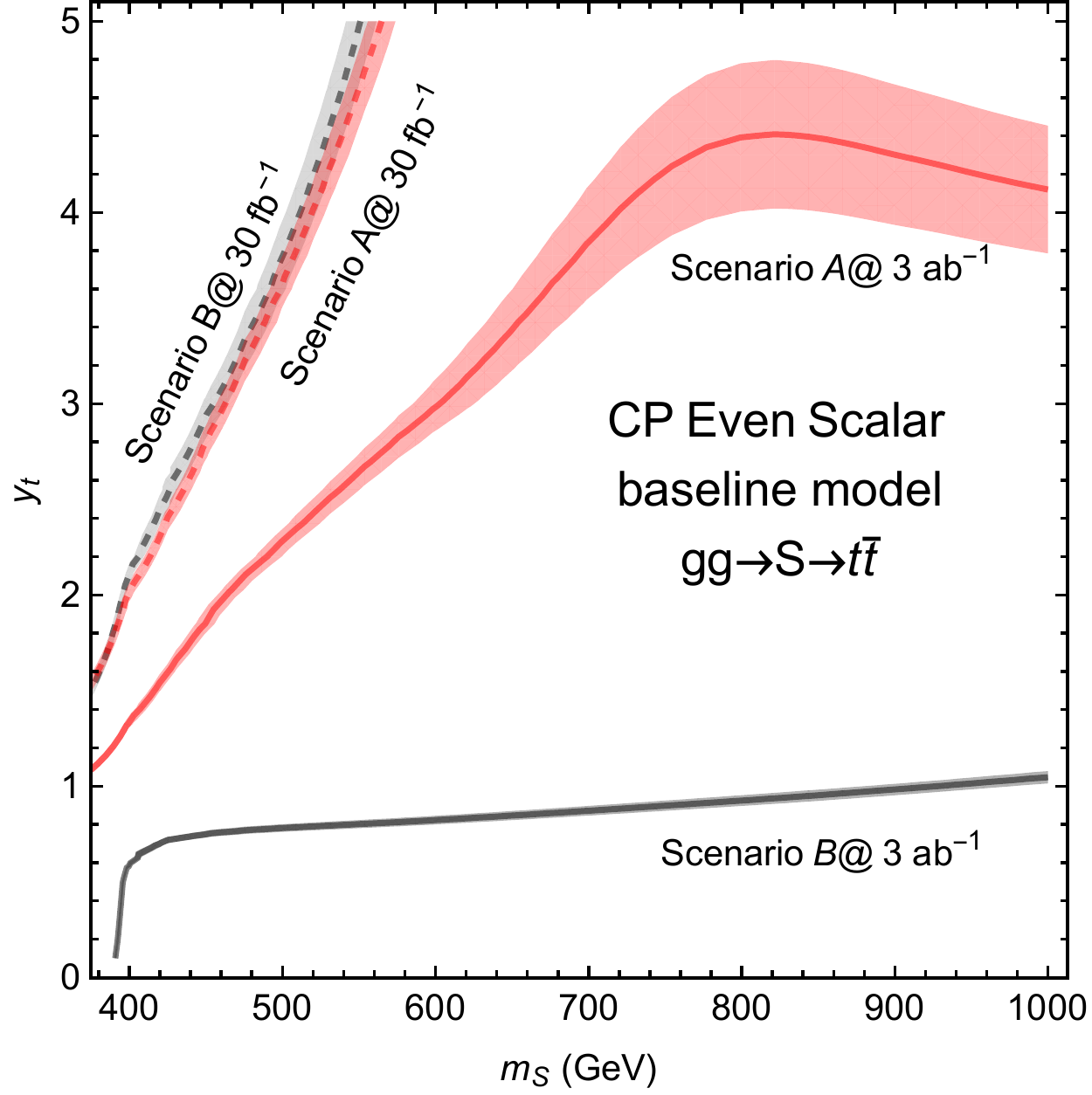}}
\subfigure{
\includegraphics[width=0.485\textwidth]{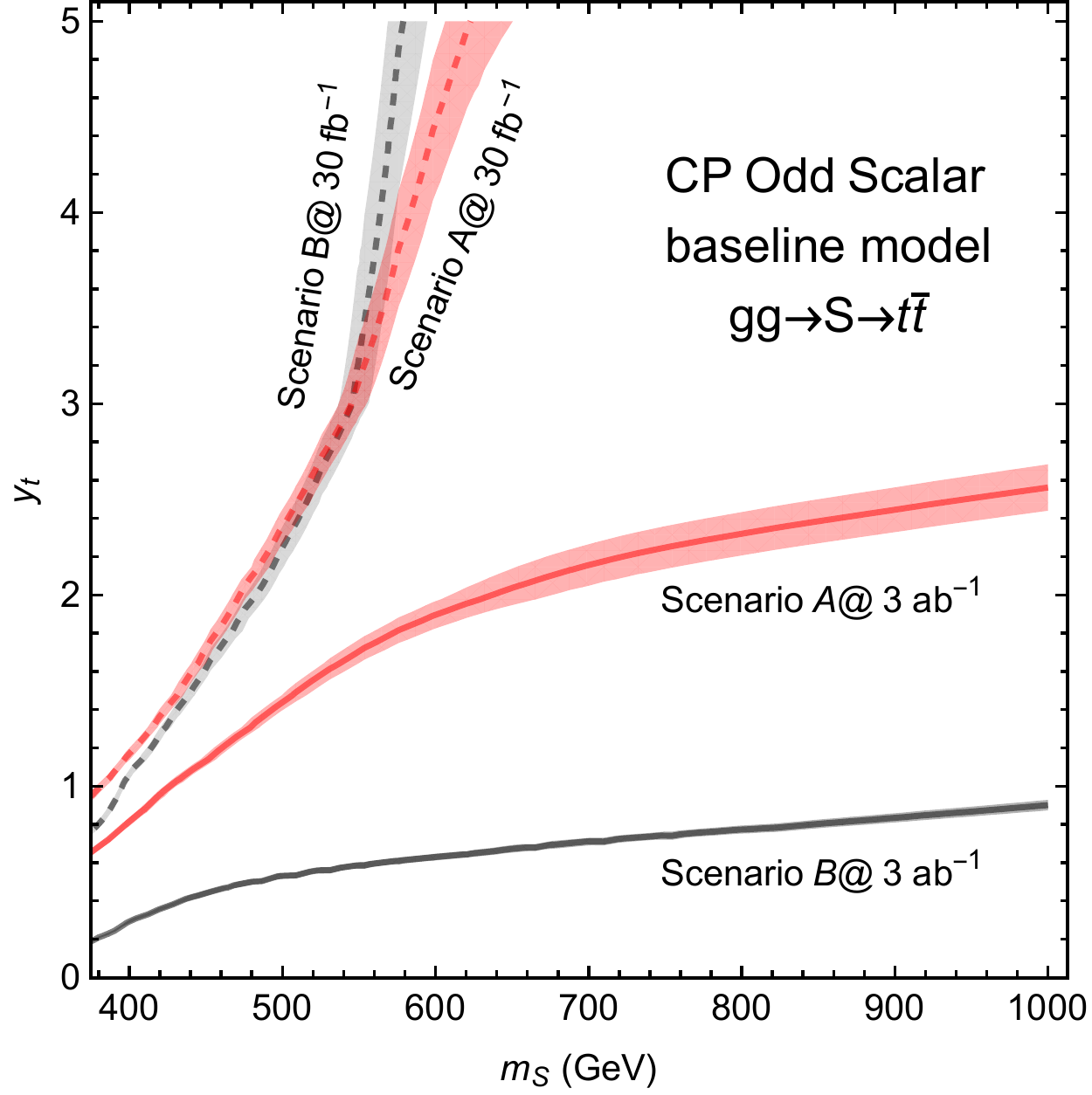}}
\caption[]{The projected 95\% C.L. exclusion limits on the top quark Yukawa coupling of the CP-even (left panel) and CP-odd (right panel)  heavy scalar $S$ at the LHC for the baseline model. 
The red and gray lines correspond to the performance scenarios A and B, tabulated in Table.~\ref{tab:LHCbenchmark}, and the regions above the curves are excluded. The solid and dashed lines show results for an integrated luminosity of $3~\abi$ and $30~\fbi$, respectively.
\Edit{As an illustration, the shaded bands indicate a variation of 5\% in the required significance to derive the limits.}
}\label{fig:sig_base_excl}
\end{figure}

In this section, we present the projected sensitivity of the $gg\to S\to t\bar t$ lineshape search in various model configurations, using the benchmark performance scenarios and statistical method depicted in the previous section. We first show the exclusions in the baseline model for a heavy CP-even or  CP-odd scalar, while later on we discuss the sensitivities in various scenarios of Type II 2HDMs. 

In Fig.~\ref{fig:sig_base_excl} we show the exclusion limit on the baseline model as a function of the  heavy scalar mass and its  Yukawa coupling to the top quark. The left panel shows the result for a CP-even scalar while the right panel is for a CP-odd scalar. The red and gray lines correspond to the $2\-\sigma$ exclusion limit in scenarios A and B as specified in Table.~\ref{tab:LHCbenchmark},  with the dashed and solid lines corresponding to LHC 13 TeV at $30~\fbi$ and $3~\abi$, respectively. The regions above the lines are excluded for each specific scenario  and  integrated luminosities, as labeled in the figure. \Edit{To illustrate the effects of  possible uncertainties in our statistical and binning treatment, we  present, as  an example,  shaded bands showing  a variation of 5\% in the required significance to derive the limits. }
In both scenarios A and B, the heavy CP-even (CP-odd) scalar in the baseline model can be excluded up to 450 (550) GeV  for a  Yukawa coupling  $y_t=3$ at $30~\fbi$. For the same value of $y_t$ and $3~\abi$, in Scenario A the reach increases to 650 GeV and beyond 1 TeV for the heavy CP-even and CP-odd scalars, respectively.
In Scenario B, the reach increases beyond 1 TeV for both a CP-even and  a CP-odd heavy scalar for a  heavy scalar-top Yukawa coupling of $y_t=1$ at $3~\abi$.  One can also consider the sensitivity for a fixed scalar mass at different luminosities and compare the exclusion reach  in the heavy scalar-top quark Yukawa coupling strength.
For example, the limit improves from 4.5 to 2.5, and from 4.5 to 0.7, for  a CP-even scalar mass of 550 GeV when  luminosity increases in scenario A and B, respectively. Comparing both performance scenarios, we observed that 
with  30 $\fbi$ of integrated luminosity, they have comparable reach, because the differences in signal efficiencies and  energy resolutions compensate each other. 
 However, the exclusion limits in the more aggressive performance scenario B at 3 $\abi$  yields a much better reach than in the conservative case of scenario A. This demonstrates again the crucial role that the systematic uncertainty plays in these projections.

\begin{figure}[t]
\centering
\subfigure{
\includegraphics[width=0.485\textwidth]{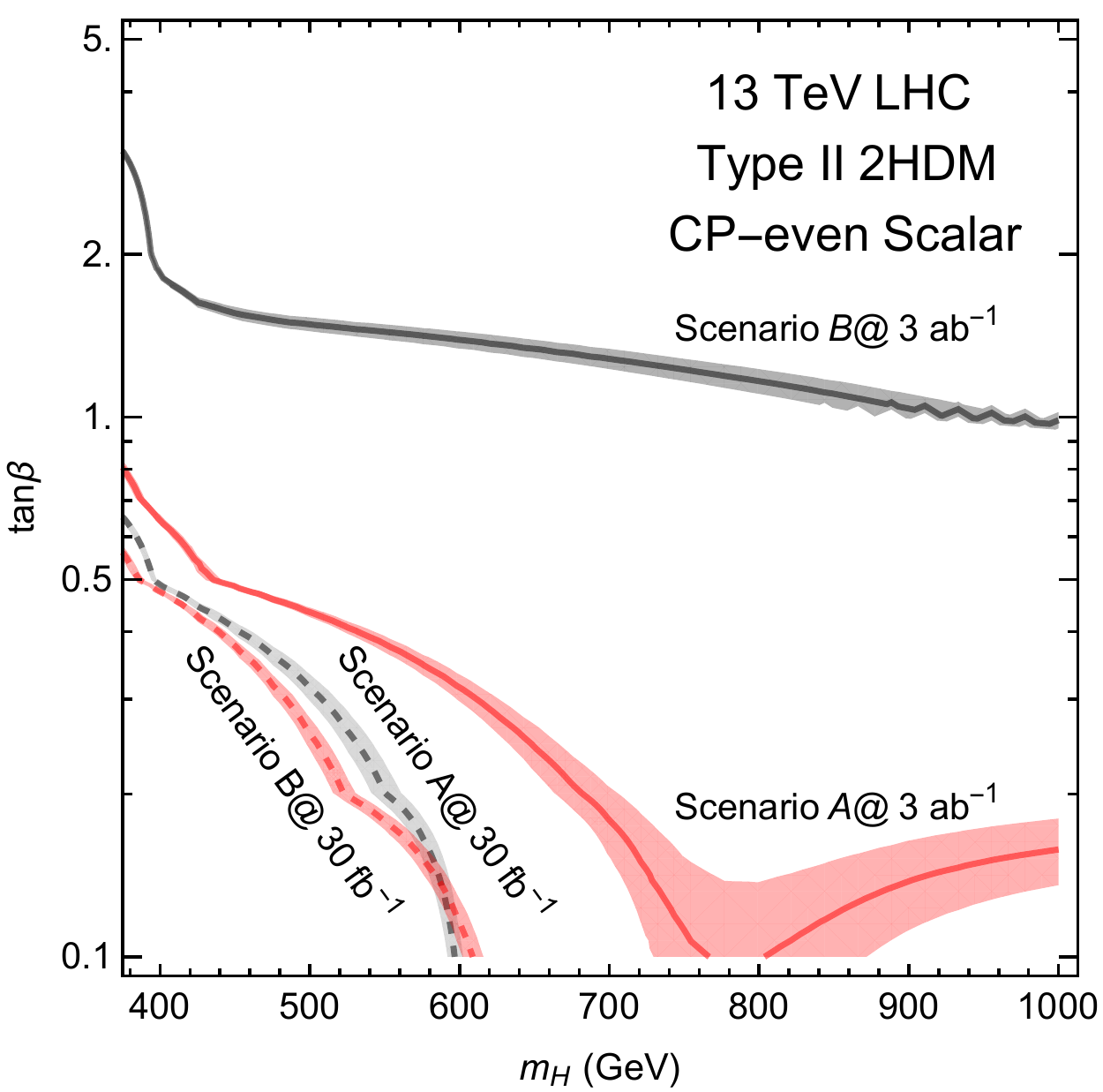}}
\subfigure{
\includegraphics[width=0.485\textwidth]{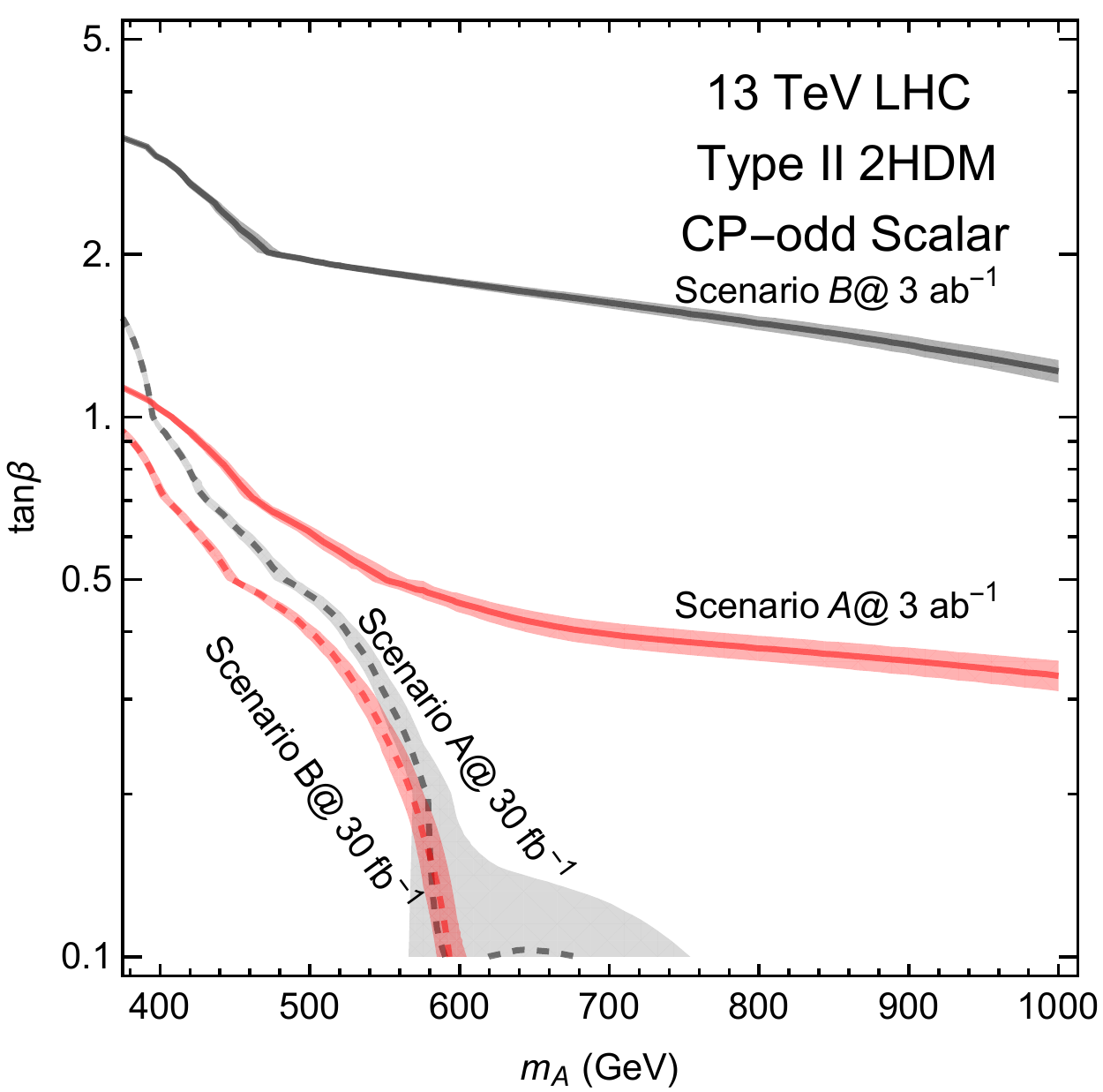}}
\caption[]{
The 95\% C.L. exclusion on the scalar mass--$\tan\beta$ plane for a type II 2HDM, including the effects of bottom quarks in the process. The regions below the curves are excluded. The result for the CP-even and CP-odd  scalars are shown in the left  and right panels, respectively. The color coding, lines  and legends are the same as in Fig.~\ref{fig:sig_base_excl}.}\label{fig:sig_2hdm_excl}
\end{figure}

\begin{figure}[t]
\centering
\subfigure{
\includegraphics[width=0.485\textwidth]{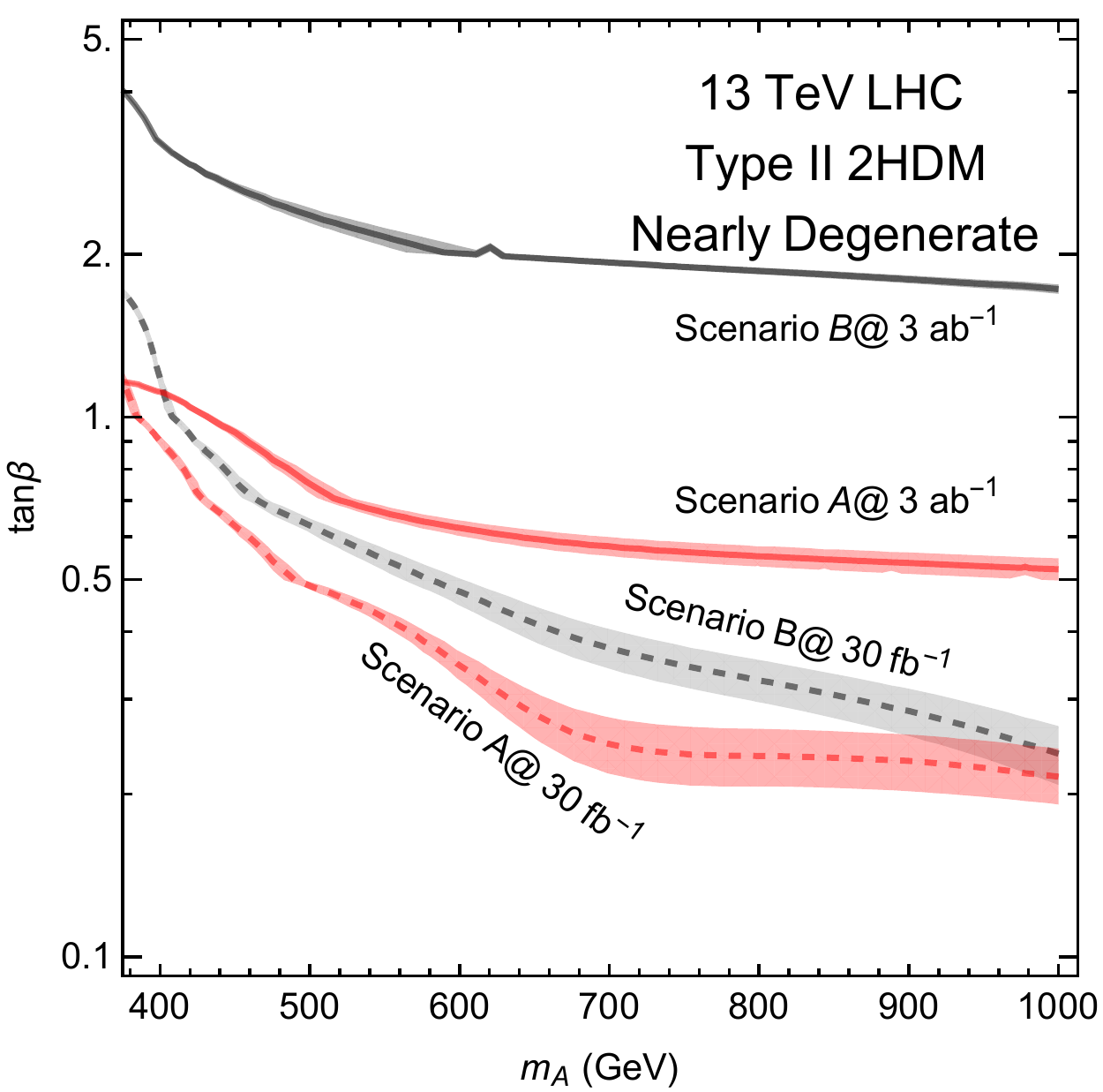}}
\subfigure{
\includegraphics[width=0.48\textwidth]{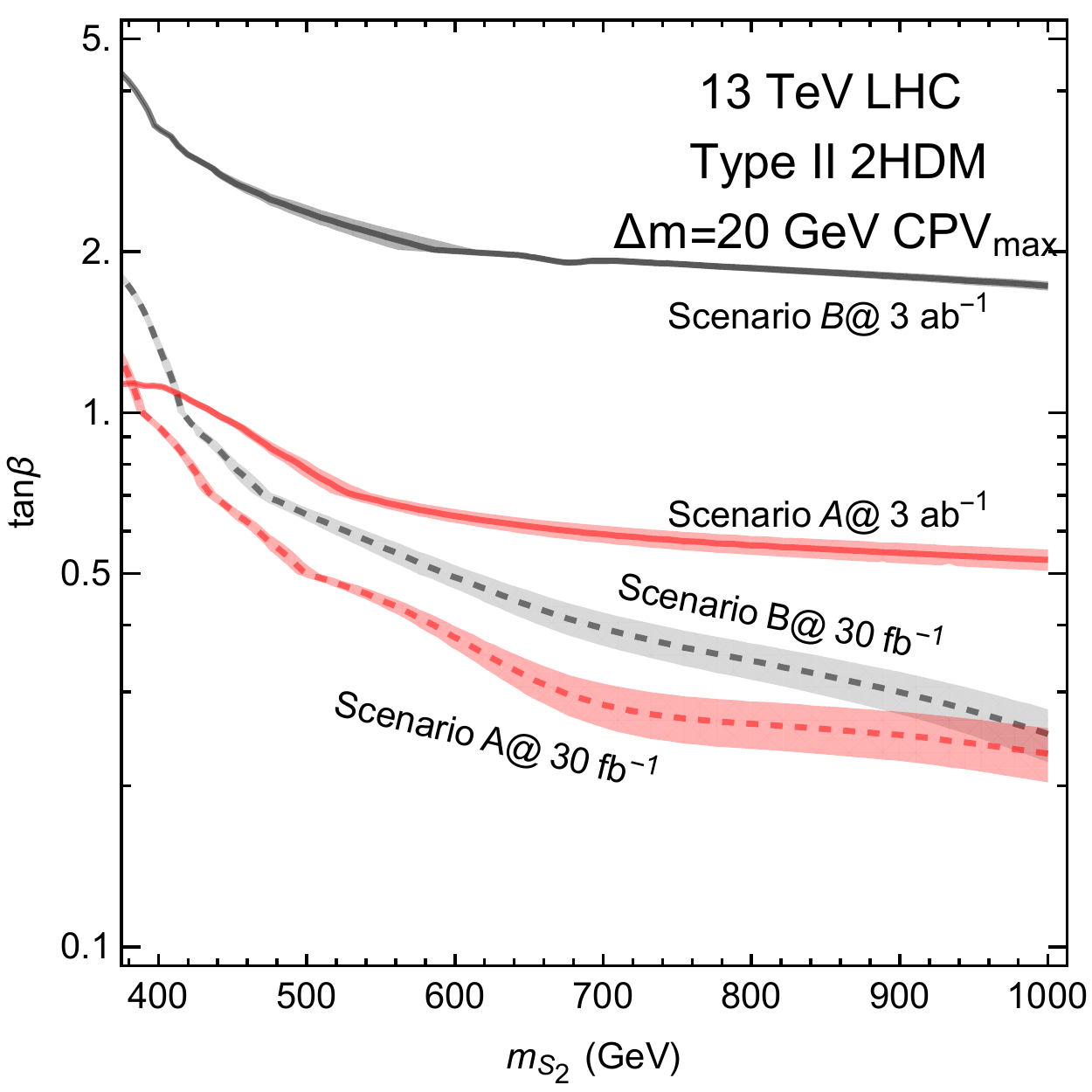}}
\caption[]{
The 95\% C.L. exclusion on the scalar mass--$\tan\beta$ plane for two nearly degenerate heavy neutral scalars in a Type II 2HDM with bottom quarks effects included. The results for  the CP-conserving  and maximal CP-violating ($\theta_{CP}$ of $\pi/4$) cases are shown in the left  and right panels, respectively. The color coding, lines  and legends are the same as in  Fig.~\ref{fig:sig_2hdm_excl}.
}\label{fig:2hdm_2s}
\end{figure}

Beyond the baseline model, we perform numerical studies for the Type II 2HDM including the bottom quark corrections in both the  production amplitudes and the decay widths. In Fig.~\ref{fig:sig_2hdm_excl} we show 95\% C.L. exclusion contours in the heavy  scalar mass--$\tan\beta$ plane. The legends are the same as in Fig.~\ref{fig:sig_base_excl}, but in this case the regions below the curves are excluded. For the CP-even scalar shown in the left panel,  the reach in mass is only around 450 GeV for most scenarios for $\tan\beta=0.5$. This moderate reach is mainly due to the $\beta^2 $ suppression factor and the smaller value of the loop function.  The restricted reach for the CP-even scalar case   is only overcome in the more aggressive scenario B at 3 $\abi$, probing mass scales up to 1 TeV.  For the CP-odd scalar shown in the right panel, the exclusion reach is much better in comparison with the previous case.  Masses up to 
450 GeV to 600 GeV can be probed  for $\tan\beta=0.5$ in most scenarios. For the more aggressive performance scenario B, masses as high as 1 TeV can be probed with  3 $\abi$ of integrated luminosity. 
In all cases considering just one new heavy scalar at a time the reach is limited to values of $\tan\beta < 2$, with the small exception of $M_A$  around 400 GeV that can reach up to $\tan\beta =3$.
For the  scalar mass near the top-quark pair threshold below 400 GeV,  the 2HDM reach for the CP-even scalar is worse than the baseline model, and this is due to the $ t \bar t$ branching fraction suppression from the scalar decays into $b\bar b$. 


Given that the two heavy scalar bosons often have  nearly degenerate masses in many 2HDMs, In Fig.~\ref{fig:2hdm_2s} we study such case and show both the CP conserving (left panel) and maximal CPV (right panel) situations for the heavy Higgs sector in Type II 2HDM. For the CP-conserving case, we assume a  mass splitting between the two scalars as in the MSSM following 
, and hence the 
 reach is equivalent to that of a CP-conserving MSSM in the limit of heavy squarks, in which both scalar signals simply add, as discussed in Sec.~\ref{sec:2scalars}. For the CPV case, we assume a constant splitting between the two scalars of 20~GeV, and a new interference effect between the two scalars emerges. This effect slightly changes the  projected limits. We show the exclusion limits in the $\tan\beta$-$m_A$ (-$m_{S_2}$) plane for the CP conserving (violating) case. The labels are the same as in Fig.~\ref{fig:sig_2hdm_excl}. The reach increases to 480~GeV and 600 GeV for $\tan\beta=0.5$ with 30~$\fbi$ of integrated luminosity, for scenarios A and B, respectively. In the HL-LHC environment masses as high as 1 TeV for $\tan\beta=0.5$ and $\tan\beta=2$  can be probed in scenarios A and B, respectively. Values of $\tan\beta \simeq 4$ may be  accessible in the restricted region of  heavy resonance masses below 500 GeV.

\section{Conclusions}
\label{sec:conclusion}

Heavy scalars are well motivated in many extensions of the standard model. The typical dominant production and decay mode of a   heavy scalar  at hadron colliders is via gluon fusion with the subsequent decay to a top quark pair, 
 $gg\to S\to t\bar t$.  
 In our baseline model for which the $ggS$ effective vertex is dominantly mediated by the top-quark triangle diagram, the signal amplitude interferes with the SM background in
  a complex way. The total signal lineshape is mainly driven by the behavior of the loop-function evaluated at $\sqrt {\hat s}$ close to the heavy scalar mass. As a result one can obtain a lineshape that behaves as a pure bump, a bump-dip, or a pure dip structure depending on the value of the  scalar masses.  
In many BSM models, additional corrections come, for example, from non-trivial CP phases associated with the heavy scalar, the existence of nearly degenerate scalars, or additional loop contributions from stops or vector-like quarks.  In this paper we study the relevant features of  top pair production from heavy scalars  and evaluate  the LHC physics potential in various BSM scenarios.

We first discuss the behavior of the loop-function and the resulting lineshapes in the baseline model for a purely CP-even or CP-odd scalar, as well as a scalar that is a mixture of CP eigenstates. 
We obtain different  behaviors of the lineshapes parametrized by the additional phases generated by the loop function of the triangle diagram.  
We consider the case of  nearly degenerate heavy scalars that may exist in a 2HDM, and show that  their contributions add to each other in the  lineshapes, resulting  in  an enhancement of the features of the lineshape structure and providing a good opportunity for detecting  the signal.
In the case where the two quasi-degenerate eigenstates are  CP admixtures, there is also a small additional interference effect between them that further modifies the lineshape structure.  

We also study BSM scenarios with additional heavy particles contributing to the gluon induced loop function, such as scalar-quarks or vector-like quarks. 
We have analyzed different illustrative scenarios: one in which the heavy particle contribution dominates over the SM top quark one, and two others in which the new heavy particle effects are  comparable or smaller to those of the top quark.
 In the case that the heavy particle contribution dominates, the lineshape is  given by the standard  Breit-Wigner resonance bump plus the off peak interference  bump-dip structure, which is proportional to the real part of the propagator. We exemplify the above behavior for a vector-like quark model with VLQs heavier than half of the heavy resonance mass.
 Examples of moderate or comparable effects to those induced by  the SM top quark loop are shown in the case of Supersymmetry.
When the stops have a negligible left-right mixing, their effects are just a small perturbation to the baseline model. In the case of sizable mixing, instead, the stop loop may yield a visible contribution and change the lineshape significantly. 

We provide a study for the search of a heavy scalar with additional contributions to the production process in the context of EFT.
The specific lineshapes could play a crucial role in interpreting the results  and projecting the discovery potential in the $t\bar t$ channel.  We find that if a scalar mass is in the 700 GeV ballpark and the gluon-gluon-fusion process is dominantly induced through top-quark loops, the resulting lineshape is a pure dip. If, instead, there are contributions from additional heavy colored particles  comparable to those of the top quark, the resulting lineshape is a bump-dip structure, where large cancellations occur once smearing effects are taken into account.  
We define a ratio of the total signal cross section, including interference effects, to the naive signal cross section without interference, that serves as a penalty factor in deriving a crude estimate of the collider limits for a heavy scalar particle decaying to top quark pairs.

In the final part of this paper we study  the LHC sensitivity to the $t\bar t$ signal from heavy scalars for two plausible LHC performance  scenarios.
The real challenge resides in the systematic uncertainties in this channel and one should make use of the large amount of accumulated data to reduce them 
through a better  detector calibration and advanced analysis techniques. 
 We propose to complement the normal bump search with a lineshape search that makes better use of the bump-dip structure by counting both the excess and deficit as part of the BSM signal. 
We present the results of our proposed lineshape search for various  BSM cases. First we  consider a heavy scalar in the baseline model and show that  
a CP-odd scalar with a top Yukawa coupling $y_t=2$ can be excluded at the  95\% C.L. up to 500 GeV in both performance scenarios  with 30~$\fbi$ of data. The reach can be  extended all the way up  to 1 TeV for both a CP-even and a CP-odd scalar, with a  top Yukawa coupling as low as $y_t=1$,  for the most aggressive performance scenario with 3~$\abi$. 
Next we consider 2HDMs for which the bottom quark effects  in  the loop-induced  production mode and  the scalar total width become relevant in the intermediate and large $\tan\beta$ regime. 
We derive the expected 95\% C.L. exclusion limits for both the CP-even and CP-odd scalars in the $\tan\beta$-scalar mass plane.  Considering one  heavy neutral Higgs  boson at a time, values of $\tan \beta$ of order 1 can be probed  for the whole mass range up to 1 TeV for the most aggressive performance scenario with 3~$\abi$ of data. In the case that the two heavy scalars are nearly degenerate in mass, we consider the combined search of both particles decaying into $t\bar t$ and show the improved  95\% C.L. exclusion limits both for the CP-conserving and CPV cases.


A few remarks before concluding: 
Other BSM searches, such as those involving  color or weakly interacting scalar  octets may also profit from the discussions in this paper.
Moreover, higher order corrections may affect the large destructive interference effects, due to the possible  reduction of the phase-space overlap  between signal and background, as well as the possible addition of new relative phases.  For example,
a next-to-leading-order study on the 2HDM~\cite{Bernreuther:2015fts} showed some distortions of the interference effects and a more detailed analysis focussing on the  specific changes due to the  higher orders corrections  will be of great interest.
Finally, there may be other observables for which the interference effects are reduced, providing additional information on the signal. For example, considering angular distributions could provide additional sensitivity for the $gg\to S\to t\bar t$ search. However, our preliminary  investigation of these observables shows very limited gain, in agreement with Ref.~\cite{Craig:2015jba}, mainly due to large systematic uncertainties and smearing effects.  Another useful handle could be to consider  top quark polarization to reduce  the background without significantly affecting the signal.\footnote{For some recent development and an overview of the top quark polarization tagger, see Ref.~\cite{Tweedie:2014yda}.} Provided higher statistics, polarization may also help to identify the CP properties of the  heavy scalar. We intend to further explore these  points  in a future study. 

\acknowledgments {We thank Y. Bai, N. Craig, S. Dawson, K. Howe, P. Fox, T. Han, R. Harnik, S. Jung, W.-Y. Keung, K.C. Kong, I. Lewis, J. Lykken, S. Martin, S. Su, B. Tweedie, L.T. Wang, C. Williams and  H. Zhang for helpful discussions.
Fermilab is operated by Fermi Research Alliance, LLC under Contract No. DE-AC02-07CH11359 with the U.S. Department of Energy. 
Z.L. thanks the Aspen Center for Physics, which is supported by National Science Foundation grant PHY-1066293,  for the hospitality during the final stage of this work. 
} 

{\flushleft{\it Notes added:} 
Upon the completion of this work, a study~\cite{Djouadi:2016ack} on  interference effects partially overlapping with our discussion in Sec.~\ref{sec:750} and a numerical study~\cite{Hespel:2016qaf} focusing on the interference effects in the $t\bar t$ channel in the framework of a 2HDM appeared. ATLAS has recently published a conference note using the LHC 8 TeV dataset to search for heavy scalars in the $t\bar t$ channel with the interference effect taken into account~\cite{ATLAS-CONF-2016-073}. The result shows comparabale limits to our projections for the baseline model, however assuming a better systematic control and a larger signal mass window.}

\appendix
\section{Additional Formulae and Benchmarks}
\label{sec:details}

The loop functions in Eq.~\ref{eq:gSgg} and Eq.~\ref{eq:ggh} $I_{\frac 1 2}(\tau_t)$ and $\tilde I_{\frac 1 2}(\tau_t)$ can be alternatively written as,
\bea
\tau_t&&=\frac {\hat s} {4 m_t^2}, ~\beta\equiv \sqrt{1-\frac {4m_t^2} {\hat s}}=\sqrt {1-1/\tau_t}\nonumber\\&&\ f(\tau) =\left\{ 
\begin{array}{lcl}
\arcsin^2(\frac 1 {\sqrt{1-\beta^2}})\ &\ ,& {\rm for\ } \tau\leq 1,\\ \nonumber
-\frac 1 4\left(\log\frac {1+\beta} {1-\beta}-i \pi\right)^2\ &\ ,& {\rm for\ } \tau> 1
\end{array}\right.\\
I_{1/2}(\tau)&&= \sqrt{1-\beta^2} (1 + \beta^2 f(\tau)),~~~\tilde I_{1/2}(\tau)= \sqrt{1-\beta^2}f(\tau) .
\eea
Since the imaginary part of the loop functions come only from $f(\tau)$ when $\tau>1$, a direct check using Cutkosky rules indicates that the  coefficient in front of $f(\tau)$ for $I(\tau)$ should have a factor $\beta^2$  suppression with respect to the $f(\tau)$  coefficient  in $\tilde I(\tau)$. Various closed-form loop functions in this paper are cross-checked using  the package {\tt Program X}~\cite{Patel:2015tea}.

The energy-dependent heavy scalar partial width is,
\beq
\Gamma_q(\hat s) (S\to q\bar q) = \frac 3 {16\pi} (y_q^2 \beta^2+\tilde y_q^2)\beta \frac {\hat s} {m_S},
\label{eq:width} 
\eeq
with $\beta\equiv \sqrt {1-\frac {4m_q^2} {\hat s}}$.
The energy dependence of the width has negligible effect for the narrow width case but becomes more relevant for the intermediate to  large width case.

The tree-level expressions for the SM QCD parton level differential cross sections for the $t\bar t$  background are
\bea
\frac {d\hat \sigma (gg\to t\bar t)} {d z}&&=\frac {\pi \alpha_s^2} {12 \hat s } \beta \left(\frac {\hat s^2} {\hat u \hat t}-\frac 9 4\right) \frac {\hat u^2+\hat t^2} {\hat s^2}\nonumber \\
\frac {d\hat \sigma (q\bar q \to t\bar t)} {d z}&&=\frac {2\pi \alpha_s^2} {9 \hat s } \beta  \frac {\hat u^2+\hat t^2} {\hat s^2}
\label{eq:bkgparton}
\eea
where $\hat s,~\hat t,~\hat u$ are the Mandelstam variables and $z$ is the cosine of the scattering angle between an incoming parton and the top quark. For collider analyses with detector acceptance, the events from different regions of the phase space cannot be used in equal manner, especially for light jets, we thus provide the differential distributions. However, as the top quark is not very boosted and even forward events with  $z=\pm 1$ can be detected, in practice, we integrate $z$ over the full range  $[-1,1]$ in our simplified analysis.

For the scalar quarks  the following abbreviations are used in the main text (in particular $X_{u,d}$ and $Y_{u,d}$ are defined in the alignment limit),
\bea
D_L^u&&= \frac 1 2 {m_W^2} (1-\frac 1  3 \tan^2 \theta_W) \cos2\beta  \nonumber\\
D_R^u&&=\frac 2 3 {m_W^2}  \tan^2 \theta_W \cos2\beta \nonumber\\
D_L^d&&=- \frac 1 2 {m_W^2}  (1+\frac 1  3 \tan^2 \theta_W) \cos2\beta  \nonumber\\
D_R^d&&=- \frac 1 3 {m_W^2} \tan^2 \theta_W \cos2\beta \label{eq:SUSYdef}\\ 
X_u&&=A_u - \frac \mu {\tan\beta}\nonumber\\
X_d&&=A_d - \mu\tan\beta \nonumber\\
Y_u&&=\frac {A_u} {\tan\beta} +\mu \nonumber\\
Y_d&&=A_b \tan\beta + \mu\nonumber,
\eea
where $\theta_W$ is the Weinberg angle.
The stop parameters used in Fig.~\ref{fig:SUSY} are,
\bea
{\rm zero~LR~mixing:} && m_{Q_3}=900~\gev, m_{t_R}=400~\gev, X_t=0 \nonumber\\
{\rm mh_{\rm max}^*:} && m_{Q_3}=900~\gev, m_{t_R}=540~\gev, Y_t= 2 X_t = 3415~\gev
\label{eq:SUSYpara}
\eea
and the corresponding stop mass eigenstates are,
\bea
{\rm zero~LR~mixing:} && m_{\tilde t_1}=436~\gev, m_{\tilde t_2}=916~\gev \nonumber\\
{\rm mh_{\rm max}^*:} && m_{\tilde t_1}=433~\gev, m_{\tilde t_2}=987~\gev \nonumber
\eea

\begin{figure}[t]
\centering
\subfigure{
\includegraphics[width=0.52\textwidth]{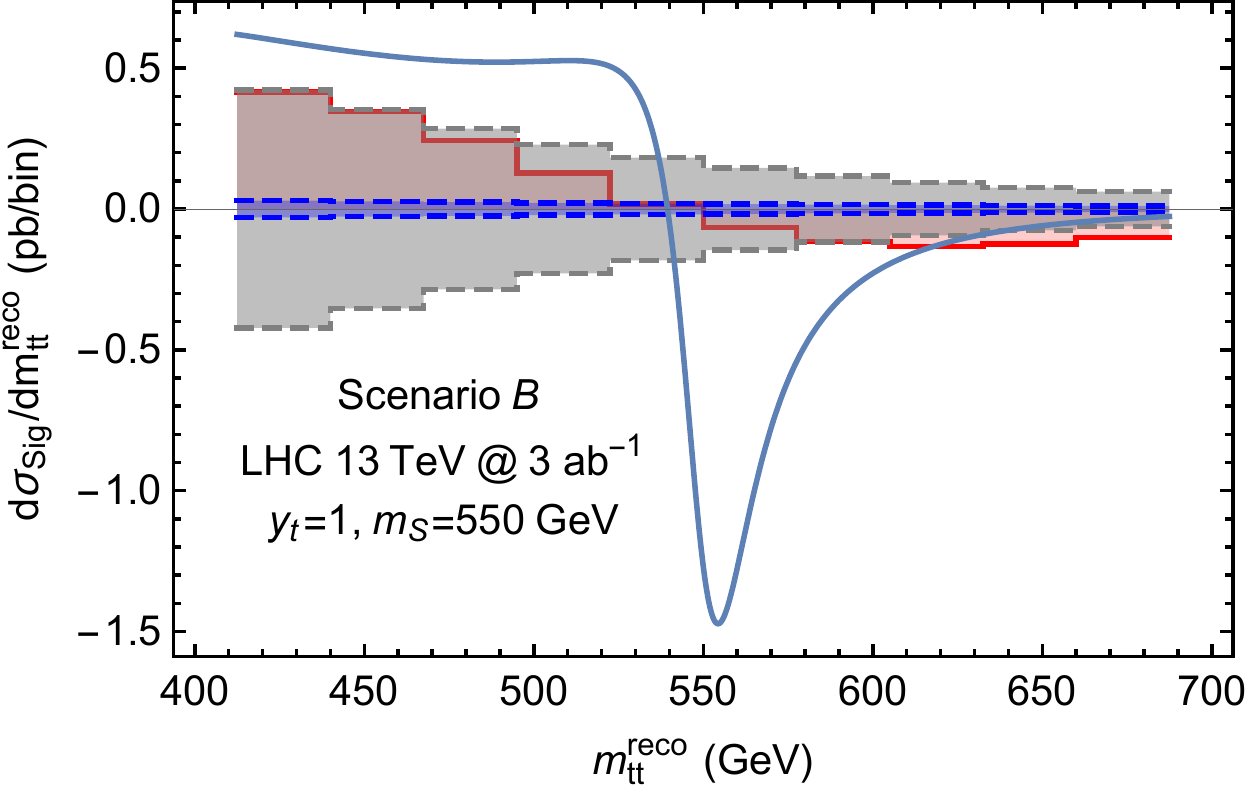}}
\caption[]{ The blue line is the differential distribution for the signal before smearing and binning, while the histograms are the same as defined in Fig. \ref{fig:mtt_smeared}.
The signal is chosen to be a 550 GeV pseudoscalar with Yukawa coupling to the  top quark, $y_t=1$.}\label{fig:binned_smeared_compare}
\end{figure}

\bibliographystyle{JHEP}
\bibliography{references1}

\providecommand{\href}[2]{#2}\begingroup\raggedright\begin{thebibliography}{10}

\bibitem{Flores:1982pr}
R.~A. Flores and M.~Sher, {\it {Higgs Masses in the Standard, Multi-Higgs and
  Supersymmetric Models}},  {\em Annals Phys.} {\bf 148} (1983) 95.

\bibitem{Gunion:1984yn}
J.~F. Gunion and H.~E. Haber, {\it {Higgs Bosons in Supersymmetric Models.
  1.}},  {\em Nucl. Phys.} {\bf B272} (1986) 1. [Erratum: Nucl.
  Phys.B402,567(1993)].

\bibitem{Djouadi:2005gj}
A.~Djouadi, {\it {The Anatomy of electro-weak symmetry breaking. II. The Higgs
  bosons in the minimal supersymmetric model}},  {\em Phys. Rept.} {\bf 459}
  (2008) 1--241, [\href{http://arxiv.org/abs/hep-ph/0503173}{{\tt
  hep-ph/0503173}}].

\bibitem{Gripaios:2009pe}
B.~Gripaios, A.~Pomarol, F.~Riva, and J.~Serra, {\it {Beyond the Minimal
  Composite Higgs Model}},  {\em JHEP} {\bf 04} (2009) 070,
  [\href{http://arxiv.org/abs/0902.1483}{{\tt arXiv:0902.1483}}].

\bibitem{ArkaniHamed:1998nn}
N.~Arkani-Hamed, S.~Dimopoulos, and G.~R. Dvali, {\it {Phenomenology,
  astrophysics and cosmology of theories with submillimeter dimensions and TeV
  scale quantum gravity}},  {\em Phys. Rev.} {\bf D59} (1999) 086004,
  [\href{http://arxiv.org/abs/hep-ph/9807344}{{\tt hep-ph/9807344}}].

\bibitem{Randall:1999ee}
L.~Randall and R.~Sundrum, {\it {A Large mass hierarchy from a small extra
  dimension}},  {\em Phys. Rev. Lett.} {\bf 83} (1999) 3370--3373,
  [\href{http://arxiv.org/abs/hep-ph/9905221}{{\tt hep-ph/9905221}}].

\bibitem{Randall:1999vf}
L.~Randall and R.~Sundrum, {\it {An Alternative to compactification}},  {\em
  Phys. Rev. Lett.} {\bf 83} (1999) 4690--4693,
  [\href{http://arxiv.org/abs/hep-th/9906064}{{\tt hep-th/9906064}}].

\bibitem{Georgi:1974sy}
H.~Georgi and S.~L. Glashow, {\it {Unity of All Elementary Particle Forces}},
  {\em Phys. Rev. Lett.} {\bf 32} (1974) 438--441.

\bibitem{Branco:2011iw}
G.~C. Branco, P.~M. Ferreira, L.~Lavoura, M.~N. Rebelo, M.~Sher, and J.~P.
  Silva, {\it {Theory and phenomenology of two-Higgs-doublet models}},  {\em
  Phys. Rept.} {\bf 516} (2012) 1--102,
  [\href{http://arxiv.org/abs/1106.0034}{{\tt arXiv:1106.0034}}].

\bibitem{Dicus:1994bm}
D.~Dicus, A.~Stange, and S.~Willenbrock, {\it {Higgs decay to top quarks at
  hadron colliders}},  {\em Phys. Lett.} {\bf B333} (1994) 126--131,
  [\href{http://arxiv.org/abs/hep-ph/9404359}{{\tt hep-ph/9404359}}].

\bibitem{Barcelo:2010bm}
R.~Barcelo and M.~Masip, {\it {Extra Higgs bosons in $t\bar{t}$ production at
  the LHC}},  {\em Phys. Rev.} {\bf D81} (2010) 075019,
  [\href{http://arxiv.org/abs/1001.5456}{{\tt arXiv:1001.5456}}].

\bibitem{Barger:2011pu}
V.~Barger, W.-Y. Keung, and B.~Yencho, {\it {Azimuthal Correlations in Top Pair
  Decays and The Effects of New Heavy Scalars}},  {\em Phys. Rev.} {\bf D85}
  (2012) 034016, [\href{http://arxiv.org/abs/1112.5173}{{\tt
  arXiv:1112.5173}}].

\bibitem{Bai:2014fkl}
Y.~Bai and W.-Y. Keung, {\it {Can vanishing mass-on-shell interactions generate
  a dip at colliders?}},  {\em Int. J. Mod. Phys.} {\bf A30} (2015), no.~20
  1550120, [\href{http://arxiv.org/abs/1407.6355}{{\tt arXiv:1407.6355}}].

\bibitem{Jung:2015gta}
S.~Jung, J.~Song, and Y.~W. Yoon, {\it {Dip or nothingness of a Higgs resonance
  from the interference with a complex phase}},  {\em Phys. Rev.} {\bf D92}
  (2015), no.~5 055009, [\href{http://arxiv.org/abs/1505.00291}{{\tt
  arXiv:1505.00291}}].

\bibitem{Craig:2015jba}
N.~Craig, F.~D'Eramo, P.~Draper, S.~Thomas, and H.~Zhang, {\it {The Hunt for
  the Rest of the Higgs Bosons}},  {\em JHEP} {\bf 06} (2015) 137,
  [\href{http://arxiv.org/abs/1504.04630}{{\tt arXiv:1504.04630}}].

\bibitem{ATLAS-CONF-2015-081}
{\it {Search for resonances decaying to photon pairs in 3.2 fb$^{-1}$ of $pp$
  collisions at $\sqrt{s}$ = 13 TeV with the ATLAS detector}},  Tech. Rep.
  ATLAS-CONF-2015-081, CERN, Geneva, Dec, 2015.

\bibitem{CMS-PAS-EXO-15-004}
{\bf CMS Collaboration} Collaboration, {\it {Search for new physics in high
  mass diphoton events in proton-proton collisions at $\sqrt{s} = 13$ TeV}},
  Tech. Rep. CMS-PAS-EXO-15-004, CERN, Geneva, 2015.

\bibitem{Pumplin:1970kp}
J.~Pumplin, {\it {Diffraction Dissociation and the Reaction gamma p
  $\rightarrow$ pi+ pi- p}},  {\em Phys. Rev.} {\bf D2} (1970) 1859.

\bibitem{Bauer:1970hk}
T.~Bauer, {\it {Anomalous real part in the t matrices of unstable particles}},
  {\em Phys. Rev. Lett.} {\bf 25} (1970) 485--488.

\bibitem{Basdevant:1977ya}
J.~L. Basdevant and E.~L. Berger, {\it {Unitary Coupled-Channel Analysis of
  Diffractive Production of the a1 Resonance}},  {\em Phys. Rev.} {\bf D16}
  (1977) 657.

\bibitem{Basdevant:1978tx}
J.~L. Basdevant and E.~L. Berger, {\it {Unstable Particle Scattering and an
  Analytic Quasiunitary Isobar Model}},  {\em Phys. Rev.} {\bf D19} (1979) 239.

\bibitem{Carena:2000yi}
M.~Carena, J.~R. Ellis, A.~Pilaftsis, and C.~E.~M. Wagner, {\it
  {Renormalization group improved effective potential for the MSSM Higgs sector
  with explicit CP violation}},  {\em Nucl. Phys.} {\bf B586} (2000) 92--140,
  [\href{http://arxiv.org/abs/hep-ph/0003180}{{\tt hep-ph/0003180}}].

\bibitem{Carena:2015uoe}
M.~Carena, J.~Ellis, J.~S. Lee, A.~Pilaftsis, and C.~E.~M. Wagner, {\it {CP
  Violation in Heavy MSSM Higgs Scenarios}},  {\em JHEP} {\bf 02} (2016) 123,
  [\href{http://arxiv.org/abs/1512.00437}{{\tt arXiv:1512.00437}}].

\bibitem{Ball:2014uwa}
{\bf NNPDF} Collaboration, R.~D. Ball et~al., {\it {Parton distributions for
  the LHC Run II}},  {\em JHEP} {\bf 04} (2015) 040,
  [\href{http://arxiv.org/abs/1410.8849}{{\tt arXiv:1410.8849}}].

\bibitem{Carena:2013ooa}
M.~Carena, I.~Low, N.~R. Shah, and C.~E.~M. Wagner, {\it {Impersonating the
  Standard Model Higgs Boson: Alignment without Decoupling}},  {\em JHEP} {\bf
  04} (2014) 015, [\href{http://arxiv.org/abs/1310.2248}{{\tt
  arXiv:1310.2248}}].

\bibitem{Dev:2014yca}
P.~S. Bhupal~Dev and A.~Pilaftsis, {\it {Maximally Symmetric Two Higgs Doublet
  Model with Natural Standard Model Alignment}},  {\em JHEP} {\bf 12} (2014)
  024, [\href{http://arxiv.org/abs/1408.3405}{{\tt arXiv:1408.3405}}].
  [Erratum: JHEP11,147(2015)].

\bibitem{Hajer:2015gka}
J.~Hajer, Y.-Y. Li, T.~Liu, and J.~F.~H. Shiu, {\it {Heavy Higgs Bosons at 14
  TeV and 100 TeV}},  \href{http://arxiv.org/abs/1504.07617}{{\tt
  arXiv:1504.07617}}.

\bibitem{Chen:2015fca}
N.~Chen, J.~Li, and Y.~Liu, {\it {LHC searches for heavy neutral Higgs bosons
  with a top jet substructure analysis}},  {\em Phys. Rev.} {\bf D93} (2016),
  no.~9 095013, [\href{http://arxiv.org/abs/1509.03848}{{\tt
  arXiv:1509.03848}}].

\bibitem{Gori:2016zto}
S.~Gori, I.-W. Kim, N.~R. Shah, and K.~M. Zurek, {\it {Closing the Wedge:
  Search Strategies for Extended Higgs Sectors with Heavy Flavor Final
  States}},  {\em Phys. Rev.} {\bf D93} (2016), no.~7 075038,
  [\href{http://arxiv.org/abs/1602.02782}{{\tt arXiv:1602.02782}}].

\bibitem{Craig:2016ygr}
N.~Craig, J.~Hajer, Y.-Y. Li, T.~Liu, and H.~Zhang, {\it {Heavy Higgs Bosons at
  Low $\tan \beta$: from the LHC to 100 TeV}},
  \href{http://arxiv.org/abs/1605.08744}{{\tt arXiv:1605.08744}}.

\bibitem{Goncalves:2016qhh}
D.~Goncalves and D.~Lopez-Val, {\it {Boosting to identify: pseudoscalar
  searches with di-leptonic tops}},
  \href{http://arxiv.org/abs/1607.08614}{{\tt arXiv:1607.08614}}.

\bibitem{Dobrescu:2013gza}
B.~A. Dobrescu and A.~D. Peterson, {\it {W${'}$ signatures with odd Higgs
  particles}},  {\em JHEP} {\bf 08} (2014) 078,
  [\href{http://arxiv.org/abs/1312.1999}{{\tt arXiv:1312.1999}}].

\bibitem{Dobrescu:2015yba}
B.~A. Dobrescu and Z.~Liu, {\it {Heavy Higgs bosons and the 2 TeV W$^{?}$
  boson}},  {\em JHEP} {\bf 10} (2015) 118,
  [\href{http://arxiv.org/abs/1507.01923}{{\tt arXiv:1507.01923}}].

\bibitem{RamseyMusolf:2006vr}
M.~J. Ramsey-Musolf and S.~Su, {\it {Low Energy Precision Test of
  Supersymmetry}},  {\em Phys. Rept.} {\bf 456} (2008) 1--88,
  [\href{http://arxiv.org/abs/hep-ph/0612057}{{\tt hep-ph/0612057}}].

\bibitem{vonGersdorff:2015fta}
G.~von Gersdorff, E.~Pont—n, and R.~Rosenfeld, {\it {The Dynamical Composite
  Higgs}},  {\em JHEP} {\bf 06} (2015) 119,
  [\href{http://arxiv.org/abs/1502.07340}{{\tt arXiv:1502.07340}}].

\bibitem{Fichet:2016xvs}
S.~Fichet, G.~von Gersdorff, E.~Pont—n, and R.~Rosenfeld, {\it {The Excitation
  of the Global Symmetry-Breaking Vacuum in Composite Higgs Models}},
  \href{http://arxiv.org/abs/1607.03125}{{\tt arXiv:1607.03125}}.

\bibitem{Fichet:2016xpw}
S.~Fichet, G.~von Gersdorff, E.~Pont—n, and R.~Rosenfeld, {\it {The Global
  Higgs as a Signal for Compositeness at the LHC}},
  \href{http://arxiv.org/abs/1608.01995}{{\tt arXiv:1608.01995}}.

\bibitem{Carena:2012xa}
M.~Carena, I.~Low, and C.~E.~M. Wagner, {\it {Implications of a Modified Higgs
  to Diphoton Decay Width}},  {\em JHEP} {\bf 08} (2012) 060,
  [\href{http://arxiv.org/abs/1206.1082}{{\tt arXiv:1206.1082}}].

\bibitem{Carena:2002qg}
M.~Carena, S.~Heinemeyer, C.~E.~M. Wagner, and G.~Weiglein, {\it {Suggestions
  for benchmark scenarios for MSSM Higgs boson searches at hadron colliders}},
  {\em Eur. Phys. J.} {\bf C26} (2003) 601--607,
  [\href{http://arxiv.org/abs/hep-ph/0202167}{{\tt hep-ph/0202167}}].

\bibitem{Carena:2005ek}
M.~Carena, S.~Heinemeyer, C.~E.~M. Wagner, and G.~Weiglein, {\it {MSSM Higgs
  boson searches at the Tevatron and the LHC: Impact of different benchmark
  scenarios}},  {\em Eur. Phys. J.} {\bf C45} (2006) 797--814,
  [\href{http://arxiv.org/abs/hep-ph/0511023}{{\tt hep-ph/0511023}}].

\bibitem{Carena:2013ytb}
M.~Carena, S.~Heinemeyer, O.~StŒl, C.~E.~M. Wagner, and G.~Weiglein, {\it {MSSM
  Higgs Boson Searches at the LHC: Benchmark Scenarios after the Discovery of a
  Higgs-like Particle}},  {\em Eur. Phys. J.} {\bf C73} (2013), no.~9 2552,
  [\href{http://arxiv.org/abs/1302.7033}{{\tt arXiv:1302.7033}}].

\bibitem{Franceschini:2015kwy}
R.~Franceschini, G.~F. Giudice, J.~F. Kamenik, M.~McCullough, A.~Pomarol,
  R.~Rattazzi, M.~Redi, F.~Riva, A.~Strumia, and R.~Torre, {\it {What is the
  $\gamma \gamma$ resonance at 750 GeV?}},  {\em JHEP} {\bf 03} (2016) 144,
  [\href{http://arxiv.org/abs/1512.04933}{{\tt arXiv:1512.04933}}].

\bibitem{Franceschini:2016gxv}
R.~Franceschini, G.~F. Giudice, J.~F. Kamenik, M.~McCullough, F.~Riva,
  A.~Strumia, and R.~Torre, {\it {Digamma, what next?}},
  \href{http://arxiv.org/abs/1604.06446}{{\tt arXiv:1604.06446}}.

\bibitem{Strumia:2016wys}
A.~Strumia, {\it {Interpreting the 750 GeV digamma excess: a review}},  2016.
\newblock \href{http://arxiv.org/abs/1605.09401}{{\tt arXiv:1605.09401}}.

\bibitem{Angelescu:2015uiz}
A.~Angelescu, A.~Djouadi, and G.~Moreau, {\it {Scenarii for interpretations of
  the LHC diphoton excess: two Higgs doublets and vector-like quarks and
  leptons}},  {\em Phys. Lett.} {\bf B756} (2016) 126--132,
  [\href{http://arxiv.org/abs/1512.04921}{{\tt arXiv:1512.04921}}].

\bibitem{Buttazzo:2015txu}
D.~Buttazzo, A.~Greljo, and D.~Marzocca, {\it {Knocking on new physics? door
  with a scalar resonance}},  {\em Eur. Phys. J.} {\bf C76} (2016), no.~3 116,
  [\href{http://arxiv.org/abs/1512.04929}{{\tt arXiv:1512.04929}}].

\bibitem{Ellis:2015oso}
J.~Ellis, S.~A.~R. Ellis, J.~Quevillon, V.~Sanz, and T.~You, {\it {On the
  Interpretation of a Possible $\sim 750$ GeV Particle Decaying into $\gamma
  \gamma$}},  {\em JHEP} {\bf 03} (2016) 176,
  [\href{http://arxiv.org/abs/1512.05327}{{\tt arXiv:1512.05327}}].

\bibitem{Bellazzini:2015nxw}
B.~Bellazzini, R.~Franceschini, F.~Sala, and J.~Serra, {\it {Goldstones in
  Diphotons}},  {\em JHEP} {\bf 04} (2016) 072,
  [\href{http://arxiv.org/abs/1512.05330}{{\tt arXiv:1512.05330}}].

\bibitem{Low:2015qep}
M.~Low, A.~Tesi, and L.-T. Wang, {\it {A pseudoscalar decaying to photon pairs
  in the early LHC Run 2 data}},  {\em JHEP} {\bf 03} (2016) 108,
  [\href{http://arxiv.org/abs/1512.05328}{{\tt arXiv:1512.05328}}].

\bibitem{Gupta:2015zzs}
R.~S. Gupta, S.~Jäger, Y.~Kats, G.~Perez, and E.~Stamou, {\it {Interpreting a
  750 GeV Diphoton Resonance}},  \href{http://arxiv.org/abs/1512.05332}{{\tt
  arXiv:1512.05332}}.

\bibitem{Kobakhidze:2015ldh}
A.~Kobakhidze, F.~Wang, L.~Wu, J.~M. Yang, and M.~Zhang, {\it {750 GeV diphoton
  resonance in a top and bottom seesaw model}},  {\em Phys. Lett.} {\bf B757}
  (2016) 92--96, [\href{http://arxiv.org/abs/1512.05585}{{\tt
  arXiv:1512.05585}}].

\bibitem{Bian:2015kjt}
L.~Bian, N.~Chen, D.~Liu, and J.~Shu, {\it {A hidden confining world on the 750
  GeV diphoton excess}},  \href{http://arxiv.org/abs/1512.05759}{{\tt
  arXiv:1512.05759}}.

\bibitem{Aloni:2015mxa}
D.~Aloni, K.~Blum, A.~Dery, A.~Efrati, and Y.~Nir, {\it {On a possible large
  width 750 GeV diphoton resonance at ATLAS and CMS}},
  \href{http://arxiv.org/abs/1512.05778}{{\tt arXiv:1512.05778}}.

\bibitem{Altmannshofer:2015xfo}
W.~Altmannshofer, J.~Galloway, S.~Gori, A.~L. Kagan, A.~Martin, and J.~Zupan,
  {\it {750 GeV diphoton excess}},  {\em Phys. Rev.} {\bf D93} (2016), no.~9
  095015, [\href{http://arxiv.org/abs/1512.07616}{{\tt arXiv:1512.07616}}].

\bibitem{Craig:2015lra}
N.~Craig, P.~Draper, C.~Kilic, and S.~Thomas, {\it {Shedding Light on Diphoton
  Resonances}},  {\em Phys. Rev.} {\bf D93} (2016), no.~11 115023,
  [\href{http://arxiv.org/abs/1512.07733}{{\tt arXiv:1512.07733}}].

\bibitem{Dev:2015vjd}
P.~S.~B. Dev, R.~N. Mohapatra, and Y.~Zhang, {\it {Quark Seesaw, Vectorlike
  Fermions and Diphoton Excess}},  {\em JHEP} {\bf 02} (2016) 186,
  [\href{http://arxiv.org/abs/1512.08507}{{\tt arXiv:1512.08507}}].

\bibitem{Dittmaier:2011ti}
{\bf LHC Higgs Cross Section Working Group} Collaboration, S.~Dittmaier et~al.,
  {\it {Handbook of LHC Higgs Cross Sections: 1. Inclusive Observables}},
  \href{http://arxiv.org/abs/1101.0593}{{\tt arXiv:1101.0593}}.

\bibitem{Jung:2015etr}
S.~Jung, J.~Song, and Y.~W. Yoon, {\it {How Resonance-Continuum Interference
  Changes 750 GeV Diphoton Excess: Signal Enhancement and Peak Shift}},  {\em
  JHEP} {\bf 05} (2016) 009, [\href{http://arxiv.org/abs/1601.00006}{{\tt
  arXiv:1601.00006}}].

\bibitem{Craig:2016iea}
N.~Craig, S.~Renner, and D.~Sutherland, {\it {Exploring Peaks and Valleys in
  the Diphoton Spectrum}},  \href{http://arxiv.org/abs/1607.06074}{{\tt
  arXiv:1607.06074}}.

\bibitem{Martin:2016bgw}
S.~P. Martin, {\it {Signal-background interference for a singlet spin-0 digluon
  resonance at the LHC}},  {\em Phys. Rev.} {\bf D94} (2016), no.~3 035003,
  [\href{http://arxiv.org/abs/1606.03026}{{\tt arXiv:1606.03026}}].

\bibitem{Aad:2015fna}
{\bf ATLAS} Collaboration, G.~Aad et~al., {\it {A search for $ t\overline{t} $
  resonances using lepton-plus-jets events in proton-proton collisions at $
  \sqrt{s}=8 $ TeV with the ATLAS detector}},  {\em JHEP} {\bf 08} (2015) 148,
  [\href{http://arxiv.org/abs/1505.07018}{{\tt arXiv:1505.07018}}].

\bibitem{Khachatryan:2015oqa}
{\bf CMS} Collaboration, V.~Khachatryan et~al., {\it {Measurement of the
  differential cross section for top quark pair production in pp collisions at
  $\sqrt{s} = 8\,\text {TeV} $}},  {\em Eur. Phys. J.} {\bf C75} (2015), no.~11
  542, [\href{http://arxiv.org/abs/1505.04480}{{\tt arXiv:1505.04480}}].

\bibitem{ATLAS-CONF-2016-014}
{\it {Search for heavy particles decaying to pairs of highly-boosted top quarks
  using lepton-plus-jets events in proton--proton collisions at $\sqrt{s} = 13$
  TeV with the ATLAS detector}},  Tech. Rep. ATLAS-CONF-2016-014, CERN, Geneva,
  Mar, 2016.

\bibitem{Kaplan:2008ie}
D.~E. Kaplan, K.~Rehermann, M.~D. Schwartz, and B.~Tweedie, {\it {Top Tagging:
  A Method for Identifying Boosted Hadronically Decaying Top Quarks}},  {\em
  Phys. Rev. Lett.} {\bf 101} (2008) 142001,
  [\href{http://arxiv.org/abs/0806.0848}{{\tt arXiv:0806.0848}}].

\bibitem{Rehermann:2010vq}
K.~Rehermann and B.~Tweedie, {\it {Efficient Identification of Boosted
  Semileptonic Top Quarks at the LHC}},  {\em JHEP} {\bf 03} (2011) 059,
  [\href{http://arxiv.org/abs/1007.2221}{{\tt arXiv:1007.2221}}].

\bibitem{Plehn:2010st}
T.~Plehn, M.~Spannowsky, M.~Takeuchi, and D.~Zerwas, {\it {Stop Reconstruction
  with Tagged Tops}},  {\em JHEP} {\bf 10} (2010) 078,
  [\href{http://arxiv.org/abs/1006.2833}{{\tt arXiv:1006.2833}}].

\bibitem{Cowan:2010js}
G.~Cowan, K.~Cranmer, E.~Gross, and O.~Vitells, {\it {Asymptotic formulae for
  likelihood-based tests of new physics}},  {\em Eur. Phys. J.} {\bf C71}
  (2011) 1554, [\href{http://arxiv.org/abs/1007.1727}{{\tt arXiv:1007.1727}}].
  [Erratum: Eur. Phys. J.C73,2501(2013)].

\bibitem{statistics}
{\it {Discovery sensitivity for a counting experiment with background
  uncertainty}},  2012.
\newblock \url{http://www.pp.rhul.ac.uk/~cowan/stat/medsig/medsigNote.pdf} and
  also talk slides
  \url{http://www-conf.slac.stanford.edu/statisticalissues2012/talks/glen_cowan_slac_4jun12.pdf}.

\bibitem{Bernreuther:2015fts}
W.~Bernreuther, P.~Galler, C.~Mellein, Z.~G. Si, and P.~Uwer, {\it {Production
  of heavy Higgs bosons and decay into top quarks at the LHC}},
  \href{http://arxiv.org/abs/1511.05584}{{\tt arXiv:1511.05584}}.

\bibitem{Tweedie:2014yda}
B.~Tweedie, {\it {Better Hadronic Top Quark Polarimetry}},  {\em Phys. Rev.}
  {\bf D90} (2014), no.~9 094010, [\href{http://arxiv.org/abs/1401.3021}{{\tt
  arXiv:1401.3021}}].

\bibitem{Djouadi:2016ack}
A.~Djouadi, J.~Ellis, and J.~Quevillon, {\it {Interference Effects in the
  Decays of 750 GeV States into $\gamma \gamma$ and $t\bar{t}$}},
  \href{http://arxiv.org/abs/1605.00542}{{\tt arXiv:1605.00542}}.

\bibitem{Hespel:2016qaf}
B.~Hespel, F.~Maltoni, and E.~Vryonidou, {\it {Signal background interference
  effects in heavy scalar production and decay to a top-anti-top pair}},
  \href{http://arxiv.org/abs/1606.04149}{{\tt arXiv:1606.04149}}.

\bibitem{ATLAS-CONF-2016-073}
{\bf ATLAS Collaboration} Collaboration, {\it {Search for heavy Higgs bosons
  A/H decaying to a top-quark pair in pp collisions at $\sqrt{s}=8$ TeV with
  the ATLAS detector}},  Tech. Rep. ATLAS-CONF-2016-073, CERN, Geneva, Aug,
  2016.

\bibitem{Patel:2015tea}
H.~H. Patel, {\it {Package-X: A Mathematica package for the analytic
  calculation of one-loop integrals}},  {\em Comput. Phys. Commun.} {\bf 197}
  (2015) 276--290, [\href{http://arxiv.org/abs/1503.01469}{{\tt
  arXiv:1503.01469}}].

\end{thebibliography}\endgroup

\end{document}